\documentclass[preprint]{aastex631}

\usepackage{tabularx}
\expandafter\let\csname singlespace\endcsname\relax
\expandafter\let\csname doublespace\endcsname\relax
\usepackage{amsmath}
\usepackage{threeparttable}
\usepackage{booktabs}
\usepackage{setspace}
\usepackage{indentfirst}
\usepackage{lipsum}
\usepackage{makecell}
\usepackage{graphics} 
\usepackage{color}
\usepackage{ulem}
\usepackage{mathrsfs}
\usepackage[figuresright]{rotating}

\begin{document}

\title{Constraints on Cosmic Rays Acceleration in Bright Gamma-ray Bursts with Observations of Fermi}

\author[0000-0001-6728-902X]{Xing-Fu Zhang}
\affiliation{School of Astronomy and Space Science, Nanjing University, Nanjing 210023, China}
\affiliation{Key laboratory of Modern Astronomy and Astrophysics (Nanjing University), Ministry of Education, Nanjing 210023, China}

\author[0000-0003-1576-0961]{Ruo-Yu Liu}
\affiliation{School of Astronomy and Space Science, Nanjing University, Nanjing 210023, China}
\affiliation{Key laboratory of Modern Astronomy and Astrophysics (Nanjing University), Ministry of Education, Nanjing 210023, China}

\author[0000-0001-6863-5369]{Hai-Ming Zhang}
\affiliation{School of Astronomy and Space Science, Nanjing University, Nanjing 210023, China}
\affiliation{Key laboratory of Modern Astronomy and Astrophysics (Nanjing University), Ministry of Education, Nanjing 210023, China}

\author[0000-0002-6036-985X]{Yi-Yun Huang}
\affiliation{School of Astronomy and Space Science, Nanjing University, Nanjing 210023, China}
\affiliation{Key laboratory of Modern Astronomy and Astrophysics (Nanjing University), Ministry of Education, Nanjing 210023, China}

\author[0000-0003-2478-333X]{B. Theodore Zhang}
\affiliation{Center for Gravitational Physics and Quantum Information, Yukawa Institute for Theoretical Physics, Kyoto University, Kyoto 606-8502, Japan}

\author[0000-0002-5881-335X]{Xiang-Yu Wang}
\affiliation{School of Astronomy and Space Science, Nanjing University, Nanjing 210023, China}
\affiliation{Key laboratory of Modern Astronomy and Astrophysics (Nanjing University), Ministry of Education, Nanjing 210023, China}

\correspondingauthor{Ruo-Yu Liu}
\email{ryliu@nju.edu.cn}

\begin{abstract}
Gamma-ray bursts (GRBs) are widely suggested as potential sources of ultrahigh-energy cosmic rays (UHECRs). The kinetic energy of the jets dissipates, leading to the production of an enormous amount of $\gamma$-ray photons and possibly also the acceleration of protons. The accelerated protons will interact with the radiation of the GRB via the photomeson and Bethe-Heitler processes, which can initiate electromagnetic cascades. This process can give rise to broadband radiation up to the GeV-TeV $\gamma$-ray regime. The expected $\gamma$-ray flux from cascades depends on properties of the GRB jet, such as the dissipation radius $R_{\rm diss}$, the bulk Lorentz factor $\Gamma$, and the baryon loading factor $\eta_p$. Therefore, observations of Fermi-LAT can impose constraints on these important parameters. In this study, we select 12 GRBs of high keV-MeV fluence and constrain the baryon loading factor, under different combinations of the bulk Lorentz factor and the dissipation radius based on Fermi-LAT's measurements. Our findings indicate a strong constraint of $\eta_p<10$ for most selected GRBs over a large parameter space except for large dissipation radii ($\gtrsim 10^{15}\rm cm$) and high bulk Lorentz factors ($\gtrsim 600$). The constraint is comparable to, and in some GRBs even stronger than, that from high-energy neutrinos for stacked GRBs. Our results suggest that for typical bulk Lorentz factor of several hundreds, the dissipation radii of GRBs need be large to avoid overshooting the GeV gamma-ray flux during the prompt emission phase of GRBs, which can be used to constrain GRBs.
\end{abstract}

\section{Introduction} \label{sec:intro}
Ultrahigh-energy cosmic rays (UHECRs) are cosmic rays with energies exceeding $10^{18}~\rm eV$. Despite extensive research, the origin of UHECRs remains an unresolved problem. The Hillas criterion suggests that potential candidates for UHECRs must have a sufficiently large product of magnetic field strength and size. Gamma-ray bursts (GRBs), active galactic nuclei (AGNs), and magnetars are among the best candidates \citep{hillas1984origin}. 

Gamma-ray bursts (GRBs) are astronomical events characterized by a sudden gigantic enhancement of $\gamma$-ray flux over a short period in a certain direction of the sky. GRBs typically occur in extreme astrophysical environments and produce ultra-relativistic jets with isotropic equivalent energies of approximately $10^{51}-10^{55}~\rm erg$ \citep{paczynski1998gamma,macfadyen1999collapsars}. 
It has been proposed that within the dissipation region of GRBs, such as internal collisions and proton/nuclei in the relativistic jet, can undergo acceleration to achieve ultrahigh energies. The underlying mechanism responsible for this acceleration is the diffusive shock acceleration, also known as Fermi acceleration. The energy production rate of GRBs is derived from $Q^{\text{GRBs}}_{\text{UHECR}}=10^{44}(\frac{\eta_{p}}{10})(\frac{f_{\text{bol}}}{0.1})(\frac{\rho_{0}}{1\,\text{Gpc}^{-3} \text{yr}^{-1}})(\frac{\mathcal{E}_{\text{iso},\gamma}}{10^{53}\,\text{erg}})\,\text{erg}\,\text{Mpc}^{-3}\,\text{yr}^{-1}$, where $\eta_{p}$ is the baryon loading factor and it is determined as the ratio of the total CR energy budget ($E_{p,\text{tot}}$) to that of the photon ($E_{\gamma,\text{tot}}$). $f_{\text{bol}}$ is the bolometric correction factor, whose value is approximately $0.1-0.2$ for a CR injection spectrum of type $E^{-2}$ \citep{2022arXiv220206480Kimura,ai2023model}. $\mathcal{E}_{\text{iso},\gamma}$ is the average isotropic $\gamma$-ray energy budget and $\Bar{\rho}_{0}$ is the average observed GRB rate. In addition, the rate of CR energy production required to explain the measured flux beyond the ankle ($\sim5\times10^{18}\,\mathrm{eV}$) is $Q^{\text{obs}}_{\text{UHECRs}}\sim 5 \times 10^{44}$\,erg\,Mpc$^{-3}$\,yr$^{-1}$ for a mixed composition \citep{2017JCAP...04..038A, 2020PhRvL.125l1106A, 2021PhRvD.104d3017J, AbdulHalim:20232A}. Consequently, if GRBs are the main factories of UHECRs, $f_{\text{bol}}\eta_{p}\ge1$ would be necessary. \citep{waxman1995cosmological,murase2008high,milgrom1995possible,2021PhRvD.104j3005Zhang}

If protons (or nuclei) are accelerated to ultrahigh energies (UHEs) in a GRB, they will interact with the intense radiation of the GRB through the photomeson and Bethe-Heitler (BH) processes. These processes produce (anti-)neutrinos and electromagnetic particles (EM), such as $\gamma$-ray photons and $e^{+}$$e^{-}$ pairs, resulting in observable neutrino spectra (ranging from TeV to EeV) and EM cascade spectra (ranging from keV to TeV). The characteristics of these spectra depend on the parameters of the GRB, such as the dissipation radius ($R_{\rm diss}$), the bulk Lorentz factor ($\Gamma_{\rm bulk}$), and $\eta_{p}$ \citep{asano2009hadronic,wang2018hadronic,2023SCPMAWang}. Consequently, the observations can be used to constrain these parameters. One such constraint arises from the neutrino observation of stacked GRBs. For example, based on five years of GRB data, the IceCube collaboration \citep{aartsen2017extending} found that for a typical bulk Lorentz factor $\Gamma_{\text{bulk}}\sim300$, the values of $\eta_{p}$ should not exceed 3, 2, and 80 for the internal shock (IS) model, photospheric fireball model, and internal collision-induced magnetic reconnection and turbulence model (ICMART), respectively. Furthermore, \cite{lucarelli2023neutrino} presented a stacking scheme based on physically motivated GRB weights with IceCube's data and found that $\eta_{p}$ is typically less than 10 for GRB prompt emissions.

Another constraint arises from observations of individual GRBs by the Fermi Gamma-Ray Space Telescope (Fermi). The observations made by the Large Area Telescope (LAT) and the Gamma-ray Burst Monitor (GBM) have revealed the multiband $\gamma$-ray behavior of GRBs \citep{abdo2009wb,meegan2009fermi,ackermann2011detection}. The spectrum of these GRBs, ranging from keV to MeV, is typically characterized by the Band function \citep{band1993batse}. In addition to this major spectral component, there is an additional GeV spectral component, usually manifesting itself as a power-law function, appearing in bright GRBs \citep{ajello2019, tang2021prevalence}. The origin of this additional component has not yet been fully determined. It could potentially originate from prompt emission electrons that undergo inverse Compton (IC) radiation, early afterglow radiation, or a proton-induced EM cascade \citep{gupta2007neutrino,kumar2009grb,beloborodov2014origin,wang2018hadronic,zhang2023external}. Nevertheless, the measured flux at the GeV band can be considered as the upper limit for any of the aforementioned components. Recently, \cite{liu2023constraints} derived the emission of EM cascades in GRB~221009A, which is the brightest GRB of all time \citep{2023ApJFrederiks,2023ApJBurns,2023ApJLesageGBM}. They demonstrated that the constraint obtained from GeV observations is more stringent than that obtained from the neutrino analysis in this GRB. Furthermore, they predicted that GeV observations can offer independent constraints on the GRB model, particularly for those with high keV-MeV fluxes. While GRB~221009A could be a unique case, the aim of this paper is to apply this method to other bright GRBs to investigate a more general constraint on key parameters of GRBs based on GeV observations, and to dig into the issue of GRBs as UHECR accelerators.

The rest of the paper is structured as follows: In Section \ref{sec:sample}, we introduce selected GRB samples and data analysis, and describe the EM cascade model in Section \ref{sec:description}. The results are shown in Section \ref{sec:result}, and we discuss the influence of some parameters in Section \ref{sec:diss}. Finally, our conclusions are summarized in Section \ref{sec:conclusion}.

\section{Sample and Data Analysis}\label{sec:sample}
In this study, we have selected several bright GRBs from the GBM \footnote{\url{https://heasarc.gsfc.nasa.gov/W3Browse/fermi/fermigbrst.html}} and LAT \footnote{\url{https://heasarc.gsfc.nasa.gov/W3Browse/fermi/fermilgrb.html}} catalogs. There are a total of 3601 GRBs identified in the GBM catalog as of August 28, 2023, and 231 GRBs recorded in the LAT catalog as of June 22, 2022. We have chosen GRBs that meet the following criteria: (i) the redshift of the GRB is measured; (ii) the GRB is within the LAT's field of view during the same time interval; (iii) the GRB has a high keV-MeV fluence during the prompt phase so that we have sufficient statistics for the subsequent studies. In total, we selected 12 GRBs of the highest fluence, including GRB~221009A, GRB~211018A, GRB~190530A, GRB~190114C, GRB~180720B, GRB~170214A, GRB~160821A, GRB~160625B, GRB~160509A, GRB~131231A, GRB~130427A and GRB~090902B.

The keV\,--\,MeV spectrum of GRBs during prompt phase can be modeled by the Band function \citep{band1993batse}: 
\begin{equation}
\begin{aligned}
N\left(E_{\gamma}\right)=\left\{\begin{array}{l}
A_{\text{nor}}\left(\frac{E_{\gamma}}{100\,\mathrm{keV}}\right)^{\alpha} \exp \left(\frac{E_{\gamma}}{-E_{\text{peak}}}\right), E_{\gamma}\le E_{b}, \\
A_{\text{nor}}(\frac{E_{b}}{100\,\mathrm{keV}})^{\alpha-\beta}\exp(\beta-\alpha)\left(\frac{E_{\gamma}}{100\,\mathrm{keV}}\right)^{\beta}, E_{\gamma}> E_{b},\\
\end{array}\right.
\label{eq:Band}
\end{aligned}
\end{equation}
where $\alpha$ and $\beta$ are the indexes for low-energy and high-energy, respectively. The peak energy $E_{\text{peak}}$ is defined as $E_{\text{peak}}$=$E_b/(\alpha-\beta)$. The normalized coefficient $A_{\text{nor}}$ is given by $A_{\text{nor}}=\Gamma_{\text{bulk}}^{2}U_{\gamma}/[\int_{E_{\gamma,\text{min}}}^{E_{\gamma,\text{max}}}N(E_{\gamma})E_{\gamma}dE_{\gamma}] $, where $U_{\gamma}=L_{\gamma}/(4\pi R_{\text{diss}}^{2}\Gamma_{\text{bulk}}^{2}c)$ is the energy density of the photon in the comoving frame, $L_{\gamma}$ is the photon luminosity integrated from $E_{\gamma,\text{min}}$=10 keV to $E_{\gamma,\text{max}}$=10 MeV with an exponential cut-off at 50 MeV. 

We use GRB~160821A as an example to demonstrate the details of the Fermi data analysis, while the analyzes of other GRBs can be found in the Appendix \ref{sec:LAT}. GBM was triggered by GRB~160821A at 20:34:30.04 UT. The GBM light curve showed a noticeable precursor emission starting from the trigger time $T_{0}$ followed by an extremely intense emission episode lasting 43\,s ($T_{90}$ ). The start time of $T_{90}$ is 118.5~s. The time interval of main flare was defined by the following the researches \citep{stanbro2016grb,arimoto2016grb,sharma2019time}. We downloaded the GBM data for this GRB from the Fermi-GBM public data archive \footnote{\url{https://heasarc.gsfc.nasa.gov/FTP/fermi/data/gbm/daily/}}. We used the Time-Tagged Event data acquired from two NaI detectors (n6 and n7) and one BGO detector (b1). Our analysis took into account the spacecraft geometry and the precise instrument viewing angles with respect to the burst location. We fitted the $\nu F_{\nu}$ spectrum for time interval, from $T_{0}+118.5$ to $T_{0}+161.5$, using the Band function. Through this approach, we obtain the corresponding flux at energies ranging from 10\,keV to 10\,MeV, denoted by $F_{\text{GBM}}$.

LAT is a pair conversion telescope that observes photons from 20 MeV to $>$300 GeV with a 2.4-steradian field of view \citep{ackermann20162fhl}.
The LAT extended-type data for bursts were taken from the Fermi Science Support Center \footnote{\url{https://fermi.gsfc.nasa.gov}}.
We performed an unbinned maximum likelihood analysis on LAT data to characterize the high-energy emission associated with GRB~160821A. We focus on a 14$^{\circ}$×14$^{\circ}$ region of interest (ROI) centered on GRB~160821A. For this purpose, we used data from the Pass 8 (P8R3) LAT processed using the Conda fermitools v2.2.0 package. To ensure data quality, we specifically selected events that fall into the ``transient'' class ($\emph{\text{P8R3\_TRANSIENT020\_V3}}$). To model the high-energy emission spectrum at GeV energies, we adopted a power law function with index, $\Gamma_{\text{LAT}}$, ranging from 0.1 to 10 GeV. We took into account potential contributions from both diffuse Galactic and extragalactic backgrounds. We considered an intensity level with a 95\% C.L. upper limit of $\rm F^{\rm UL}_{\rm LAT}=1.3\times10^{-7}~\text{erg cm}^{-2}~\text{s}^{-1}$.

\begin{figure}[htb]
\centering
\includegraphics[width=0.32\textwidth]{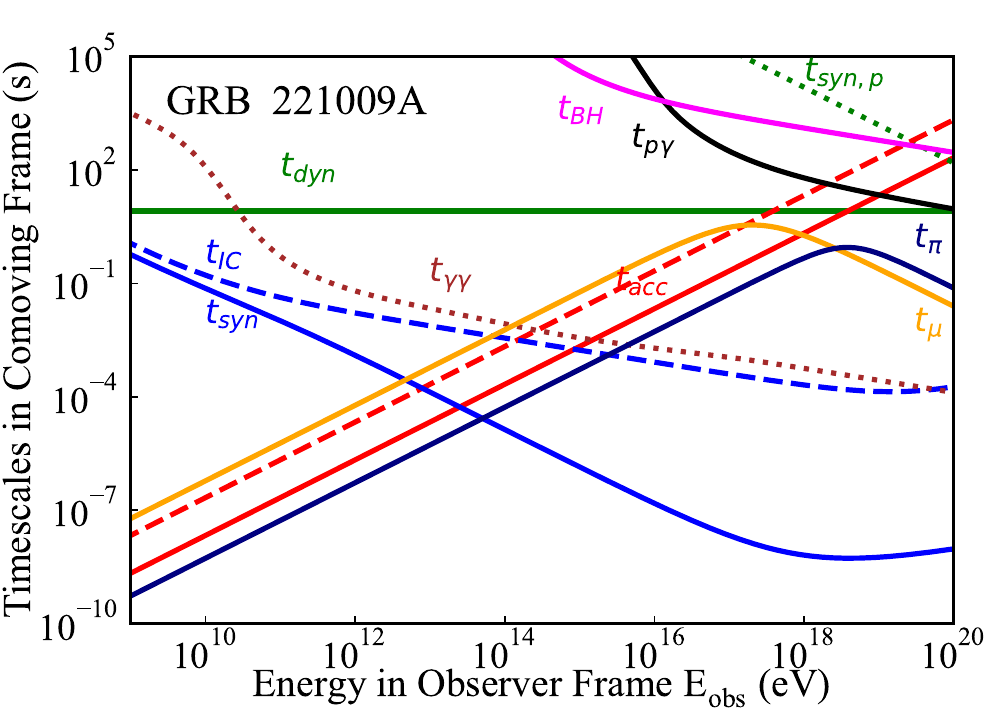}
\includegraphics[width=0.32\textwidth]{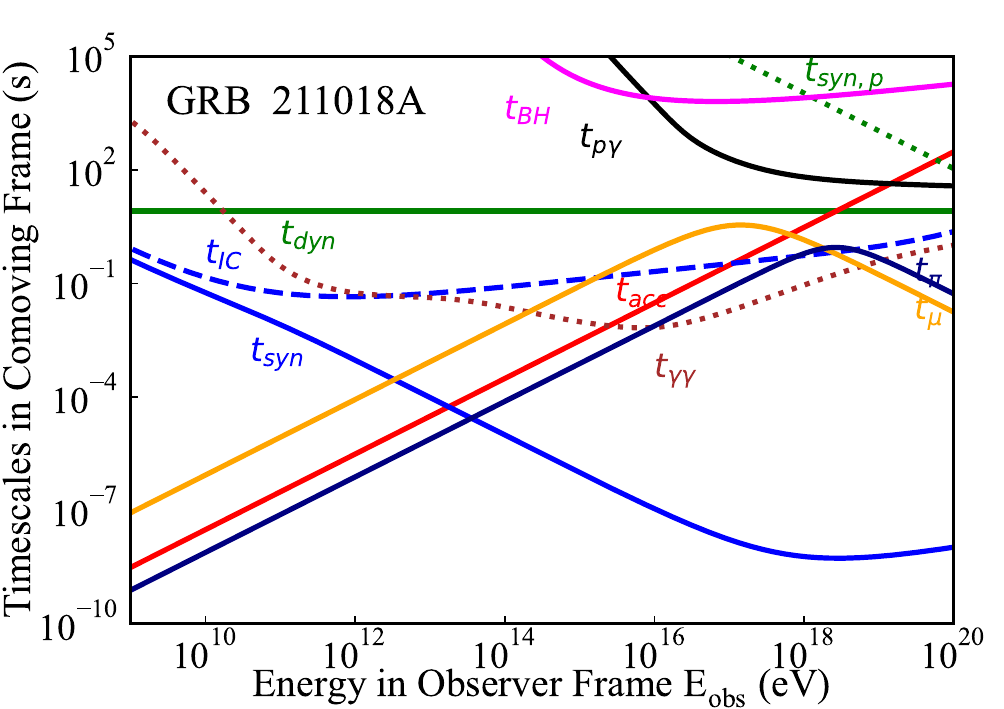}
\includegraphics[width=0.32\textwidth]{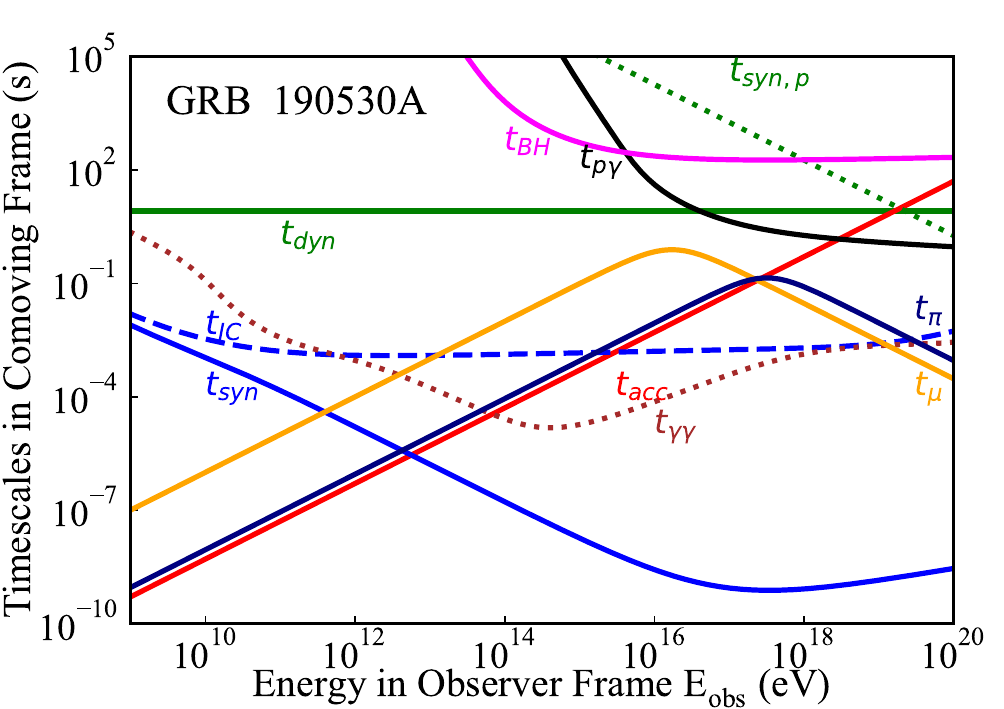}
\includegraphics[width=0.32\textwidth]{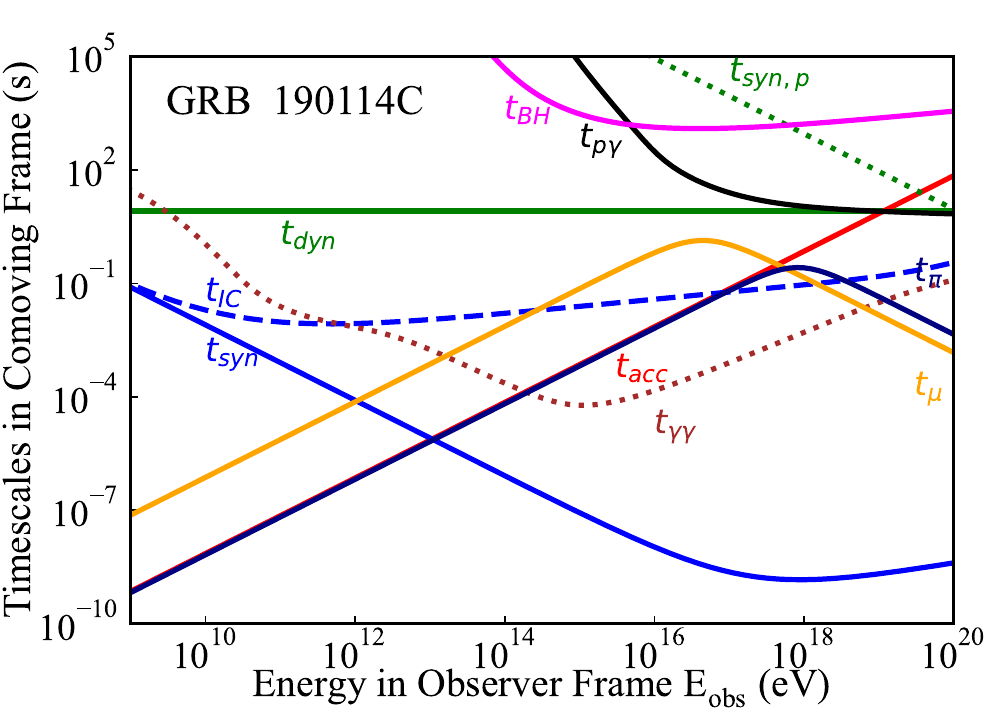}
\includegraphics[width=0.32\textwidth]{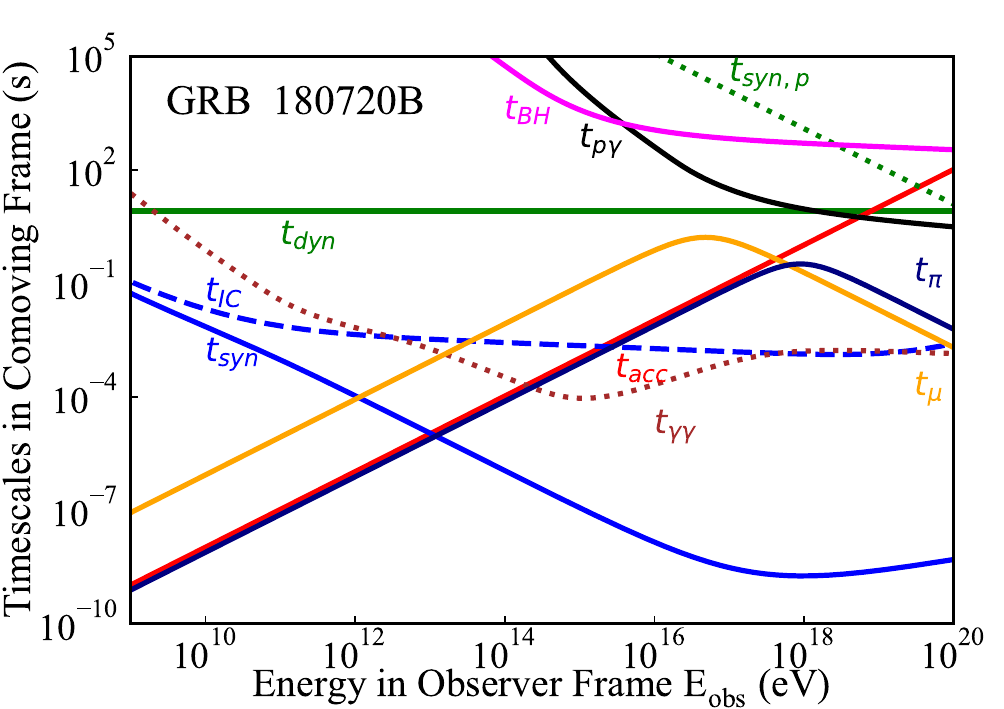}
\includegraphics[width=0.32\textwidth]{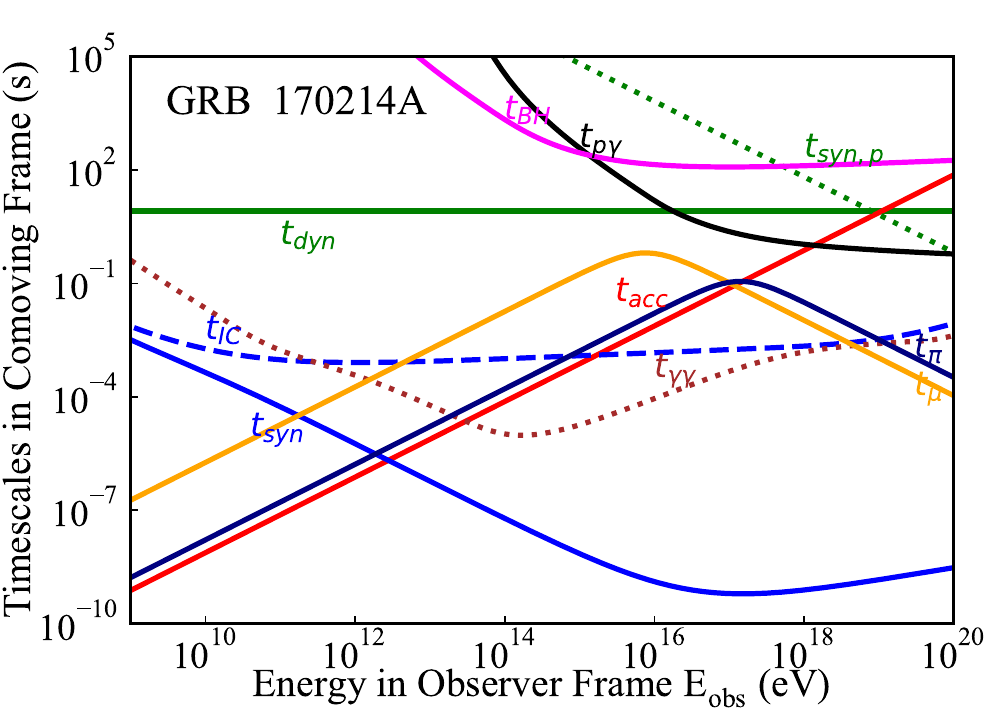}
\includegraphics[width=0.32\textwidth]{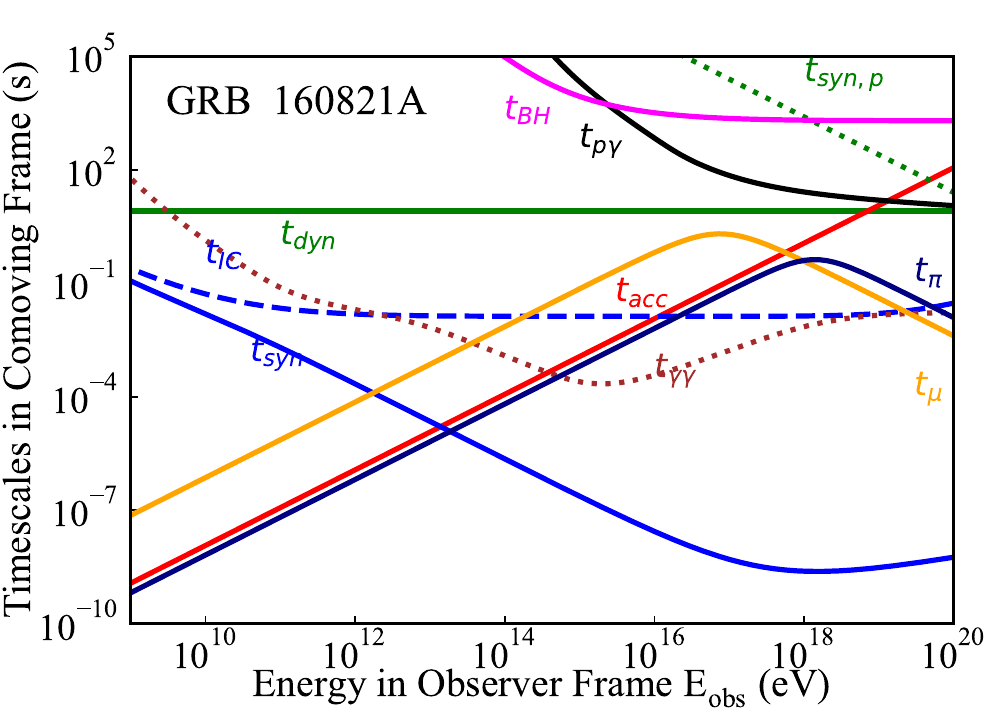}
\includegraphics[width=0.32\textwidth]{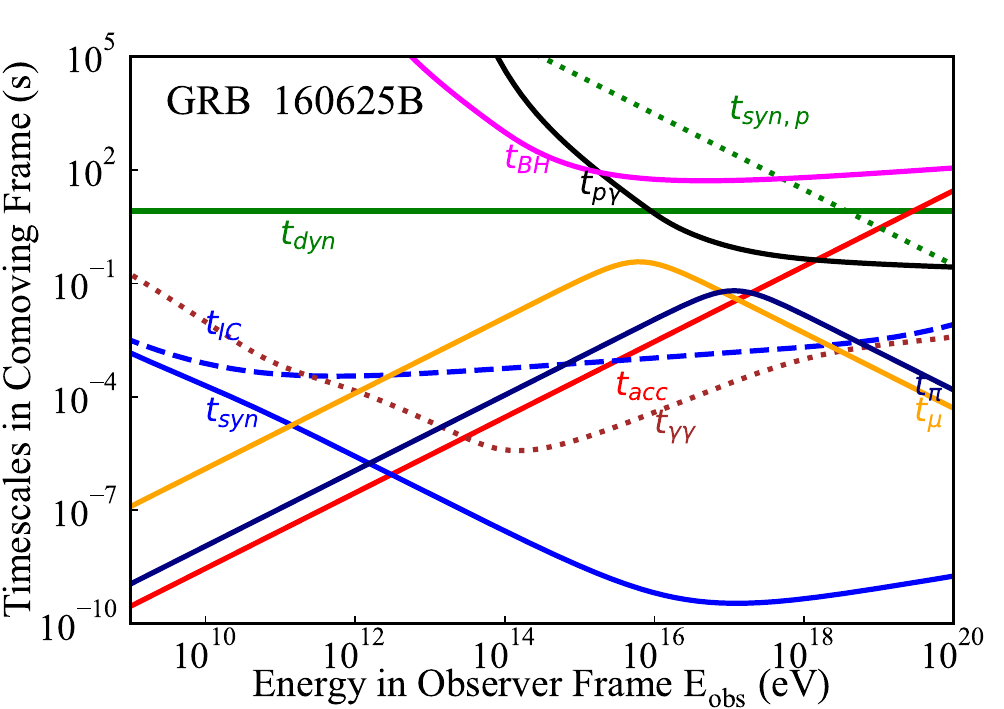}
\includegraphics[width=0.32\textwidth]{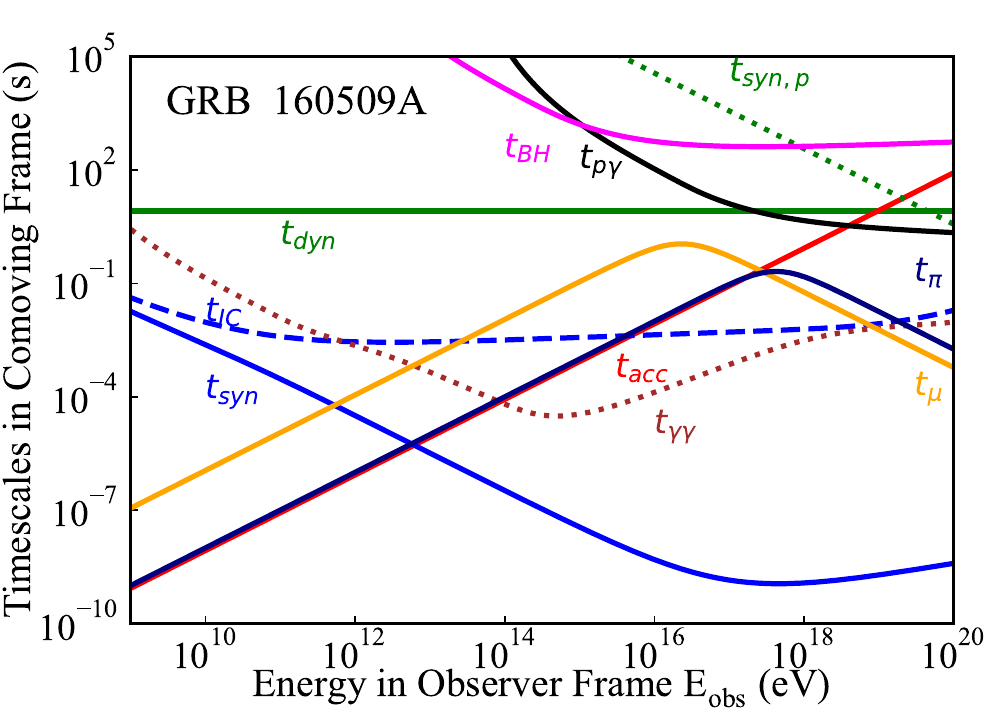}
\includegraphics[width=0.32\textwidth]{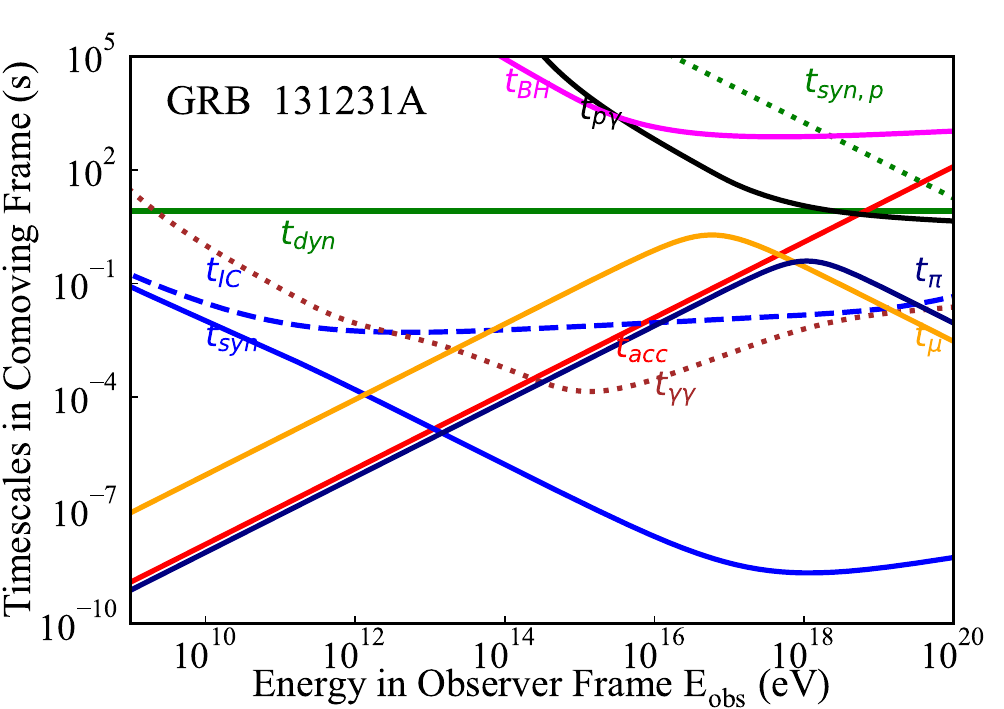}
\includegraphics[width=0.32\textwidth]{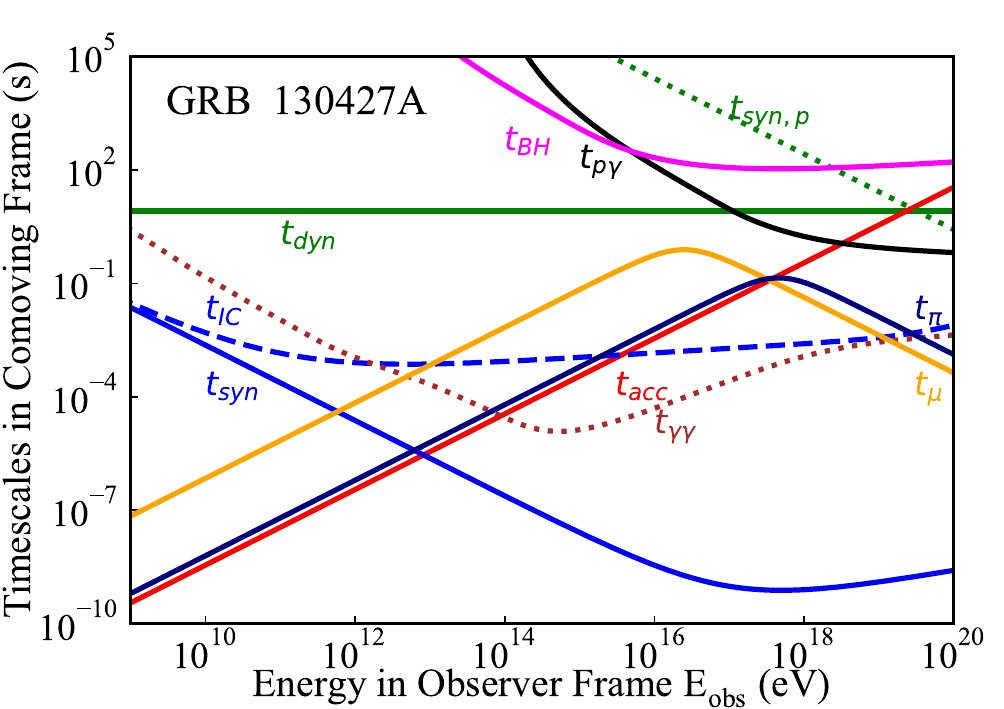}
\includegraphics[width=0.32\textwidth]{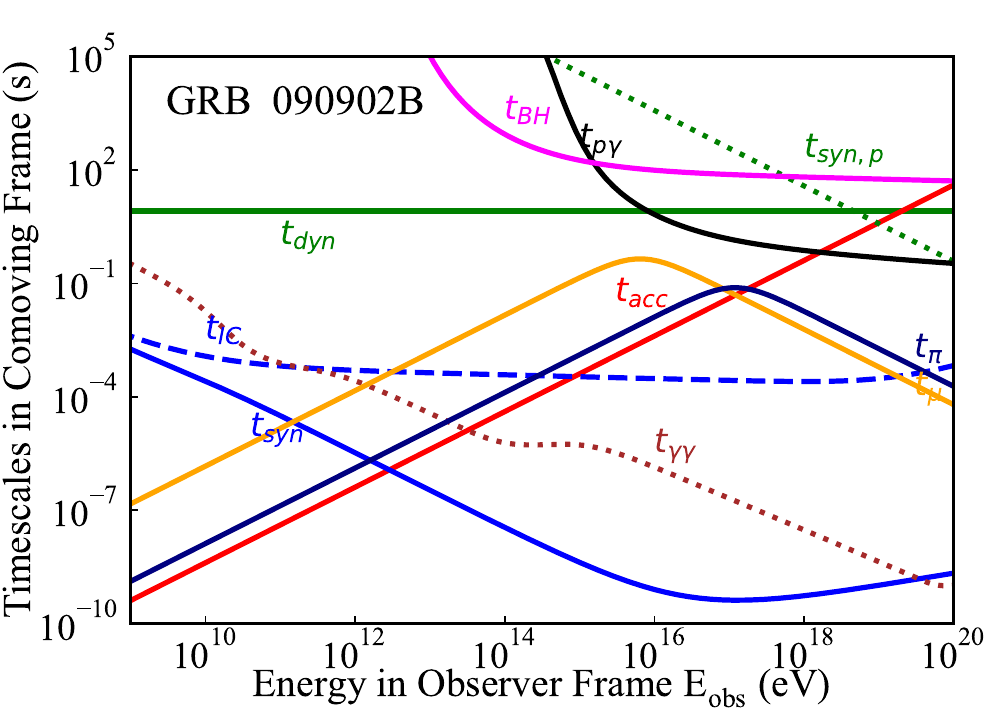}
\caption{Timescales in the comoving frame of various processes. Note that the results of GRB~221009A and GRB 130427A are from the $T_{0}+[177,~219]~\rm s$ and $T_{0}+[11.4,~18]~\rm s$, respectively. The black and pink solid curves represent the energy loss timescales of protons via the photomeson and BH processes, respectively. The blue solid and dashed curves show the cooling timescales of electrons via synchrotron and IC radiation, respectively. The dotted brown curves show the timescale of the fifth generation $\gamma\gamma$ absorption. The solid green curves present the dynamical timescale of the GRBs. The navy and orange solid curves are the cooling timescales of pion and muon, respectively. The red soiled and dashed curves are the acceleration timescales for $\xi=10\%$, and $\xi=1\%$, respectively. We plot all panels in the reference case. The best fitting parameters of the Band function and the corresponding luminosity for GRBs are shown in Table~\ref{tab:fermi}.}
\label{fig:timescale}
\end{figure} 

\section{DESCRIPTIONS OF MODELS}\label{sec:description}

To simulate the EM cascade emission initiated by UHE protons during the prompt emission phase of GRBs, we have incorporated six physical processes as outlined in \cite{asano2009hadronic}. These processes include: (i) synchrotron radiation that encompasses synchrotron self-absorption (SSA) and inverse Compton scattering (IC) for electron/positron pairs; (ii) $\gamma\gamma$ annihilation leading to pair production; (iii) radiation from proton synchrotron; (iv) photomeson process; (v) BH process; (vi) radiation from synchrotron and decay processes of pions and muons. Quantum effects at extremely high energies are considered in the synchrotron radiation processes for all particles \citep{Aharonian2003, Wangj2018}. As a starting point, a set of standard GRB parameters is used to establish a reference scenario. These reference parameters are detailed in Table~\ref{tab:reference}, and the timescales for the processes mentioned are shown in Figure~\ref{fig:timescale}. Our model is discussed in more detail as follows.

We assume a proton injection spectrum\footnote{We do not consider acceleration of heavier nuclei during the prompt emission phase of GRBs because they cannot survive the photo-disintegration process in the intense radiation field of these GRBs \citep{WangXY08, murase2008high}.} of $Q_{p}(E_{p})=Q_{0}E_{p}^{-s_{p}}$exp($-E_{p}/E_{p,\text{max}}$) in the comoving frame, where $Q_{0}$ is the normalization factor. This factor is related to photon luminosity by $\Gamma_{\text{bulk}}^{2}\int_{E_{p,min}}^{E_{p,\text{max}}}E_{p}Q_{p}(E_{p})dE_{p}=\eta_{p}L_{\gamma}$. Here, $E_{p,\text{min}}$ is set to 1 GeV, slightly higher than the rest-frame proton energy. The maximum proton energy, $E_{p,\text{max}}$ is determined by the balance between acceleration and cooling or escape processes. More specifically, we determine $E_{p,\rm max}$ by setting $t_{\text{acc}}^{-1}=t_{\text{dyn}}^{-1}+t_{p\gamma}^{-1}+ t_{\text{BH}}^{-1}+t_{\text{syn},p}^{-1}$, where $t_{\text{dyn}}$ represents the dynamical timescale or adiabatic cooling timescale, given by $t_\text{dyn}=R_{\text{diss}}/(\Gamma_{\text{bulk}}c)$, whereas $t_{p\gamma}$, $t_{\text{BH}}$ and $t_{\text{syn},p}$ denote the energy loss timescale via the photomeson process, the BH process and the proton synchrotron, respectively. Phenomenologically, the acceleration time in the comoving frame can be written as $t_\text{acc}=\frac{20}{3}\xi^{-1} E_{p}/(ZeB_{\mathrm{jet}}c)$, where $\xi(\le1)$ is the acceleration efficiency, $B_{\mathrm{jet}}$ represents magnetic field in the jet and $Z$ is the charge, which is set to 1 for the pure proton in this work. We estimate that the magnetic field in comoving frame is approximately $B_{\mathrm{jet}}\simeq6.5\times10^{4}\,\text{G}(\frac{\chi_BL_{\gamma}}{10^{53}~\text{erg}/\text{s}})^{1/2}(\frac{R_\text{diss}}{10^{14}~\text{ cm}})^{-1}(\Gamma_{\text{bulk}}/400)^{-1}$, where $\chi_B$ indicates the ratio between the energy attributed to the magnetic field and that to nonthermal electrons. With reference parameters as listed in Table~\ref{tab:reference}, we see that the maximum protons energy is primarily determined by the photomeson process (represented by the black solid curves in Figure~\ref{fig:timescale}), while the adiabatic cooling becomes more important if the acceleration efficiency is relatively small (see the top-left panel of Figure~\ref{fig:timescale}). Pions are produced in the photomeson process. Neutral pions decay into photons, while charged pions decay into muons and neutrinos. Muons will further decay into electrons/positrons and neutrinos. Charged pions and muons will also radiate via the synchrotron radiation in the magnetic field of GRBs before they decay. The timescales for pions and muons are given by $t_{\pi/\mu}^{-1}=t_{\pi/\mu,\text{syn}}^{-1}+t_{\pi/\mu,\text{dec}}^{-1}+t_\text{dyn}^{-1}$. Here, $t_{\pi/\mu,\text{syn}}$ represents the synchrotron cooling timescales \footnote{$t_{\pi/\mu,\text{syn}}=7.7\times10^{8}\left(\frac{E_{\pi/\mu}}{m_{\pi/\mu}c^{2}B_{\rm jet}^{2}}\right)(m_{\pi/\mu}/m_{e})^{3}\left[1+\left(\frac{E_{\pi/\mu}B_{\rm jet}}{m_{\pi/\mu}c^{2}B_{\text{cri},\pi/\mu}}\right)^{2/3}\right]^{2}~\rm s$, where $m_{\pi/\mu}$ is the masses of pions or muons and $B_{\text{cri},\pi/\mu}$ is the critical magnetic field of pions or muons.}, and $t_{\pi/\mu,\text{dec}}$ denotes the decay time \footnote{$t_{\pi/\mu,\text{dec}}=\frac{E_{\pi/\mu}\tau_{\pi/\mu}}{m_{\pi/\mu }c^{2}}~\rm s$, where $\tau_{\pi/\mu}$ is the lifetimes of pions or muons in the rest frame.}. We see the breaks in the timescales of pions and muons in  Figure~\ref{fig:timescale}, as represented by the navy and orange solid curves, respectively. Decays of these particles are dominant processes before breaks while the synchrotron loss become important above breaks. The IC cooling timescales of muons and pions are negligible at the energies after breaks due to the KN effect so we do not consider them in our calculation.

\begin{table}
\addtolength{\tabcolsep}{0pt}
    \centering
    \caption{The reference parameters in this work.}
    \begin{tabular}{CCC}
    \hline
    \hline
        \text{Descriptions}&\text{Parameters} & \text{Values}  \\ 
        \hline
        \text{bulk Lorentz Factor}&$\Gamma_{\rm bulk}$ & 400\\
        \text{Dissipation radius}&$R_\text{diss}$& 10^{14}\,\text{cm}\\
        \text{The index of proton injection spectrum}&$s_{p}$&2\\
        \text{Acceleration efficiency}&$\xi$&10\%\\
        \text{The ratio between the energy attributed to the}&$\chi_{B}$&1\\
        \text{magnetic field and non-thermal electron}&&\\
        \text{The baryon loading factor}&$\eta_{p}$&10\\
    \hline
    \end{tabular}
    \label{tab:reference}
\end{table}

\begin{figure}[htb]
\centering
\includegraphics[width=0.6\textwidth]{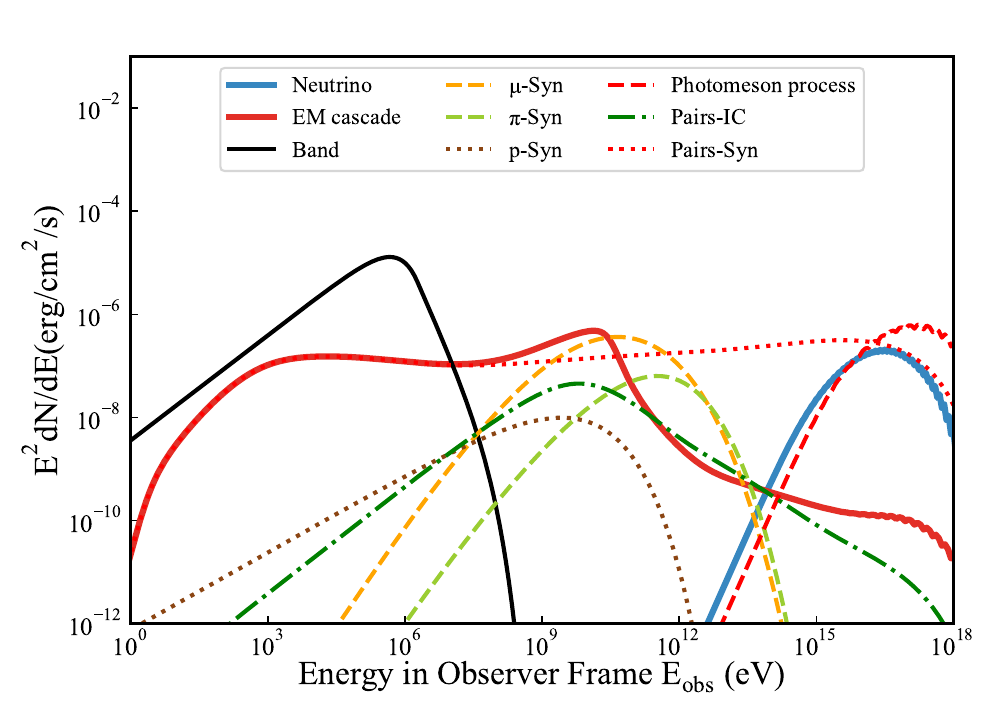}
\caption{The example of GRB 221009A during $T_{0}+[177,~219]~\rm s$ in the case of reference parameters. The thick red and blue curves are the predicted spectral energy distributions (SEDs) of the total EM cascade emission and the high-energy neutrino emission with $R_{\text{diss}}=10^{14}\,\text{cm}$, and $\Gamma_{\text{bulk}}=400$. The black curves represent the SED of the Band component with $\alpha=-1.32\,,\beta=-3.98\,,E_{\text{peak}}=0.8\,\text{MeV}\,,L_{\gamma}=2.9\times10^{51}$~erg/s. Synchrotron radiation and IC radiation of pairs generated in the cascade (before considering $\gamma\gamma$ absorption) are shown with dotted red curve and dash-dotted green curve respectively. Synchrotron radiation of muons, pions and protons are represented by dashed yellow, dashed green and dotted brown curves, respectively. The red dashed curve represents the photon flux from the photomeson process (before considering $\gamma\gamma$ absorption).} 
\label{fig:sample}
\end{figure}

 Since the electron cooling timescale is much shorter than the dynamical timescale, the EM cascade process has developed sufficiently and we may consider a quasi-stable state reached in the cascade process. We consider the one-zone model (i.e., the same zone for particle acceleration and emission) and employ the same method described in \citet{liu2020neutrino}, but additionally take into account the radiations of muons and pions, as well as the synchrotron self-absorption (SSA) of electrons. The secondary products in the BH process are calculated following \citet{kelner2008energy} and the photomeson process are calculated with SOPHIA\citep{Mucke2000}. We sum up the photon flux of the first five generations of the EM cascade, which is sufficient to describe the radiation spectrum generated in cascade. The cascade radiation is also included as the target radiation field for relevant processes such as $\gamma\gamma$ annihilation and the IC process. We present the corresponding EM cascade spectrum as well as the neutrino spectrum of GRB~221009A during $T_{0}+[177,~219]~\rm s$ in Figure~\ref{fig:sample} for reference. The dotted red and green curves represent the synchrotron and IC emission of electron/positron pairs. The dotted brown, dashed orange, and dashed green curves represent the synchrotron radiation from protons, pions, and muons, respectively. The solid red curve shows the final cascade spectrum including the SSA effect and the intrinsic $\gamma\gamma$ annihilation, which reduce the flux below $\sim 100~\rm eV$ and above $\sim 10~\rm GeV$ respectively.

\section{Results} \label{sec:result}
\subsection{The predicted SEDs for selected GRBs} \label{sec:predicted SEDs}

\begin{figure}[htb]
\centering
\includegraphics[width=0.32\textwidth]{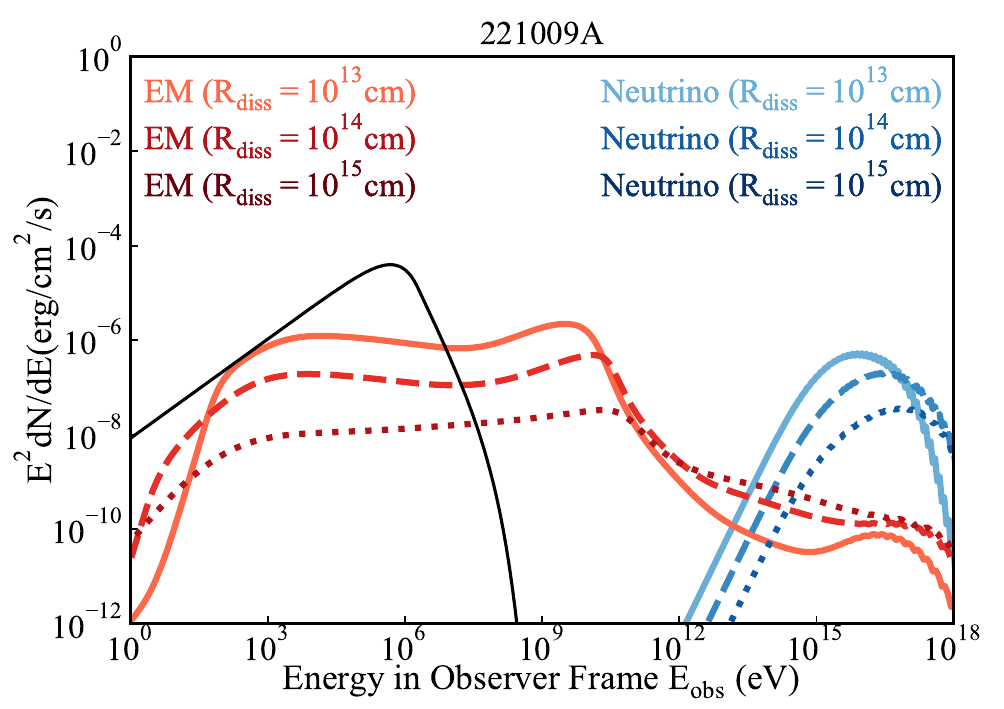}
\includegraphics[width=0.32\textwidth]{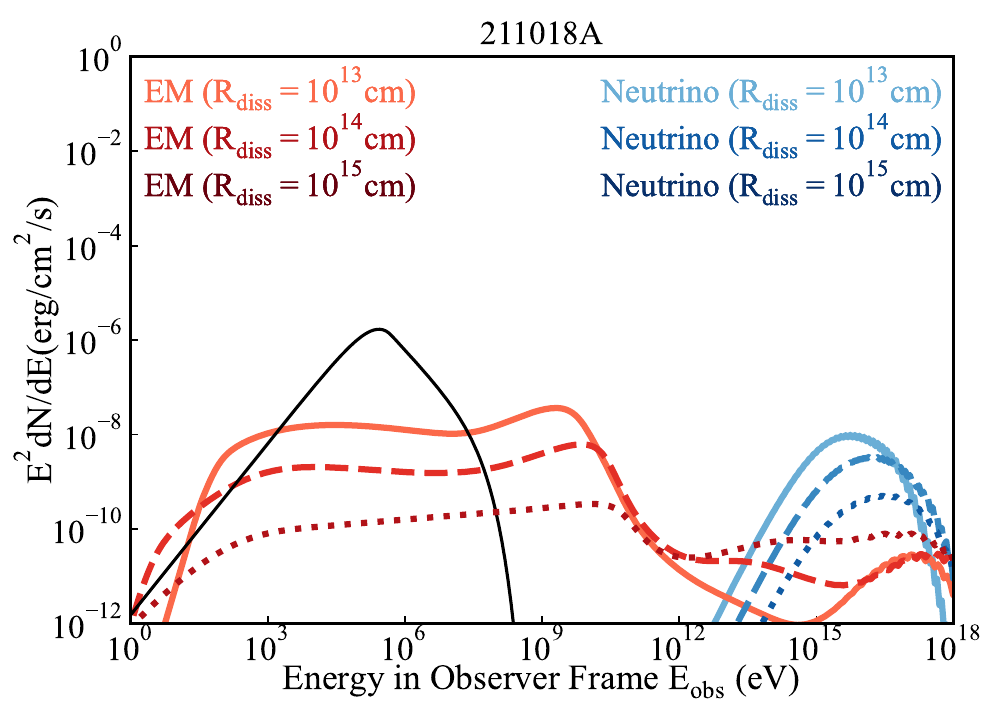}
\includegraphics[width=0.32\textwidth]{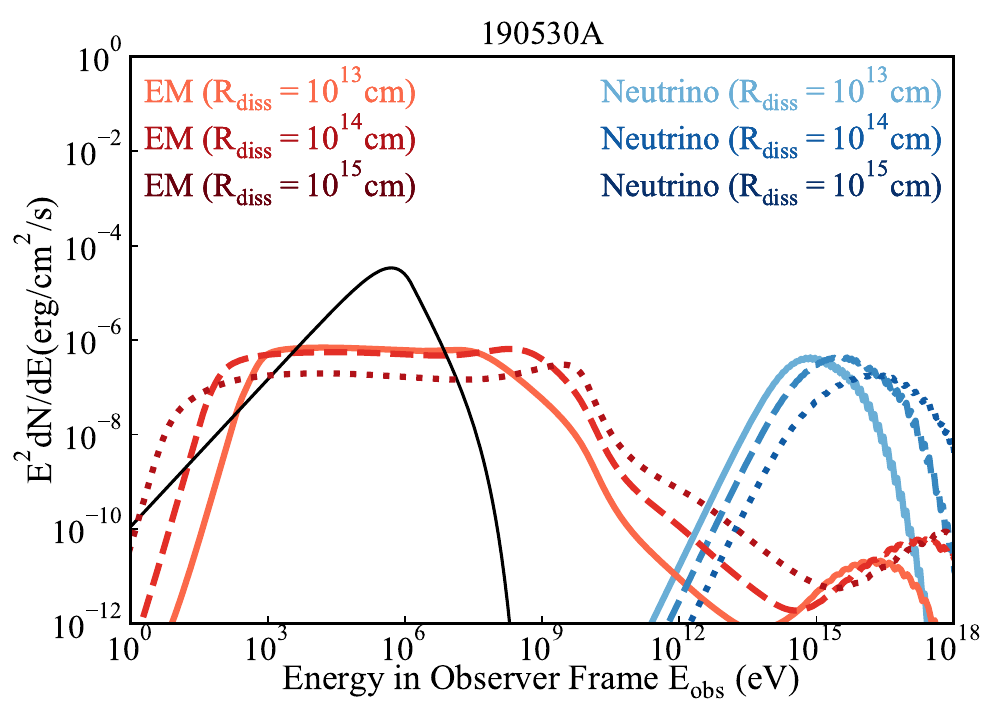}
\includegraphics[width=0.32\textwidth]{ 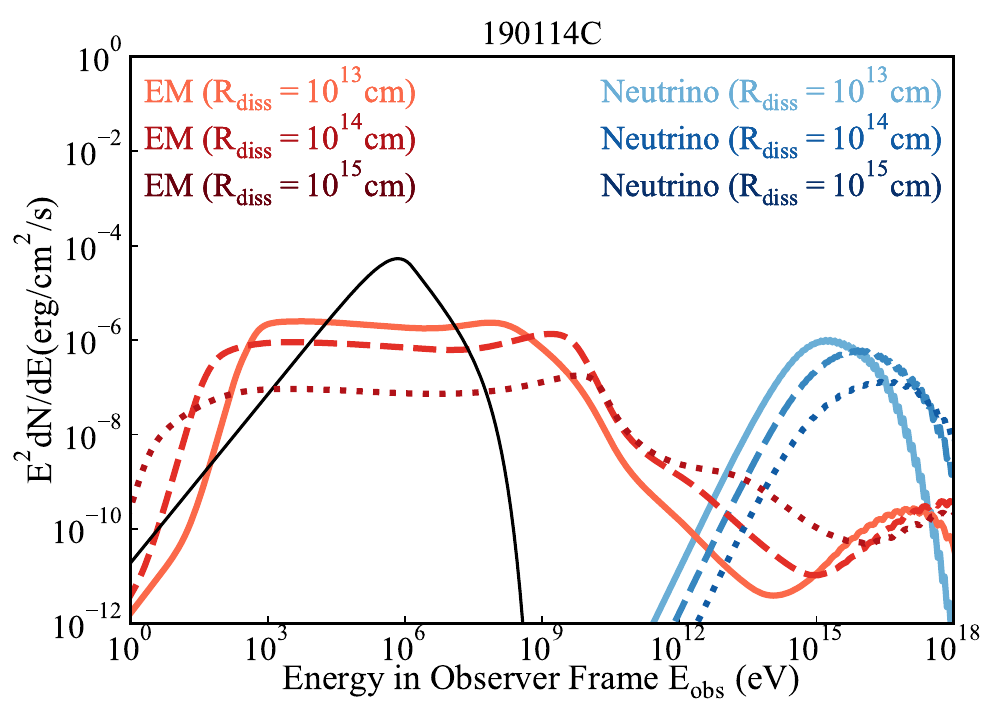}
\includegraphics[width=0.32\textwidth]{ 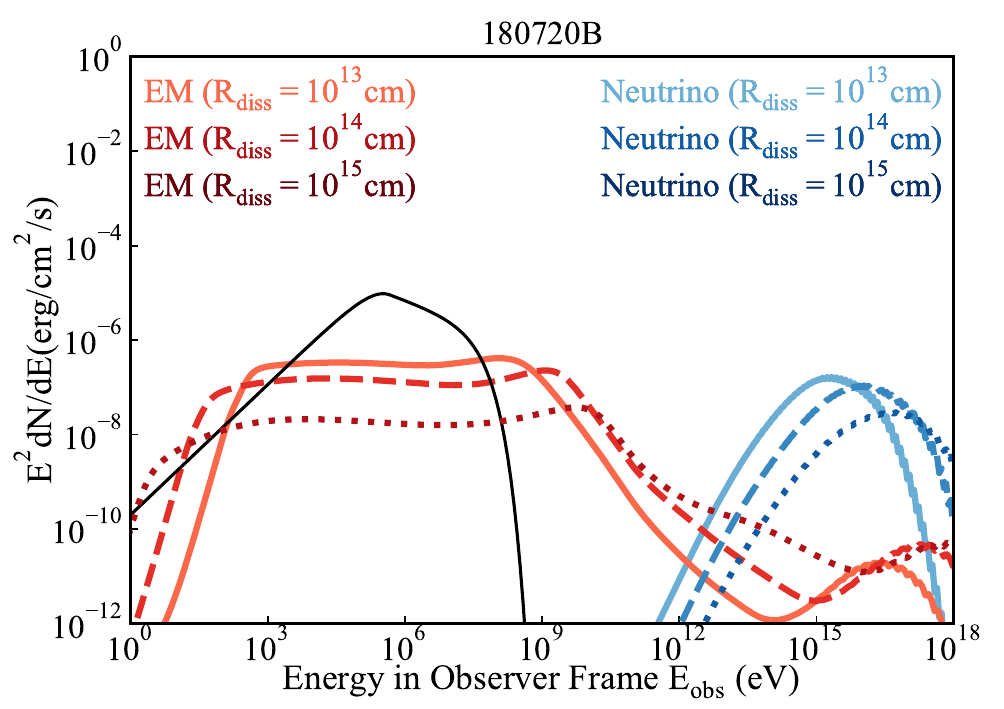}
\includegraphics[width=0.32\textwidth]{ 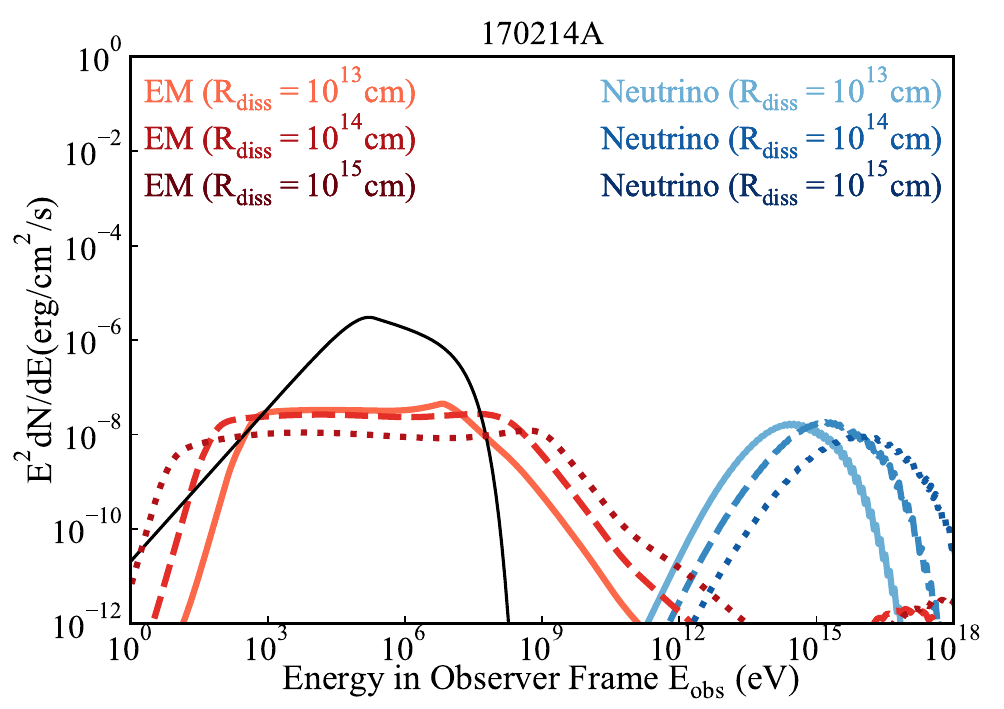}
\includegraphics[width=0.32\textwidth]{ 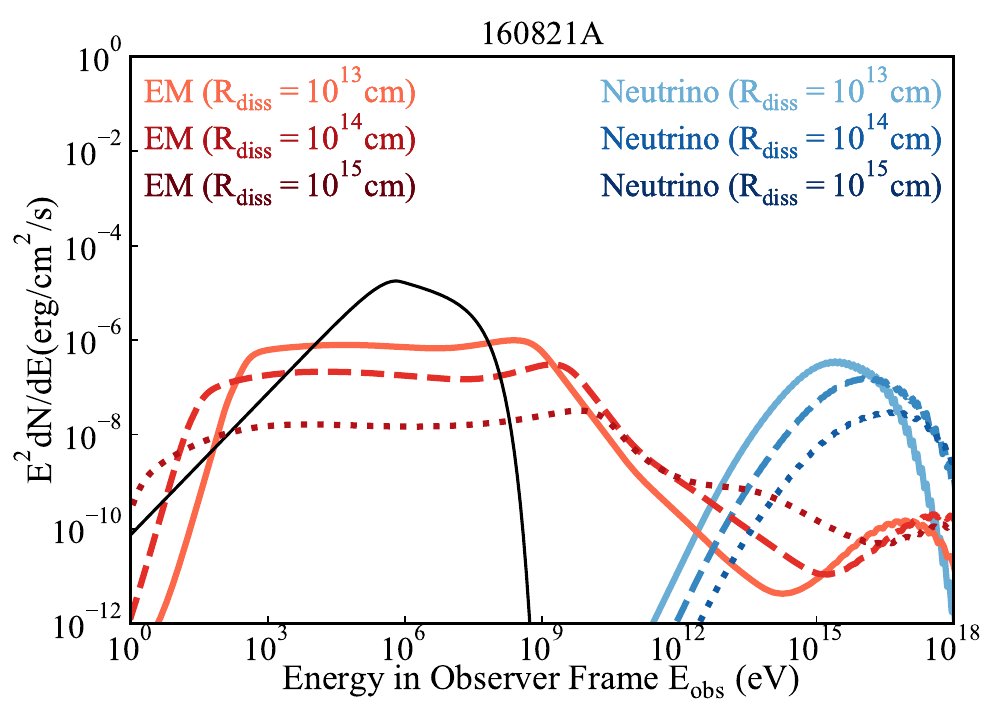}
\includegraphics[width=0.32\textwidth]{ 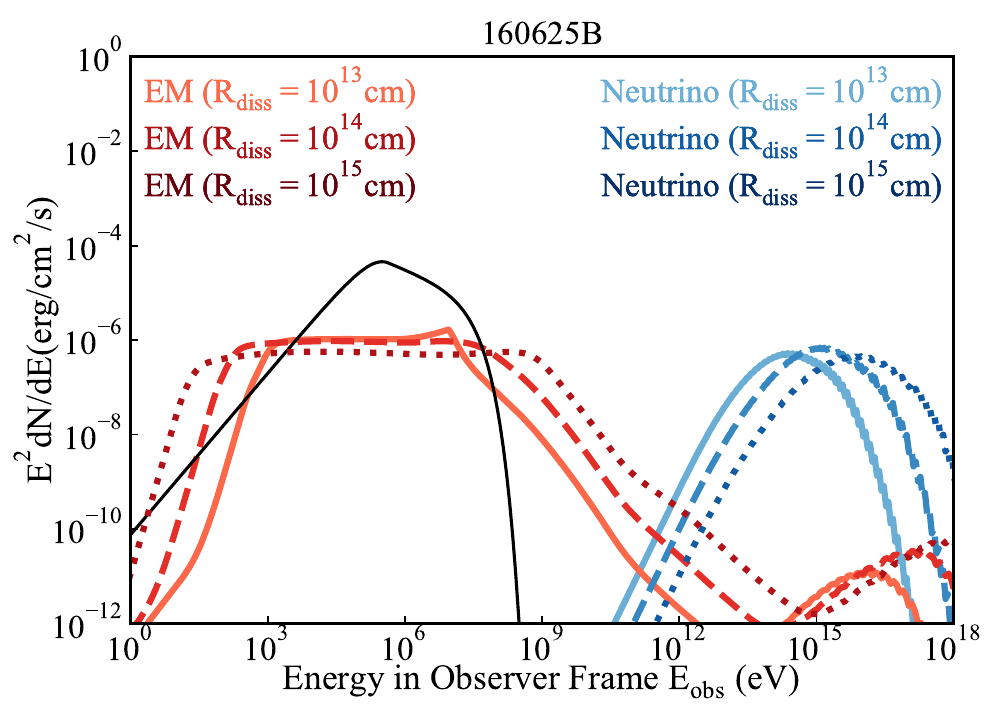}
\includegraphics[width=0.32\textwidth]{ 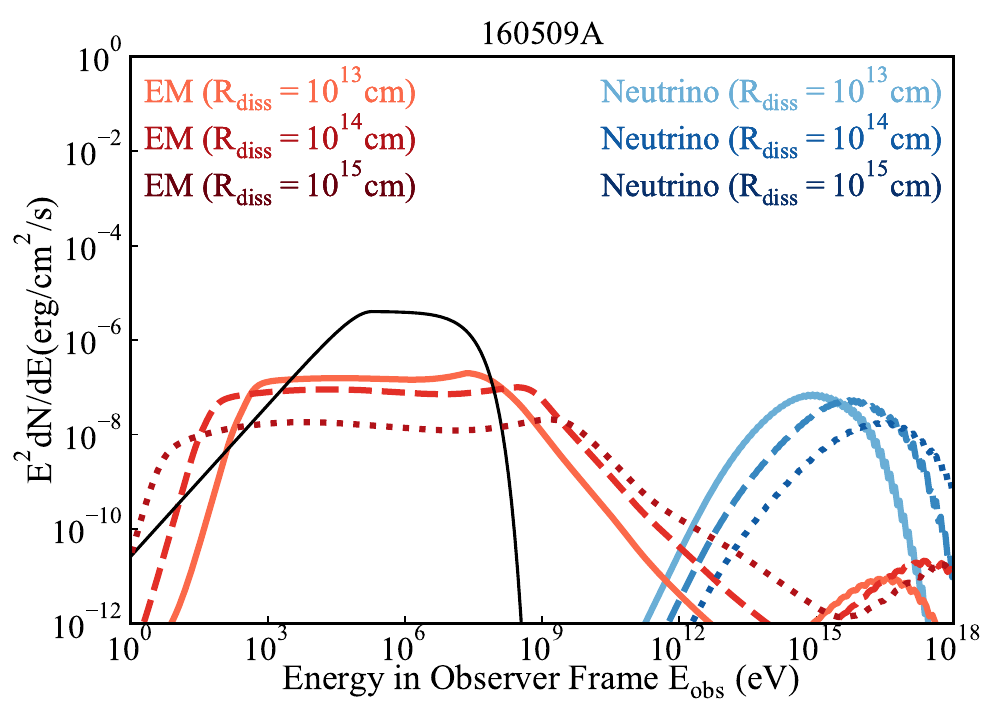}
\includegraphics[width=0.32\textwidth]{ 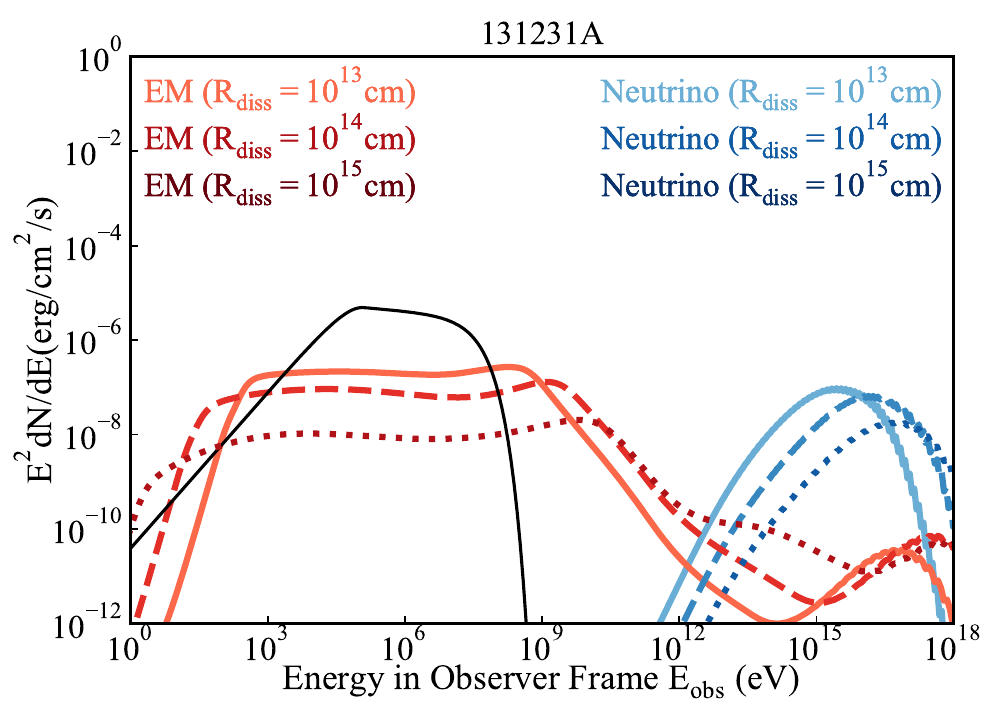}
\includegraphics[width=0.32\textwidth]{ 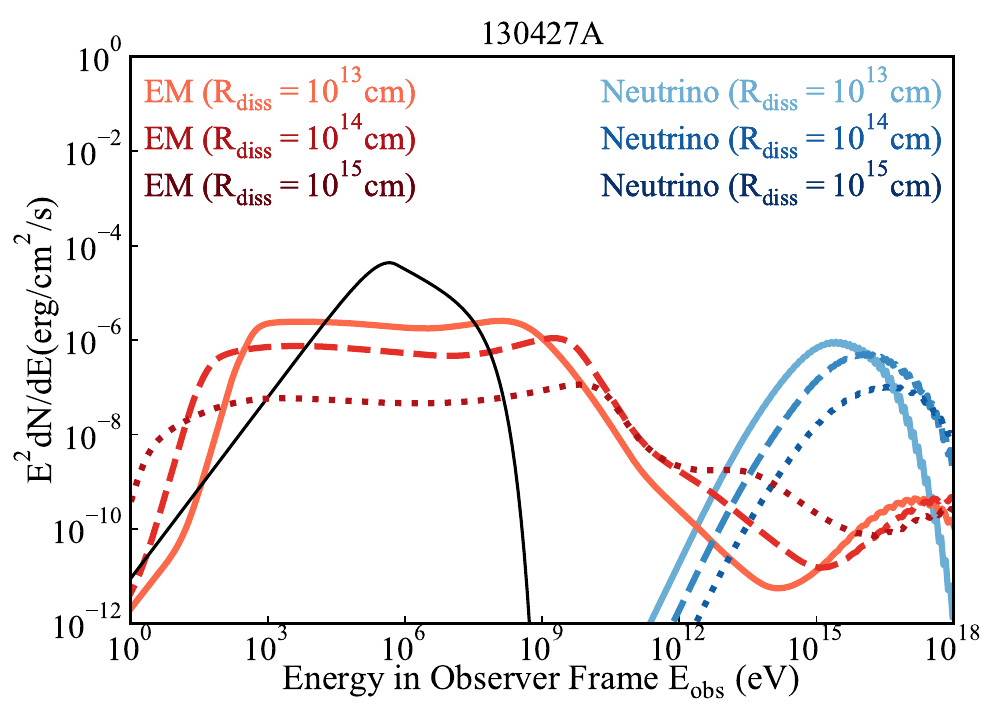}
\includegraphics[width=0.32\textwidth]{ 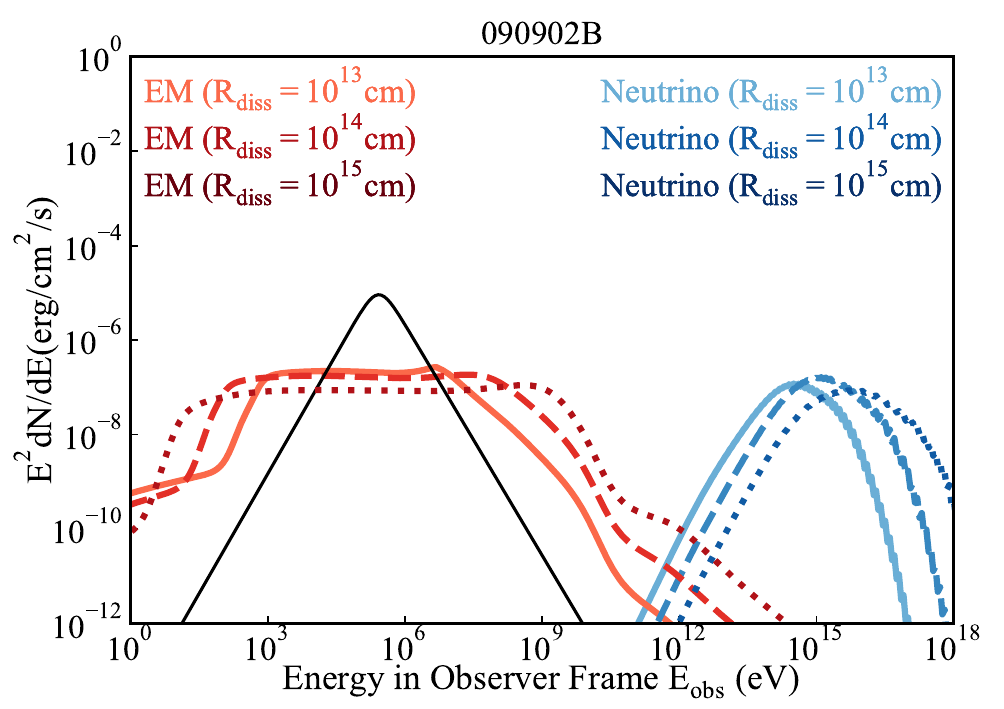}
\caption{The red curves represent the predicted SEDs of the EM cascade emission, while the blue curves represent the high-energy neutrino emission for the reference parameters, except for $R_{\text{diss}}=10^{13}~\mathrm{cm}$~(solid curves), $10^{14}~\mathrm{cm}$~(dashed curves), $10^{15}~\mathrm{cm}$~(dotted curves). The black curves correspond to the Band spectra with parameters from Table~\ref{tab:fermi}. For GRB~190114C and GRB~090902B, we exclusively consider the Band component as the soft photon field. Although certain studies have reported additional components being detected in some time bins \citep{ajello2019,2009ApJ...706L.138A}, their inclusion does not significantly impact the flux of the cascade radiation spectrum due to their low number density.}
\label{fig:SED}
\end{figure}

Since our aim is to constrain the baryon loading factor of GRBs (or the energy budget of UHECRs in GRBS), we take priority of dealing with data over the entire prompt emission phase of GRBs such as $T_{90}$\footnote{The time interval between the epochs when 5\% and 95\% of the total fluence is collected by the detector.}. However, in some GRBs, the low activity intervals (about a few second to a few hundreds of seconds) occur after the precursor radiation, and early GeV afterglow might arise well before the end of the prompt emission phase. We exclude these time intervals in our analyses in order to get better constraints. For example, $T_{90}$ of GRB~221009A is $289\pm1~\rm s$, and it contained a triggering pulse (starting from $T_{0}-1$~s to $T_{0}+43$~s), a quiet period (starting from $T_{0}+121$~s to $T_{0}+163$~s) and pre-main pulse (starting from $T_{0}+177$~s to $T_{0}+210$~s) \citep{2023ApJLesageGBM}. The data of GBM from $T_{0}+219$~s to $T_{0}+277$~s and $T_{0}+508$~s to $T_{0}+514$~s was piled-up \citep{2023ApJLesageGBM} and the early afterglow had appeared at $T_{0}+226$~s \citep{Science2023LHAASO}. Hence, we selected the interval from $T_{0}+177$~s to $T_{0}+219$~s for our study.  The employed time interval of data analysis for other GRBs follow the same criteria. For GRB~190114C, GRB~180720B, GRB~170214A, GRB~131231A, GRB~090902B, some previous studies show that their GeV emission are contributed significantly by early afterglows, and thus we picked up intervals $T_{0}+[0, 7]$~s for GRB~190114C \citep{ajello2019}, $T_{0}+[4.3, 35]$~s for GRB~180720B  \citep{2020A&A636A55R}, $T_{0}+[12.5, 52]$~s for GRB~170214A \citep{2017ApJ...844...56T}, $T_{0}+[13.3, 30]$~s for GRB~131231A \citep{2014ApJ...787L...6L}, and $T_{0}+[2.8, 9]$~s for GRB~090902B \citep{2009ApJ...706L.138A,2011ApJ...730..141Z}.  We excluded the precursor and the follow-up low activity interval of GRB~190530A and GRB~160509A, and picked up the interval $T_{0}+[7.8, 20.2]$~s \citep{gupta2022probing} of the former and the interval $T_{0}+[8.2, 40]$~s of the latter \citep{2017ApJ...844L...7T}. For GRB~160625B, we excluded the precursor radiation, low activity and early afterglow intervals, and picked up $T_{0}+[188.5, 200]$~s \citep{2017ApJ...848...15F,zhang2018transition}. For GRB~130427A, the early GeV afterglow already became important at $T_0+18$\,s \citep{2013ApJ...779L...1K,2013ApJ...773L..20L,2017ApJ...844...92F},  and the data of GBM was influenced by piled-up during $T_{0}+[4.9, 11.4]$~s. We therefore combine the data in two time intervals, $T_{0}+[4.1, 4.9]$~s and $T_{0}+[11.4, 18]$~s in our analyses for this GRB. For GRB~211018A and GRB~160821A, without relevant studies and clear evidence for the early afterglow contribution, we employed the data in their entire $T_{90}$ duration for our analyses.

The parameters of the best-fit spectra with the Band function in corresponding time intervals of each GRB are summarized in Table~\ref{tab:fermi}, and the spectra are shown as the black curves in Figure~\ref{fig:SED}. Similar to Figure~\ref{fig:sample}, Figure~\ref{fig:SED} also includes the predicted SEDs of the EM cascade emission (red curves) and the neutrino emission (blue curves) for three typical dissipation radii: $R_{\text{diss}}=10^{13}~\mathrm{cm}$~(solid curves), $10^{14}~\mathrm{cm}$~(dashed curves), $10^{15}~\mathrm{cm}$~(dotted curves), while maintaining a constant bulk Lorentz factor of $\Gamma_{\text{bulk}}=400$. We see that the value of the $R_{\rm diss}$  has a significant impact on the spectrum of the EM cascade emission. On one hand, the SSA becomes weaker with a larger $R_{\rm diss}$ , due to the decreasing electron column density and the magnetic field. On the other hand, a larger $R_{\rm diss}$  reduces the reaction efficiency of $p\gamma$ processes, resulting in lower cascade flux and neutrino flux.

\subsection{\texorpdfstring{\label{sec:influence_chiB}The influence of $\chi_{\rm B}$}{}}
The values of $\chi_{\rm B}$ of GRBs during the prompt phase are important to the predicted flux of the cascade emission. The parameter $\chi_{\rm B}$ mainly influences the strength of the magnetic field, thereby determining the timescales for proton acceleration and cooling. More specifically, it plays a crucial role in setting the maximum proton energy \footnote{Note that $t_{\rm p \gamma}$ relies on the photon field and we here assume a Band function with $\alpha=-1$, $\beta=-2$, and $E_{\rm peak}=1~\rm MeV$ for simplicity.}:
\begin{equation}
E_{p,\rm max}\simeq \left\{\begin{matrix}
1.4\times10^{19}(\frac{L_{\gamma}}{10^{53}~\rm erg/s})^{1/2}(\frac{\xi}{10\%})(\frac{\Gamma_{\rm bulk}}{400})^{-1}(\frac{1+z}{2})^{-1} \chi_{\rm B}^{1/2}{\, \rm eV}, &{\rm for}~ t_{\rm acc}=t_{\rm dyn}\\
5.6\times10^{18}(\frac{L_{\gamma}}{10^{53}~\rm erg/s})^{-1/2}(\frac{\xi}{10\%})(\frac{R_{\rm diss}}{10^{14}~\rm cm})(\frac{\Gamma_{\rm bulk}}{400})(\frac{E_{\rm peak}}{1~\rm MeV})^{1/2}(\frac{1+z}{2})^{-1} \chi_{\rm B}^{1/2}{\, \rm eV}, &{\rm for}~t_{\rm acc}=t_{p\gamma}\\
1.9\times10^{19}(\frac{L_{\gamma}}{10^{53}~\rm cm})^{-1/4}(\frac{\xi}{10\%})^{1/2}(\frac{R_{\rm diss}}{10^{14}~\rm cm})^{1/2}(\frac{\Gamma_{\rm bulk}}{400})^{3/2}(\frac{1+z}{2})^{-1} \chi_{\rm B}^{-1/4}{\, \rm eV}, &{\rm for}~ t_{\rm acc}=t_{\rm syn,p}\\
\end{matrix}.\right.
\label{eq:Epmax}
\end{equation}
The maximum proton energy is determined by the minimum one among the three values. If GRBs are efficient accelerators of UHECRs, the maximum proton energy should reach the energy of the `ankle’ (i.e., $\sim 10^{18.5}\,$eV) or above. With the reference parameters, it would impose a constraint on the parameter $\chi_{\rm B}$. If $\chi_B$ is too small, the acceleration of UHE proton cannot be accomplished within the dynamical timescale, or the acceleration rate cannot overcome the energy loss rate of photopion production, this thus sets a lower limit of $\chi_B\gtrsim0.1$. On the other hand, too high a value of $\chi_{\rm B}$ (i.e., $\gg 10$) would lead to an extreme requirement of the energy budget of GRBs.

\begin{figure}
\centering
\includegraphics[width=0.48\textwidth]{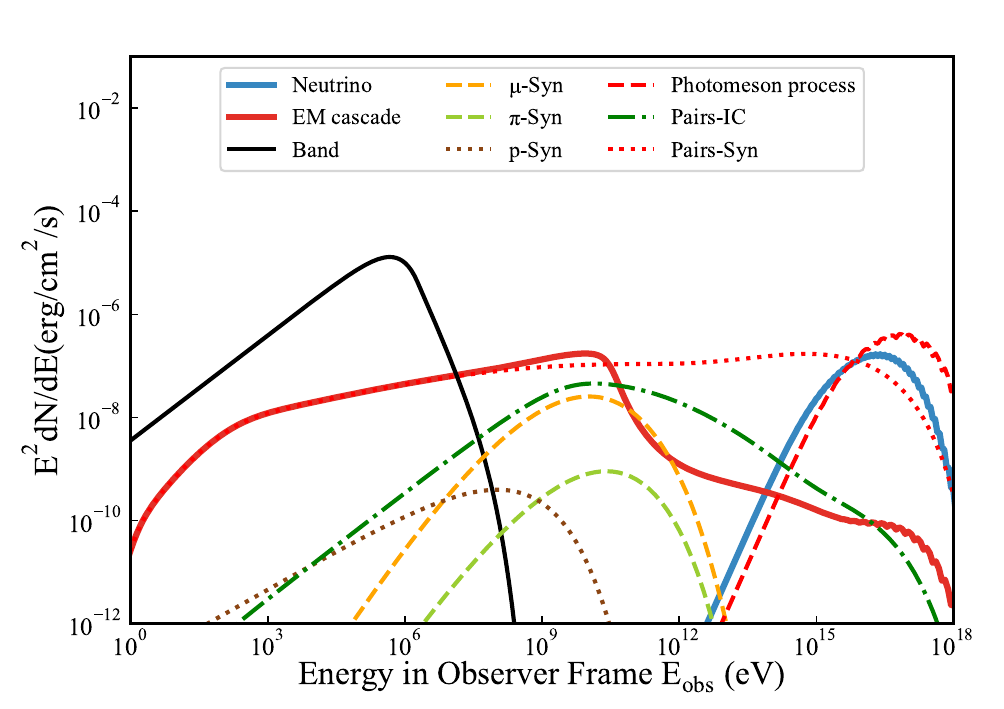}
\includegraphics[width=0.48\textwidth]{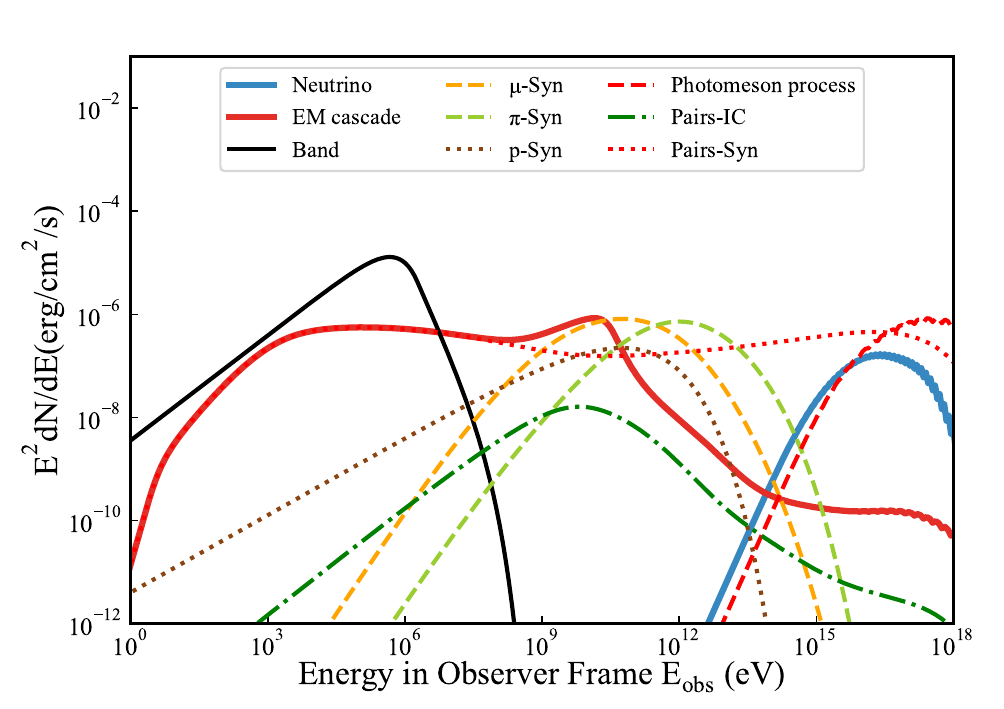}
\caption{Same with Figure \ref{fig:sample}, but $\chi_{\rm B}=0.1$ (left panel), and $\chi_{\rm B}=10$ (right panel).}
\label{fig:etaB}
\end{figure}

Therefore, we discuss the influence of $\chi_{\rm B}$ on the SEDs of the EM cascade within a reasonable range of $\chi_{\rm B}\in(0.1,10)$ with other model parameters following the reference case. We take the observation of GRB~221009A during $T_{0}+[177,~219]~\rm s$ as the example. The expected EM cascade spectrum and the neutrino spectrum during this time interval are presented in Figure \ref{fig:etaB}. In the left panel, $\chi_{\rm B}$ is set to 0.1, while in the right panel, $\chi_{\rm B}$ is set to 10. Comparing with the GeV flux in the case of $\chi_{\rm B}=1$ (i.e., the reference case), the IC flux of the cascade pairs is enhanced, while the synchrotron flux of muons and pions is reduced for $\chi_{\rm B}=0.1$. On the other hand, the IC flux of cascade pairs decreases and the synchrotron flux of muons and pions increases for $\chi_{\rm B}=10$. As a result, the flux within $0.1-10$\,GeV are more or less the same for different values of $\chi_{\rm B}$. More specifically, we have $F_{0.1-10}=6.2\times10^{-7}~\text{erg cm}^{-2}~\text{s}^{-1}$ for $\chi_{\rm B}=0.1$, $F_{0.1-10}=2.2\times10^{-6}~\text{erg cm}^{-2}~\text{s}^{-1}$ for $\chi_{\rm B}=10$, and $F_{0.1-10}=1.2\times10^{-6}~\text{erg cm}^{-2}~\text{s}^{-1}$ for $\chi_{\rm B}=1$. Apparently, for a different value of $\chi_{\rm B}$, the changes in the GeV flux produced by different radiation processes cancel each other, and hence the total predicted GeV flux is not sensitive to the value of $\chi_{\rm B}$ as long as the latter is within a reasonable range.

\subsection{Gamma-Ray Constraints}\label{sec:constraints}
\subsubsection{Constraints in the reference case} \label{sec:reference}
Generally, the prediction of cascade flux aligns linearly with the value $\eta_{p}$ of GRBs. This enables us to determine an upper limit on the baryon loading factor, denoted as $\eta_{p}^{\text{UL}}$, by equating the predicted flux in the range of $0.1-10~\rm GeV$ in the selected time interval to the observational flux UL at the 95\% CL in the same energy range. Our results are summarized in Table~\ref{tab:binsize}. In general, the constraint on $\eta_p$ would be stronger if the ratio between $F_{\rm GBM}$ and $F^{\rm UL}_{\rm LAT}$ is higher. Except for GRB~211018A and GRB~090902B, the derived values of $\eta_p^{\rm UL}$ for all selected GRBs are $\lesssim10$, stronger than the constraint obtained from the stacking analysis of neutrino data measured by IceCube \citep{2016PhRvL.117x1101A, aartsen2017extending,lucarelli2023neutrino}. Moreover, our results indicate that $\eta_p^{\rm UL}$ can be further constrained to be less than unity for GRB~180720B, GRB~160625B, and GRB~131231A. In the case of GRB~090902B, constraining $\eta_p$ with GeV observations yields looser results compared to those reported in \cite{2010ApJ...725L.121A}, suggesting $\eta_p\sim3$. The discrepancy observed can be attributed to the utilization of different energy bands, specifically those below $10~\rm keV$ and between $10-100~\rm MeV$, in their study. This implies that if data from a lower energy band is available, it may allow for more stringent constraints on $\eta_p$. Another contributing factor is the relatively lower ratio between $F_{\rm GBM}$ and $F^{\rm UL}_{\rm LAT}$ within the selected interval compared to other chosen GRBs. A similar reason was found for GRB~211018A (Ratio =6), however, we discovered a remarkably steep power-law index of $\Gamma_{\rm LAT}=-1.8$ confirmed by a more comprehensive time-resolved analysis; see Table \ref{tab:lat-211018A}) during this interval, suggesting the presence of an additional component masking it. Further detailed analyses of the afterglow are necessary to better constrain $\eta_p$ during the prompt phase.

We may compare the results with those from neutrino measurements. Previously, the main constraints on the baryon loading factor of GRBs come from the stacking analysis of the neutrino data measured by IceCube \citep{2016PhRvL.117x1101A, aartsen2017extending,lucarelli2023neutrino}. For a similar dissipation radius and the jet's Lorentz factor to our reference parameters, 90\% C.L. upper limit of $\eta_p$ obtained by the neutrino measurements is $\gtrsim 10$, which is weaker than the constraint obtain here with GeV gamma-ray measurements. There are also reports of non-detection of muon neutrino events from individual GRBS, i.e., GRB~221009A \citep{2022GCN.32665IceCube} and GRB~130427A \citep{2013GCN.14520Blaufuss}, based on which the constraint of $\eta_p$ can be derived. Following \citet{liu2023constraints}, we set the expected number of muon neutrino events from 1\,TeV to 1\,EeV to three, which corresponds to a 5\% chance probability for the Poisson distribution of the detection probability and will result in a 95\% C.L. upper limit. Consequently, we find $\eta_{p}^{\rm UL, \nu}=7.9$ for GRB~221009A and $\eta_{p}^{\rm UL, \nu}=69$ for GRB~130427A in the reference case, which are also weaker than the constraint from GeV gamma-ray observations. Note that \citet{liu2023constraints} reported a lower value of $\eta_{p}^{\rm UL, \nu}=2.4$ for GRB~221009A because their calculations did not account for the synchrotron cooling of muons and pions, leading to an overestimation of the neutrino flux.

\begin{sidewaystable}
\tabcolsep=3pt
\caption{The properties of selected GRBs. For conservation, we choose the 95\% C.L UL of the LAT observation ($F^{\rm UL}_\text{LAT}$). Each entry includes the GRBs names, the redshift ($z$), the selected time intervals, best fitting parameters of Band function, the fluxes of GBM ($F_\text{GBM}$), isotropic photon luminosities ($L_{\gamma}$), the ratios of $F_\text{GBM}$ and $F^{\rm UL}_\text{LAT}$, and the upper limit of the baryon loading factor ($\eta_{p}^{\mathrm{UL}}$) in reference case. }
\begin{flushleft}
\begin{tabularx}{\textwidth}{cccccccccccc}
\hline
\hline
 Name&Redshift& $\alpha$ & $\beta$ & $E_{\text{peak}}$ & $F_{\text{GBM}}$ & $L_{\gamma}$ & $\Gamma_{\text{LAT}}$&$F^{\rm UL}_{\rm LAT}$ &Ratio&Selected & $\eta_{p}^{\text{UL}}$\\ 
 &&&&(keV)&($\text{erg/cm$^{2}$/s}$)&($\text{erg/s}$)&&($\text{erg/cm$^{2}$/s}$)&(=$\frac{F_{\text{GBM}}}{F_{\text{LAT}}}$)&intervals$^{a}$(s)&\\
\hline
221009A&0.151   & -1.321$\pm0.020$ &  -3.98$\pm0.01$  &  800$\pm6$  &   $4.3\times10^{-5}$ &    $2.9\times10^{51}$ &2.0&  $2.2\times10^{-7}$&195&[177,219]&1.8 \\
211018A &0.64  & -0.792$\pm0.010$ &  -3.97$^{+0.89}_{-6.12}$  &  388$\pm5$  &   $1.4\times10^{-6}$ &    $2.7\times10^{51}$ &1.8$\pm0.2$&  $2.5\times10^{-7}$&6&[4.3,128.2]&113 \\
190530A &0.936  & -0.953$\pm0.010$ &  -4.55$\pm0.92$  &  961$\pm9$  &   $4.0\times10^{-5}$ &    $1.9\times10^{53}$ &3.9$\pm0.6$&  $2.4\times10^{-7}$&142&[7.8,20.2]&1.2 \\
190114C$^{\bigtriangledown}$ &0.42 & -0.311$\pm0.011$ &  -4.58$\pm1.81$ &  729$\pm5$  &   $4.7\times10^{-5}$ &   $3.3\times10^{52}$ & 2.3$\pm0.1$& $4.2\times10^{-6}$&11&[0,7]&8.8\\
180720B &0.653 & -1.082$\pm0.005$ & -2.35$\pm0.05$   &   605$\pm7$   &   $1.4\times10^{-5}$ &   $2.6\times10^{52}$  &  3.9$\pm0.7$ &$4.7\times10^{-8}$&298&[4.3,35]&0.6\\
170214A &2.53 & -0.923$\pm0.001$ &  -2.28$\pm0.09$  &  547$\pm10$ &    $4.1\times10^{-6}$  &   $2.3\times10^{53}$ &6.7$\pm2.0$& $1.5\times10^{-8}$&273& [12.5,52]&4.6 \\
160821A &0.4 & -1.001$\pm0.004$ &  -2.25$\pm0.03$  &  915$\pm9$ &    $2.4\times10^{-5}$  &   $1.5\times10^{52}$ &1.9$\pm0.1$&  $4.5\times10^{-8}$&533& [118.5,161.5]& 0.4\\
160625B & 1.406& -0.862$\pm0.005$ &  -2.38$\pm0.03$  &  680$\pm5$ &    $5.5\times10^{-5}$  &   $7.3\times10^{53}$ &  3.0$\pm0.2$&$5.7\times10^{-7}$&96&[188.4,200]&11 \\
160509A &1.17 & -0.932$\pm0.010$ &  -2.00$\pm0.01$  &  389$\pm5$ &    $8.0\times10^{-6}$  &   $6.7\times10^{52}$ &3.4$\pm0.2$&  $2.4\times10^{-7}$ &33 & [8.2,40] &9.7 \\
131231A&0.642  & -0.903$\pm0.007$ &  -2.08$\pm0.02$  &  175$\pm1$ &    $7.7\times10^{-6}$  &   $1.8\times10^{52}$ &5.2$\pm1$&  $2.2\times10^{-8}$  &350& [13.3,30] &0.6 \\
130427A$^{\triangle}$ &0.34 &  -0.502$\pm0.010$  &    -2.79$\pm0.06$   &  583$\pm5$ &    $1.7\times10^{-4}$ &   $7.2\times10^{52}$ & 2.0 &$5.8\times10^{-7}$&295&[4.1,4.9]&0.3\\
130427A$^{\triangle}$ &- &  -0.883$\pm0.004$  &    -2.07$\pm0.01$   &  158$\pm1$ &    $2.9\times10^{-4}$ &   $1.2\times10^{53}$ & 2.4$\pm0.1$ &$9.4\times10^{-7}$&308&[11.4,18]&0.3 \\
090902B $^{*}$&1.82& -- & -- & --& $2.0\times10^{-5}$ &   $5.0\times10^{53}$ & 2.4$\pm0.1$ &$8.2\times10^{-7}$&24&[2.8,9]&42 \\
\hline
\end{tabularx}
\end{flushleft}
$^a$ The main flares after $T_{0}$.\\
$^{\bigtriangledown}$  We used Band pulsing additional power-law (PL) function to fit the soft photon field for GRB~190114C and the best fitting index of PL was $\Gamma_{\rm GBM}=1.64\pm 0.02$ (The specific functional forms can be found in \cite{zhang2018physics}). \\
$^{\triangle}$ The data of GBM from $T_{0}+4.9~\rm s$ to $T_{0}+11.4~\rm s$ was piled-up, we picked up two main flares of the prompt phase of GRB~130427A.\\
$^{*}$ We used smoothly broken power-law (SBPL) pulsing PL function to fit the soft photon field for GRB~090902B and the best fitting parameters are $\lambda_{1}=-0.31\pm0.01$, $\lambda_{2}=-3.63\pm0.1$, $\Lambda=0.27\pm 0.01$, $E_{b}=754\pm13~\rm keV$ and $\Gamma_{\rm GBM}=1.85\pm 0.02$. 
\label{tab:binsize}
\end{sidewaystable}

\subsubsection{The 2D constraint maps of the baryon loading factor}\label{sec:map}
\begin{figure}[htb]
\centering
\includegraphics[width=1\textwidth]{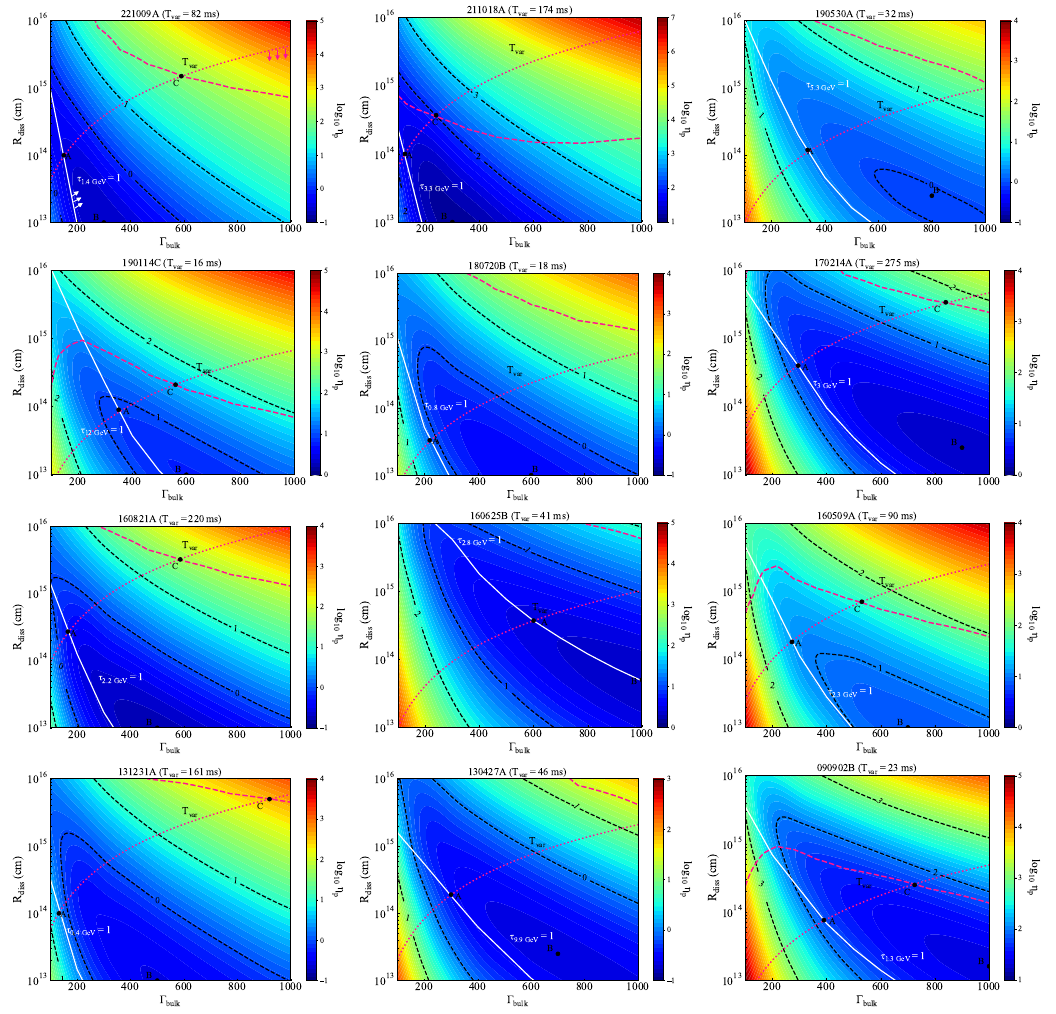}
\caption{Upper limits of the baryon loading factor in logarithmic form (log$_{10}\eta_{p}^{\mathrm{UL}}$) as the function of $R_{\rm diss}$  and $\Gamma_{\rm bulk}$ for selected GRBs. Note that the results of GRB~221009A and GRB 130427A are from the $T_{0}+[177,~219]~\rm s$ and $T_{0}+[11.4,~18]~\rm s$, respectively. The magenta dotted curves represent the maximum dissipation radii given certain $\Gamma_{\rm bulk}$, which are constrained by the observed variability timescale $T_{\rm var}$. The gray solid curves depict the photospheric radii for the highest energy photons in the selected intervals. The black dots $A$ are the intersection of above two curves. The black dots $B$ are the locations of the strongest constraint on $\eta_{p}$ in the considered parameter space. The magenta dashed curves represent the limits of $f_{\rm bol}\eta_{p}\ge1$, and the black dots $C$ are the intersection of magenta dotted curves and magenta dashed curves.}
\label{fig:map}
\end{figure}

The values of $\Gamma_{\rm bulk}$ and $R_{\rm diss}$ of GRBs during the prompt phase are important to the predicted flux of the cascade emission. Depending on GRB models, they are usually thought to be in the range $100 \leq \Gamma_{\text{bulk}} \leq 1000$ and $10^{13} \leq R_{\text{diss}} \leq 10^{16}$ cm.  Thus, we scan a large range of $\Gamma_{\rm bulk}$ and $R_{\rm diss}$  and obtain $\eta_{p}^{\text{UL}}$ in the two-dimensional parameter space for selected GRBs, as shown in Figure~\ref{fig:map}. The constraint of $\eta_{p}^{\text{UL}}$ imposed by Fermi data exhibits a discernible dependence on these two parameters, as the cascade flux is restricted by the $\gamma\gamma$ absorption effect and the production rate of EM particles. Larger values of $\Gamma_{\rm bulk}$ and $R_{\rm diss}$  result in inefficient interactions of protons and subsequently low cascade flux. On the other hand, smaller values of $\Gamma_{\rm bulk}$ and $R_{\rm diss}$  can lead to a higher production rate of EM particles but also result in a stronger $\gamma\gamma$ absorption.

However, not all combinations of $R_{\mathrm{diss}}$ and $\Gamma_{\mathrm{bulk}}$ are reasonable. Firstly, the dissipation radius is limited by the observed variability timescale as $R_{\text{diss}}\leq 4.8\times10^{14}(T_{\text{var}}/0.1\,\text{s})(\Gamma_{\text{bulk}}/400)^{2}(1+z)^{-1}~\mathrm{cm}$. $T_{\text{var}}$ for each GRB considered here is analyzed with the Bayesian block method, and details can be found in Appendix \ref{sec:variability}. In Figure~\ref{fig:map}, magenta dotted curves represent the maximum dissipation radii for different bulk Lorentz factors, inferred from the variability timescale $T_{\mathrm{var}}$. On the other hand, the emission region must be transparent to the high-energy gamma-ray photon observed by Fermi-LAT. White solid curves in  Figure~\ref{fig:map} showcase the characteristic photospheric radii for different bulk Lorentz factors, where the optical depth of the highest-energy photons in the selected time intervals (see Appendix \ref{sec:LAT} for details) is unity. Thus, the allowable parameter space for $R_{\rm diss}-\Gamma_{\rm bulk}$ combinations is between the two curves (see the example in the top-left panel of Figure~\ref{fig:map}). Within the allowable region, we discuss three representative combinations of $R_{\rm diss}$ and $\Gamma_{\rm bulk}$, termed as Case A, B, C, apart from the reference case.

Case A is the point (labeled with $A$ in Figure \ref{fig:map}) where the magenta dotted curve (limit derived from variability) and the white solid curve (limit derived from transparency) intersects. This case gives the minimum bulk Lorentz factor that can satisfy both conditions. Resulting $R_{\rm diss}$ of Case A are roughly a few $10^{14}~\mathrm{cm}$ which is the typical IS
radii for GRB population and the corresponding $\eta_{p}^{\text{UL}}$ are mostly found to be $\lesssim 10$, except for GRB~211018A and GRB~090902B (see the second column of Table \ref{tab:parameter-space}), and they are more stringent than the results divided by the IceCube observation \citep{lucarelli2023neutrino} and such low $\eta_{p}^{\text{UL}}$ disfavors a baryonic-dominated jet composition.

Case B is the point (labeled with $B$ in Figure \ref{fig:map}) where the minimum $\eta_{p}^{\rm UL}$ is in the allowable region. The corresponding $\eta_{p}^{\text{UL}}$ values are several times more stringent than Case A and are found at $\sim10^{13}~\rm cm$, the minimum $R_{\rm diss}$, and $\Gamma_{\rm bulk}\gtrsim500$, except for GRB~221009A and GRB~211018A(see the third column of Table \ref{tab:parameter-space}). Subsequently, inference from the solid white curves indicates that $R_{\rm diss}$ values less than approximately $10^{13}~\rm cm$ are not considered, as illustrated by GRB~160625B shown in Figure \ref{fig:map}. On the trend of the contours, it can be inferred that $\eta_{p}^{\rm UL}$ are still strongly constrained within $\sim10^{11-13}~\rm cm$, the typical photospheric radii range for GRB population. Besides, the contribution of primary electrons are neglect in this study, if keV-MeV emission are explained by the their synchrotron radiation, their synchrotron self-Compton (SSC) may contribute to GeV emission (see GRB~180720B and GRB~190114C in \cite{wang2019synchrotron}), then less room are left for hadronic process and more stringent constraint can be expected.  

In order to generate enough energy production rate of UHECRs, we find the region that satisfies $f_{\text{bol}}\eta_{p}\ge1$ (the region above the magenta dashed curves in Figure \ref{fig:map}). Then, Case C is the point (labeled with $C$) where the magenta dashed curve and magenta dotted curve intersects. This case gives the minimum bulk Lorentz factor that can satisfy both conditions. Corresponding result of Case C are summarized in Table~\ref{tab:parameter-space}, implying large values of $\Gamma_{\mathrm{bulk}}$ and $R_{\rm diss}$  are required for selected GRBs. It means that if these selected GRBs are representative of the overall population, they can be largely excluded as primary sources of UHECRs during the prompt phase within a large parameter space (such as the typical parameter space for the photosphere model and the internal shock model). 

\begin{sidewaystable}
\caption{The values of $\Gamma_{\rm bulk}$, $R_{\rm diss}$  and $\eta^{\text{UL}}_{\text{p}}$ for different cases. The last three rows are the cases where only the acceleration efficiency and spectral index are changed, respectively. If the combination of $\Gamma_{\rm bulk}$ and $R_{\rm diss}$  can not be found in Case C for the selected GRBs, we pick up the values of $\Gamma_{\rm bulk}$ and $R_{\rm diss}$ which can generate the maximum of $Q^{\text{GRBs}}_{\text{UHECR}}$ in the allowable region (labeled with *). Note that the results of GRB~221009A and GRB 130427A are from the $T_{0}+[177,~219]~\rm s$ and $T_{0}+[11.4,~18]~\rm s$, respectively.}
\begin{tabular}{CCCC}
    \hline
    \hline
        $\text{Name}$ & \text{Case}\,\text{A}$\,$($\Gamma_{\rm bulk}$,$\,$R$_{\text{diss}}$,$\,$$\eta^{\text{UL}}_{p}$) &\text{Case}\,\text{B} 
        &\text{Case}\,\text{C}
       \\ 
        \hline
        \text{221009A}& (150,$\,$1.0$\times10^{14}$$\,\text{cm}$,$\,$0.4)  & (300,$\,$1.0$\times10^{13}$$\,\text{cm}$,$\,$0.2)  & (590,$\,$1.5$\times10^{15}$$\,\text{cm}$,$\,$130)    \\
        \text{211018A}& (125,$\,$1.0$\times10^{14}$$\,\text{cm}$,$\,$41)  & (300,$\,$1.0$\times10^{13}$$\,\text{cm}$,$\,$15)  & (240,$\,$3.7$\times10^{14}$$\,\text{cm}$,$\,$102)    \\
          \text{190530A}^{*} & (335,$\,$1.2$\times10^{14}$$\,\text{cm}$,$\,$1.4) & (800,$\,$2.5$\times10^{13}$$\,\text{cm}$,$\,$0.7) & (1000,$\,$1$\times10^{15}$$\,\text{cm}$,$\,$28) 
          \\ 
          \text{190114C} & (350,$\,$9.1$\times10^{13}$$\,\text{cm}$,$\,$9.0)  & (600,$\,$1.0$\times10^{13}$$\,\text{cm}$,$\,$6.4)  & (560,$\,$2.1$\times10^{14}$$\,\text{cm}$,$\,$36)
          \\    
        \text{180720B}^{*} & (220,$\,$3.3$\times10^{13}$$\,\text{cm}$,$\,$1.5) & (600,$\,$1.0$\times10^{13}$$\,\text{cm}$,$\,$0.4) & (1000,$\,$6.5$\times10^{14}$$\,\text{cm}$,$\,$53)
        \\ 
          \text{170214A} & (295,$\,$4.0$\times10^{14}$$\,\text{cm}$,$\,$5.7) & (900,$\,$2.5$\times10^{13}$$\,\text{cm}$,$\,$1.9) & (840,$\,$3.4$\times10^{15}$$\,\text{cm}$,$\,$53)
          \\
         \text{160821A} & (165,$\,$2.7$\times10^{14}$$\,\text{cm}$,$\,$0.5) & (500,$\,$1.0$\times10^{13}$$\,\text{cm}$,$\,$0.2) & (585,$\,$3.2$\times10^{15}$$\,\text{cm}$,$\,$49)
         \\
        \text{160625B}^{*} & (600,$\,$3.7$\times10^{14}$$\,\text{cm}$,$\,$2.9)  & (1000,$\,$5.0$\times10^{13}$$\,\text{cm}$,$\,$2.1) &(1000,$\,$1.6$\times10^{15}$$\,\text{cm}$,$\,$11)
        \\ 
        \text{160509A} & (270,$\,$1.8$\times10^{14}$$\,\text{cm}$,$\,$14) & (700,$\,$1.0$\times10^{13}$$\,\text{cm}$,$\,$7) & (530,$\,$7.0$\times10^{14}$$\,\text{cm}$,$\,$40)
        \\ 
        \text{131231A} & (120,$\,$8.3$\times10^{13}$$\,\text{cm}$,$\,$2.3)  & (500,$\,$1.0$\times10^{13}$$\,\text{cm}$,$\,$0.3)  & (900,$\,$5.2$\times10^{15}$$\,\text{cm}$,$\,$361)
        \\ 
         \text{130427A}^{*} & (300,$\,$1.9$\times10^{14}$$\,\text{cm}$,$\,$0.4) & (700,$\,$2.5$\times10^{13}$$\,\text{cm}$,$\,$0.2) & (1000,$\,$2.0$\times10^{15}$$\,\text{cm}$,$\,$4) 
         \\   
       \text{090902B} & (390,$\,$7.6$\times10^{13}$$\,\text{cm}$,$\,$69) & (1000,$\,$1.6$\times10^{13}$$\,\text{cm}$,$\,$24) & (725,$\,$2.5$\times10^{14}$$\,\text{cm}$,$\,$52) \\
     \hline
        \text{221009A}\,($\xi=1\%$) & (150,$\,$1.0$\times10^{14}$$\,\text{cm}$,$\,$3.1)  & (530,$\,$1.0$\times10^{13}$$\,\text{cm}$,$\,$0.4)  & (490,$\,$1.0$\times10^{15}$$\,\text{cm}$,$\,$124)    \\ 
        \text{221009A}\,($s_{p}$ = 2.5) & (140,$\,$8.7$\times10^{13}$$\,\text{cm}$,$\,$40)  & (530,$\,$1.0$\times10^{13}$$\,\text{cm}$,$\,$8.3)  & (490,$\,$1.0$\times10^{15}$$\,\text{cm}$,$\,$92360)    \\ 
        \text{221009A}\,($s_{p}$ = 1.5) & (165,$\,$1.2$\times10^{14}$$\,\text{cm}$,$\,$0.3)  & (530,$\,$1.0$\times10^{13}$$\,\text{cm}$,$\,$0.1)  & (490,$\,$1.0$\times10^{15}$$\,\text{cm}$,$\,$8)    \\ 
    \hline
    \end{tabular}
    \label{tab:parameter-space}
\end{sidewaystable}

\section{Discussions} \label{sec:diss}
\subsection{The influence of the chosen width of time interval} \label{sec:sizes}
The width of the time interval chosen in the data analysis can influence the average GBM/LAT flux in the interval and consequently affect the constraints on the baryon loading factor. Generally speaking, the obtained $\eta_p^{\rm UL}$ by selecting a wide interval such as $T_{90}$ would reflect the constraint on the baryon loading over the entire period of the prompt emission, which is relevant with our discussion on the energy budget of UHECRs. However, the energy of the GRB may be released in the form of multiple individual flares within $T_{90}$, and the spectrum from MeV to GeV (i.e., the measured flux ratio between MeV emission and GeV emission) may vary from time to time significantly, even within a single flare.

Hence, we took GRB~190114C, which contained the additional components on certain time intervals as an example. We used the same time interval with \citet{ajello2020fermi}, i.e., $T_0+[0,2.3]$\,s, $T_0+[2.3,2.8]$\,s, $T_0+[2.8,3.8]$\,s, $T_0+[3.8,4.8]$\,s, and $T_0+[4.8,7.0]$\,s and considered the same photon field parameters, which were summarized in Table \ref{tab:lat-190114C}. Then, we defined the weighted baryon loading factor by $\bar{\eta}_{p}=\frac{E_{p,\rm tot}}{E_{\gamma,\rm tot}}=\frac{\Sigma\eta_{p,\rm i}L_{\gamma,\rm i}T_{\rm i}}{\Sigma_i L_{\gamma,i}T_i}$, where $\eta_{p,\rm i}$, $L_{\gamma,\rm i}$ and $T_{\rm i}$ are the baryon loading factor, gamma-ray luminosity, and the duration in the $i$th time bin in order to describe the tendency of $\eta_p$. After doing the Fermi-LAT data analysis, we used the reference model and obtained the results, as shown in Table \ref{tab:lat-190114C}. Basically, the average baryon factor is consistent with the previous one presented in Table~\ref{tab:binsize}, indicating no significant disparity between dividing into multiple sub-bins and utilizing the entire interval to constrain $\eta_{p}$.

We repeated the analysis of the Fermi data for each burst with equally cutting into three time intervals except for GRB~130427A (because of the events piled-up of the instrument). The corresponding results are summarized in Table~\ref{tab:fermi}. For most selected GRBs, the constraints are generally weaker by about $\lesssim1.5$ than reference case, except for GRB~221009A, GRB~190530A, GRB~180720B, GRB~170214A, GRB~160821A and GRB~131231A. These loosen constraints can attributed to $F^{\rm UL}_{\rm LAT}$ in the sub-bins. Utilizing shorter time intervals, the collecting photons maybe few, leading to low TS values and high $F^{\rm UL}_{\rm LAT}$ used in the model calculating (see the Fermi analyses tables in Appendix \ref{sec:LAT}). Therefore, dividing more sub-intervals of prompt phase should consider the influence of collecting the number of high energy photon, otherwise appearing a weaker constraints.

\begin{sidewaystable}
\tabcolsep=2pt
\caption{Spectral fitting to GBM and LAT data of GRB~190114C for various time intervals. Note that in this table, $\alpha$, $\alpha_{\rm CPL}$ (the index of a power law with an exponential cutoff) and $E_{\text{peak}}$ were from \cite{ajello2020fermi}. The average baryon factor is 9.8.}
\begin{flushleft}
\begin{tabularx}{\textwidth}{ccccccccccc}
\hline
\hline
Intervals & $\Gamma_{\text{LAT}}$$^{\mathrm{a}}$ & $F^{\rm UL}_{\text{LAT}}$$^{\mathrm{b}}$& TS&$\alpha$/&$\beta$&$E_{\text{peak}}$&$F_{\text{GBM}}$&$E_{\gamma}$&$\eta_{p}^{\mathrm{UL}}$\\ 
(s, from $T_{0}$)&&($10^{-7}$ erg~cm$^{-1}$~s$^{-1}$)&&$\alpha_{\rm CPL}$&&(keV)&($10^{-5}$ erg~cm$^{-1}$~s$^{-1}$)&(erg)&\\
\hline 
0 - 2.3&$2.0^{*}$&4.8 &$\sim0$& -0.73$\pm0.01$ & -4.00$\pm0.27$ & 548.6$_{-7.7}^{+7.6}$&5.0$_{-0.3}^{+0.2}$&$(8.0^{+0.3}_{-0.5})\times10^{52}$&0.9\\
2.3 - 2.8&$2.0^{*}$&22.4&$\sim0$&-0.36 $\pm$ 0.03&-&730.0$_{-15.5}^{+16.2}$&9.3$_{-0.6}^{+0.7}$&$(3.2^{+0.2}_{-0.2})\times10^{52}$&2.3\\
2.8 - 3.8&$2.4\pm0.9$&22.2&25&-0.04 $\pm$ 0.03&-&814.9$_{-13.0}^{+13.4}$&9.2$_{-0.7}^{+0.8}$&$(6.4^{+0.6}_{-0.5})\times10^{52}$&2.3\\
3.8 - 4.8&$3.3\pm0.4$&81.5&343&-0.05 $\pm$ 0.03&-3.63$_{-0.26}^{+0.21}$&563.1$_{-9.6}^{+8.8}$&9.2$_{-0.8}^{+0.9}$&$(6.4^{+0.7}_{-0.6})\times10^{52}$&8.8\\
4.8 - 7.0&$2.1\pm0.2$&117.1&986&-0.30 $\pm$ 0.04&-&425.4$_{-7.4}^{+7.7}$&2.2$_{-0.1}^{+0.2}$&$(4.9^{+0.4}_{-0.3})\times10^{52}$&50.0\\
\hline
\end{tabularx}
\end{flushleft}

$^{\mathrm{a}}$ The photon index. ULs are calculated with a photon index with 2.0 (labeled with *)\\
$^{\mathrm{b}}$ TS value of each interval; the significance of the GRB is approximate to $\sqrt{\mathrm{TS}}~\sigma$ ; the TS value of the interval that is less than 25 will be estimated as a 95\% C.L. UL (labeled with UL)\\
\label{tab:lat-190114C}
\end{sidewaystable}

\begin{table}
\addtolength{\tabcolsep}{0pt}
    \centering
    \caption{The results of $\eta_{p}$ with different methods. $\eta_{\rm p}^{\mathrm{UL}}$ is the constraint with selected interval  (see Table \ref{tab:binsize}). $\bar{\eta}^{\rm UL}_{\rm p}$ is the weighted average result during selected interval. Note that $\bar{\eta}^{\rm UL}_{\rm p}$ of GRB~130427A is from $T_{0}+[4.1, 4.9]$~s and $T_{0}+[11.4, 18]$~s, the available intervals.}
    \begin{tabular}{CCC}
    \hline
    \hline
     \rm Name & $\eta_{p}^{\mathrm{UL}}$ & $\bar{\eta}^{\rm UL}_{\rm p}$ \\ 
\hline
\rm 221009A  &1.8&23.2\\
\rm 211018A  &113&131\\
\rm 190530A  &1.2&2.8\\
\rm 190114C  & 8.8&9.8\\ 
\rm 180720B  & 0.6&1.6\\
\rm 170214A  & 4.6&10.9\\
\rm 160821A &  0.4&0.6\\
\rm 160625B &  11&12.8\\
\rm 160509A & 9.7&11.2\\
\rm 131231A &  1.8&2.8\\
\rm 130427A &  0.3&0.3\\
\rm 090902B &  42&59\\
    \hline
    \end{tabular}
    \label{tab:fermi}
\end{table}

\subsection{The energy production rate of GRBs} \label{sec:production}
The energy production rate required for GRBs to account for the observed flux beyond the ankle is estimated to approximately be $10^{44}$ ~erg~Mpc$^{-3}$~yr$^{-1}$. Additionally, observations from the Auger observatory have revealed a predominance of light composition in UHECRs at an energy level of around $10^{18.5}$~eV. Therefore, we assume that this composition primarily consists of protons, and the proportion of accelerated protons above $10^{18.5}$~eV can be determined using Equation \ref{eq:bol}:

\begin{equation}
f_{\text{bol}}=\frac{\int_{10^{18.5}(1+z)/\Gamma_{\text{bulk}}}^{10^{20}(1+z)/\Gamma_{\text{bulk}}}E_{p}Q_{p}(E_{p})dE_{p}}{\int_{E_{p,\text{min}}}^{E_{p,\text{max}}}E_{p}Q_{p}(E_{p})dE_{p}},
\label{eq:bol}
\end{equation}

The energy budget of GRBs can be described by Equation \ref{eq:budget}:
\begin{equation}
\mathcal{E}_{\text{iso},\gamma}=\frac{\int^{L_{\text{max}}}_{L_{\text{min}}}\mathcal{E}_{\text{iso},\gamma}(L_{\text{iso},\gamma}){\frac{d \rho_{0}}{dL}}d{L}}{\int^{L_{\text{max}}}_{L_{\text{min}}}{\frac{d \rho_{0}}{d L}}d{L}},
\label{eq:budget}
\end{equation}
where the luminosity function is described by $\frac{d \rho_{0}}{d L}=A_{0}\left[\left(\frac{L}{L_{b}}\right)^{\alpha_{1}}+\left(\frac{L}{L_{b}}\right)^{\alpha_{2}}\right]^{-1}$, with parameters of A$_{0}$, $\alpha_{1}=0.65$, $\alpha_{2}=2.3$ and $L_{b}=10^{52.5}$~erg~s$^{-1}$ \citep{2007ApJLiang662.1111L}. The lower and upper limits of luminosity ($L_{\text{min}}$ and $L_{\text{max}}$) are set to $10^{49}$~erg~s$^{-1}$ and $10^{54}$~erg~s$^{-1}$, respectively. Hence, the average observed rate of GRB is determined as $\Bar{\rho}_{0}=\int^{L_{\text{max}}}_{L_{\text{min}}}{\frac{d \rho_{0}}{d L}}d{L}=1.2~\rm Gpc ^{-3}~\rm yr^{-1}$. Statistical analysis of the correlation between isotropic energy and isotropic luminosity ($L_{\text{iso},\gamma}$), based on well-measured GRBs detected by the Fermi and Konus-Wind detectors, yields $(0.94\pm0.03)\text{log}(\frac{\mathcal{E}_{\text{iso},\gamma}}{{10^{52}~\text{erg}}})=\text{log}(\frac{L_{\text{iso}},\gamma}{{10^{52}~\text{erg/s}}})+(0.51\pm0.05)$ \citep{xue2019characteristics}. Using these established correlations, the calculated value of $\mathcal{E}_{\text{iso},\gamma}$ in Equation \ref{eq:budget} is $1.3\times10^{53}~\rm erg$, leading to $Q_{\rm UHECR}^{\rm GRB}$. Alternative luminosity functions and empirical relationships $L_{\rm iso, \gamma}-\mathcal{E}_{\text{iso}}$ have been suggested \citep[e.g.,][]{2010MNRASWanderman1944W, Ghirlanda2010}; however, utilizing these alternatives would not significantly alter the results.

We examine the values of $f_{\text{bol}}$ for the selected GRBs and find the minimums of $\Gamma_{\rm bulk}$ in the allowed regions (Case C in Table \ref{tab:parameter-space}) that satisfy $f_{\text{bol}}\eta_{p}\ge1$. Note that the condition of $f_{\text{bol}}\eta_{p}\ge1$ falls outside the allowable region (unless $\Gamma_{\rm bulk}>1000$) for GRB~190530A, GRB~180720B, GRB~160625B and GRB~130427A. The reasons are as follows. GRB~130427A and GRB~160625B stand out as the most luminous GRBs among all considered GRBs in the selected time intervals. The high luminosity results in a high flux from the EM cascade, leading to a strong constraint of $\eta^{\text{UL}}_{p}= 4$ and $\eta^{\text{UL}}_{p}= 11$ in Case C for GRB~130427A and GRB~160625B, respectively. For GRB~190530A and GRB~180720B, we have strong constraint on $R_{\rm diss}$ and $\Gamma_{\rm bulk}$ given a short variability timescale $T_{\rm var}=32\,$ms and $T_{\rm var}=18\,$ms, respectively. Therefore, we select $\Gamma_{\rm bulk}$ and $R_{\rm diss}$ to maximize $Q^{\text{GRBs}}_{\text{UHECR}}$ within the allowable region for these three GRBs as Case C (labeled with the asterisk symbol in Table.~\ref{tab:parameter-space}), resulting in $f_{\rm bol}\eta_p=$ 0.82, 0.33, 0.37, and 0.14 for Case C of GRB~190530A, GRB~180720B, GRB 160625B, and GRB~130427A, respectively. For other GRBs, large values of $\Gamma_{\mathrm{bulk}}$ ($\gtrsim600-1000$) and $R_{\rm diss}$ ($\gtrsim 10^{15}$ cm) are also suggested in Case C.

In addition to employing high values of $\Gamma_{\rm bulk}$ or $R_{\rm diss}$, we also explore whether modifications of some other parameters could result in a higher value of $f_\mathrm{bol}\eta_p$. Firstly, reducing the constraint on $\eta_{p}$ can be achieved by considering a lower acceleration efficiency (see Table~\ref{tab:parameter-space}), which will lead to a smaller $E_{p,\rm max}$. As such, there would be much fewer protons that can efficiently participate in photomeson or BH processes, and the anticipated EM cascade emission. However, GRBs cannot be UHECR accelerators in this case and result in a significant reduction of $f_{\rm bol}$. Hence, we do not look further into this direction. Note that if the acceleration efficiency is higher, $E_{\rm p,max}$ will be higher and lead to a few times greater $f_{\rm bol}$ and $p\gamma$ reaction efficiency. In this case, EM cascade flux will be higher and the resultant constraint on the baryon loading factor will be stronger, and our conclusion will not be changed. 

Subsequently, we examine the impact of the spectral index of injected protons by taking GRB~221009A as our sample. We conduct a comparative analysis for two variations of $s_p$ and present the results in Table~\ref{tab:density}. When adopting a harder index ($s_{p}=1.5$), it increases the ratio of particle energy in the UHE band (i.e., a larger $f_{\rm bol}$. Since higher energy protons have larger $p\gamma$ interaction efficiency, it consequently deposits more energy of protons into EM cascade and pose more stringent constraints on $\eta_p^{\rm UL}$. Conversely, injecting protons with a softer spectral index ($s_{p}=2.5$) relaxes the constraint on $\eta_p$, but results in significantly smaller values for $f_\mathrm{bol}$. In other words, changing the spectral index does not alter our conclusion significantly. As we can see in the Table, in Case A and B, the constraint on the energy production rate of UHECRs are not sufficient to explain the observation. If GRBs have large $\Gamma_{\rm bulk}$ and/or $R_{\rm diss}$ (as represented by Case C), the allowed energy production rate $Q_{\rm UHECR}^{\rm GRBs}$ can reach $\sim 10^{44}~\rm erg~Mpc^{-3}yr^{-1}$, and we see that the value is not sensitive to $s_p$\footnote{For GRB~221009A, $s_{p}=2.5$ leads to a very small $f_\mathrm{bol}$ of $1.3\times10^{-5}$ in Case C and allows a very loose constraint on $\eta^{\text{UL}}_{p}$ with a value of $\sim 92360$. $Q_{\rm UHECR}^{\rm GRBs}$ can reach $\sim 10^{44}~\rm erg~Mpc^{-3}yr^{-1}$ only at this unrealistically limiting value, challenging to scenario of the soft index of injecting protons in GRBs.}.

\begin{table}
\addtolength{\tabcolsep}{0pt}
    \centering
    \caption{The values of $\eta^{\text{UL}}_{p}$, $f_{\text{bol}}$ and $Q^{\text{GRBs}}_{\text{UHECR}}$ in different cases for GRB~221009A. $E^{\text{obs}}_{p,\text{max}}$ is the maximum energy of protons in observer frame.}
    \begin{tabular}{CCCCCCC}
    \hline
    \hline
        \text{Cases}& \text{Index} &E^{\text{obs}}_{p,\text{max}} ($\mathrm{EeV}$)& $\eta^{\text{UL}}_{p}$ & $f_{\text{bol}}$ & $Q^{\text{GRBs}}_{\text{UHECR}}\,(\text{erg}\,\text{Mpc}^{-3}\,\text{yr}^{-1})$  \\ 
        \hline
\text{Reference}&s_{p}=1.5&3.5&0.4&1.1\times10^{-1}&7.2\times10^{42}\\
&s_{p}=2.0&3.5&1.8&9.5\times10^{-3}&2.7\times10^{42}\\
 &s_{p}=2.5&3.5&70&1.8\times10^{-5}&7.4\times10^{41}\\
 \hline
\text{Case A}&s_{p}=1.5&3.1&0.2&8.6\times10^{-2}&4.0\times10^{42}\\
&s_{p}=2.0&2.5&1.8&5.3\times10^{-3}&1.5\times10^{42}\\
 &s_{p}=2.5&2.5&21&2.7\times10^{-6}&9.3\times10^{40}\\
 \hline
\text{Case B}&s_{p}=1.5&1.6&0.1&1.9\times10^{-2}&3.0\times10^{41}\\
&s_{p}=2.0&1.2&0.2&6.3\times10^{-4}&2.0\times10^{40}\\
 &s_{p}=2.5&0.8&8.3&7.8\times10^{-8}&1.1\times10^{38}\\
 \hline
\text{Case C}&s_{p}=1.5&3.6&8.0&1.3\times10^{-1}&1.8\times10^{44}\\
 &s_{p}=2.0&2.9&130&8.2\times10^{-3}&1.7\times10^{44}\\
 &s_{p}=2.5&2.3&92360&1.3\times10^{-5}&2.0\times10^{44}\\
    \hline
    \end{tabular}
\label{tab:density}
\end{table}

\subsection{The influence of the IC emission of primary electron} \label{sec:IC}
There have been suggestions that the prompt keV-MeV emission of GRBs may be attributed to synchrotron emission of relativistic electrons \citep{2020NatAs...4..174B,ryde2022onset,2018ApJ...853...43X,2019A&A...625A..60R}. These electrons will inevitably produce IC emission, contributing to the GeV-TeV gamma-ray band and initiating EM cascade. Considering the potential contribution from primary electrons will reduce the room left for the hadronic processes and hence make the constraints on $\eta_p$ even more stringent. To evaluate their impact on our results, we assume the presence of a population of quasi-stable electrons during the prompt emission phase,  following a broken power-law distribution in the dissipative region. The power-law indexes and the total energy of the electrons are determined by fitting the Band function with their synchrotron radiation. Then, we calculate the corresponding IC emission and the consequent cascade emission. To make a conservative estimation of the IC emission, we employ $\chi_{\rm B}=10$. This is because $\chi_{\rm B}$ basically determines the ratio between the magnetic field energy density and the radiation energy density in the dissipative region, the predicted IC flux roughly scales with $\chi_{\rm B}^{-1}$. The values of $\Gamma_{\rm bulk}$ and $R_{\rm diss}$ of each GRB follow  
those of Case C, in which the required energy production rate of UHECRs can be achieved for selected GRBs.

\begin{figure}
\centering
\includegraphics[width=0.5\textwidth]{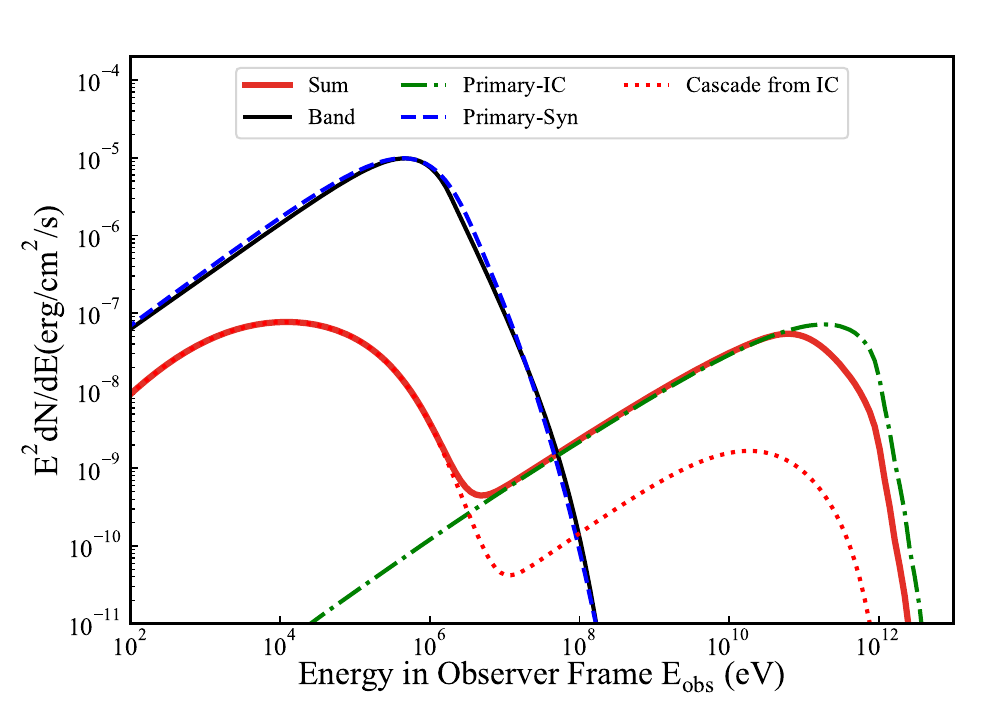}
\caption{The synchrotron (blue dash curve) and IC emission (dotted-dashed curve) of primary electrons for GRB~221009A with $R_{\rm diss}=10^{15}\rm cm$, $\Gamma_{\rm bulk}=470$ and $\chi_{\rm B}=10$. The black solid curve is the Band function. The red dotted curve is cascade induced by IC, and the red solid curve is the sum for them.}
\label{fig:IC}
\end{figure}

We show the IC emission and corresponding cascade emission of GRB~221009A in Figure~\ref{fig:IC} as an example. The black curve shows the Band function and the dashed blue curve represents the synchrotron emission of primary electrons. The dot-dashed green curve is the intrinsic IC emission of primary electrons. The high-energy part of the IC emission will be absorbed by the keV-MeV radiation field and reprocessed into lower energies via cascade processes, as shown with the dotted red curve. The solid red curve is the sum of the unabsorbed IC emission and cascade emission. In $0.1-10\,$GeV, the flux from primary electrons $F_{0.1-10}$ reaches 23\% of $F_{\rm LAT}^{\rm UL}$, and therefore would lower the obtained $\eta_p^{\rm UL}$ by a factor of 0.77. The results for other selected GRBs are shown in Table~\ref{tab:IC}. We see that the influence of the IC emission of primary electrons varies in a large range, mainly due to different fluxes or flux upper limits measured by Fermi-LAT and different levels of suppression on the IC flux by the KN effect for different GRBs. Note that we aim to pose upper limits on $\eta_{\rm p}$ and UHECR energy production rate, so the constraints obtained in previous sections remain valid even if primary electrons have important contributions to the prompt emission at the GeV band.

\begin{table}
\addtolength{\tabcolsep}{0pt}
    \centering
    \caption{The value of each variable in the case of $\chi_{\rm B}=10$. $F_{0.1-10}$ is the flux within 0.1-10~GeV. If the combination of $\Gamma_{\rm bulk}$ and $R_{\rm diss}$ can not be found in Case C for the selected GRBs, we pick up the values of $\Gamma_{\rm bulk}$ and $R_{\rm diss}$ which can generate the maximum of $Q^{\text{GRBs}}_{\text{UHECR}}$ in the allowable region (labeled with *).}
    \begin{tabular}{CCCCCC}
    \hline
    \hline
     \rm Name &$\Gamma_{\rm bulk}$ & $R_{\rm diss}$&$F_{0.1-10}$&$F^{\rm UL}_{\rm LAT}$&\text{Ratio}=($\frac{F_{0.1-10}}{F^{\rm UL}_{\rm LAT}}$) \\ 
    &&\rm (cm)&(\rm erg~cm$^{-1}$~s$^{-1}$)&(\rm erg~cm$^{-1}$~s$^{-1}$)&(\%) \\ 
\hline
\rm 221009A&470&1.0\times10^{15}&5.2\times10^{-8}&2.2\times10^{-7}&23.1\\
\rm211018A&180&2.1\times10^{14}&2.2\times10^{-10}&2.5\times10^{-7}&0.09\\
\rm 190530A^{*}&1000&1.0\times10^{15}&8.7\times10^{-9}&2.4\times10^{-7}&3.6\\
\rm 180720B^{*}&1000&6.5\times10^{14}&4.1\times10^{-9}&4.7\times10^{-8}&8.3\\
\rm 170214A& 670&2.1\times10^{15}&2.1\times10^{-10}&1.5\times10^{-8}&1.4\\
\rm 160821A &480&2.2\times10^{15}&6.1\times10^{-10}&4.5\times10^{-8}&1.4\\
\rm 160625B^{*} &1000&1.6\times10^{15}&6.1\times10^{-9}&5.7\times10^{-7}&1.1\\
\rm 160509A &375&3.5\times10^{14}&5.6\times10^{-10}&2.4\times10^{-7}&0.2\\
\rm131231A&640&2.4\times10^{15}&7.6\times10^{-10}&2.2\times10^{-8}&3.5\\
\rm 130427A^{*}&1000&2.0\times10^{15}&4.9\times10^{-8}&9.4\times10^{-7}&7.0\\
    \hline
    \end{tabular}
    \label{tab:IC}
\end{table}

\section{CONCLUSION} \label{sec:conclusion}
If GRBs are the main factories of ultrahigh-energy cosmic rays, the baryon loading factor, $\eta_{\rm p}$ should be exceed 10 in order to satisfy the effective energy production rate. Then, these UHECRs will undergo interactions with the photon radiation of the GRB through the photomeson and BH processes, resulting in a series of EM cascade extending up to the GeV-TeV $\gamma$-ray regime. Besides, the expected $\gamma$-ray from cascades rely on the properties of GRB jet, such as $R_{\rm diss}$, $\Gamma_{\rm bulk}$ and $\eta_{p}$. Thus, observations of Fermi-LAT can impose constraints on these important parameters.

We selected GRBs with both high keV–MeV fluence and known redshift information within the LAT’s field of view from the GBM catalogs, including GRB~221009A, GRB~211018A, GRB~190530A, GRB~190114C, GRB~180720B, GRB~170214A, GRB~160821A, GRB~160625B, GRB~160509A, GRB~131231A, GRB~130427A and GRB~090902B. We did Fermi analyse and obtained their GBM and LAT light curves. We assume protons are accelerated in GRBs and their total energy budget is related to the keV-MeV radiation energies via $\eta_p$. We deal with the interactions between protons and GRB radiation fields, including the EM cascade emission assuming a quasi-stable state is reached. The anticipated cascade flux in the energy range of $0.1-10$\,GeV generally exhibit a linear correlation with the value of $\eta_{p}$. Thus, we may pose constraints on the upper limit of $\eta_{p}$, $\eta_{p}^{\text{UL}}$, by ensuring the expected flux in each time interval not to overshoot the 95\% C.L. UL of the LAT flux in the same energy range. The main results are as follows.

\begin{itemize}
	\item We found that stringent constraints on the baryon loading factors can be obtained for most selected GRBs, with upper limits of $\eta_{p}$ ranging from 0.3 to $\sim10$ in the reference case, which is more restrictive than the neutrino stacking observation of IceCube, except for GRB~211018A and GRB~090902B.
	\item We scanned a large parameter space for $R_{\rm diss}$ and $\Gamma_{\rm bulk}$, and obtained the 2D constraint map of $\eta_{p}^{\text{UL}}$. In the meanwhile, considering the requirement from the variability timescales and the transparency of the highest observed photons, we defined the allowable region in the parameter space and obtained the constraints.
    \item We discussed the energy production rate of UHECRs based on the obtained $\eta_p^{\rm UL}$. Generally, large values of $\Gamma_{\mathrm{bulk}}$ ($\gtrsim 600$) and $R_{\rm diss}$ ($\gtrsim 10^{15}$ cm) are needed to satisfy $\eta_p^{\rm UL}>10$ which provide a sufficient energy production rate of $\rm 10^{44}~erg~Mpc^{-3}yr^{-1}$ to account for the measured UHECR flux. The required parameters is inconsistent with the photosphere model but follow the prediction of the ICMART model. The IS model is not favored but cannot be excluded neither.
    \end{itemize}

We also examined the impact of the IC emission and its cascade from a primary population of quasi-stable electrons, assuming that their synchrotron radiation accounts for the Band spectrum in the keV-MeV band. We found that the contribution of primary electrons may range from  0.1\% up to 20\% of $F_{\rm LAT}^{\rm UL}$ within 0.1-10~GeV for considered GRB samples under a conservative assumption of $\chi_B=0.1$. A smaller $\chi_B$ would result in a higher IC flux. Considering the radiation of primary electrons would leave a smaller room for GeV emission of hadronic origin, and thus leads to more stringent constraints on $\eta_p^{\rm UL}$.

Finally, it is worth mentioning that the results of this study is based on the single-zone model of GRBs. If the radiation zone of the keV-MeV  photons is different from proton acceleration zone, our constraint cannot apply. Besides, our results only apply to the prompt emission phase of GRBs, and the afterglow phase of GRBs could be independent UHECR accelerators \citep[e.g.,][]{vietri1995acceleration, waxman1995cosmological, Asano16,2021PhRvD.104j3005Zhang}.

\section*{Acknowledgements}
We thank Bing Zhang for helpful discussion. This study is supported by National Natural Science Foundation of China under grants No. 12393852, 12333006, 12121003.

\appendix
\section{Details of Fermi Data Analysis} \label{sec:LAT}
 $\emph{\text{GRB~221009A}}$: GRB~221009A triggered the GBM instrument 13:16:59.99 UT on 2022 October 9. Detailed analysis of the GBM data for energies 50–300 keV yields a formal $T_{90}$ duration of $\sim289$~s. \cite{2023ApJLesageGBM} plotted its light curve in the 20 keV to 40 MeV energy range, containing a triggering pulse (starting from $T_{0}-1$~s to $T_{0}+43$~s), a quiet period (starting from $T_{0}+121$~s to $T_{0}+163$~s) and pre-main pulse (starting from $T_{0}+177$~s to $T_{0}+210$~s). However, the data of GBM from $T_{0}+219$~s to $T_{0}+277$~s and $T_{0}+508$~s to $T_{0}+514$~s was piled-up \citep{2023ApJLesageGBM} and the early afterglow had appeared at $T_{0}+226$~s \citep{Science2023LHAASO}. Hence, we selected detectors n7, n8, and b1 for the data processing procedures of GRB~221009A. We picked up the interval from $T_{0}+177$~s to $T_{0}+219$~s divided into 3 intervals. The hightest energy photon is 1.4~GeV during entire interval. The result of $\emph{\text{Fermi}}$ analysis can be obtained in Table~\ref{tab:lat-221009A}.

 $\emph{\text{GRB~211018A}}$: GRB~211018A triggered the GBM instrument 22:28:13.94 UT on 18 October 2021. Detailed analysis of the GBM data for energies 50–300 keV yields a formal $T_{90}$ duration of $\sim123.9$~s, starting from $T_{0}+4.3$~s \citep{2021GCN.30947....1R}. Hence, we selected detectors n0, n1, and b0 for the data processing procedures of GRB~211018A. We picked up the interval from $T_{0}+4.3$~s to $T_{0}+128.2$~s divided into 3 intervals. The hightest energy photon is 3.3~GeV during entire interval. The result of $\emph{\text{Fermi}}$ analysis can be obtained in Table~\ref{tab:lat-211018A}.
 
$\emph{\text{GRB~190530A}}$: GRB~190530A triggered GBM at 10:19:08 on 2019 May 30 \citep{team2019grb}. Its redshift was set to 0.9386 \citep{gupta2022probing}. Detailed analysis of the GBM data for energies 50–300 keV yields a formal $T_{90}$ duration of $\sim18.4$~s, starting from $T_{0}+1.8$~s to $T_{0}+20.2$~s. Similar to the methodology used for GRB~160821A, but we employed detectors n1, n3, and b0 for this event. The $\gamma$-ray emission light curve of GRB~190530A exhibits a quiet period ($T_{0}+1.8$~s to $T_{0}+7.8$~s) \citep{gupta2022probing}. We picked up interval $T_{0}+[7.8,20.2]$~s and further divided into 3 intervals. The hightest energy photon is 5.3~GeV during entire interval. The results of $\emph{\text{Fermi}}$ analysis can be obtained in Table~\ref{tab:lat-190530A}.

$\emph{\text{GRB~190114C}}$: On 14th January 2019 at 20:57:02.63 UTC, the GBM instrument triggered and precisely localized GRB~190114C \citep{hamburg2019grb}. The GBM team reported a main emission duration, $T_{90}$, of 116~s in the energy range of 50–300 keV and they determine the redshift of z = 0.42 \citep{hamburg2019grb}. For the data processing procedures of GRB~190114C, we followed a method similar to that used for GRB~160821A, with the selection of detectors n3, n4, and b0 for the analysis. \cite{ajello2020fermi} reported that at about $\sim7$ s GRB~190114C had been transferred to afterglow-dominated emission from the prompt. We divided this interval into the same 5 intervals. The hightest energy photon is 12~GeV during entire interval. The result of $\emph{\text{Fermi}}$ analysis can be obtained in Table~\ref{tab:lat-190114C}.

$\emph{\text{GRB~180720B}}$: On 20th July 2018 at 14:21:39 UT, Fermi/GBM detected GRB~180720B \citep{bissaldi2018grb}. The redshift of the burst was precisely measured using the X-shooter instrument on the VLT UT2 telescope, yielding a value of z = 0.654 \citep{vreeswijk2018grb}. For the data processing procedures of GRB~180720B, we followed a similar approach as employed for GRB~160821A, but selecting detectors n6, n7, and b1. The duration of $T_{90}$ was measured as 49~s (starting from $T_{0}+4.3$~s) within the energy range of 50–300 keV \citep{roberts2018grb}. Besides, \cite{2020A&A636A55R} reported that the emission detected by LAT up to 35 s was characterised by a soft spectrum with spectral index $\Gamma_{\rm LAT}\ge3$ and hard photon indexes were needed from 35 s onwards. They had interpreted this as the emission produced by the external forward shock. Hence, we picked up the time interval from $T_{0}+4.3$~s to $T_{0}+35$~s, which we subsequently divided into 3 sub-intervals for a more detailed examination. The hightest energy photon is 0.8~GeV during entire interval. The result of $\emph{\text{Fermi}}$ analysis can be obtained in Table~\ref{tab:lat-180720B}.

$\emph{\text{GRB~170214A}}$: GRB~170214A triggered the GBM instrument at 15:34:26.92 UTC. A distinctive characteristic of GRB~170214A is its composition of multiple sub-bursts, and its $T_{90}$ is about 123~s (starting from $T_{0}+12.5$~s) \citep{2017GCN.20675....1M}. \cite{2017ApJ...844...56T} plotted the LAT and GBM light curves of GRB~170214A and found the fast rising and fast decaying behaviors after $T_{90}+52$~s, which were usually considered as the potential early afterglow component \citep{2017ApJ...848...15F}. Hence, we selected detectors n0, n1, and b0 for the data processing procedures of GRB~170214A. We picked up the time interval from $T_{0}+12.5$~s to $T_{0}+52$~s and divided it into 3 intervals for meticulous examination and analysis. The hightest energy photon is 3~GeV during entire interval. The result of $\emph{\text{Fermi}}$ analysis can be obtained in Table~\ref{tab:lat-170214A}.

$\emph{\text{GRB~160625B}}$: GRB~160625B triggered the GBM instrument at 22:40:16.28 UTC. A distinctive characteristic of GRB~160625B is its composition of three sub-bursts, separated by two quiescent times of approximately 180~s and 339~s, respectively \citep{troja2017significant,zhang2018transition}. Its $T_{90}$ is about 483~s, starting from 188~s. We employed a method akin to that used for GRB~160821A, selecting detectors n6, n9, and b1 for the data processing procedures of GRB~160625B. \cite{2017ApJ...848...15F} reported the early afterglow arose from $T_{0}+\sim153~\rm s$ and dominated the emission from $T_{0}+200~\rm s$, so we picked up the intervals $T_{0}+[188.4,~200]~\rm s$ and divided into 3 intervals for meticulous examination and analysis. The hightest energy photon is 2.8~GeV during entire interval. The result of $\emph{\text{Fermi}}$ analysis can be obtained in Table~\ref{tab:lat-160625B}.

$\emph{\text{GRB~160509A}}$: On 9th May 2016 at 08:59:07.16 UT, GBM detected the photon flux from GRB~160509A, leading to its triggering \citep{roberts2016grb}. The redshift of this event was determined to be z = 1.17 \citep{tanvir2016grb}. For GRB~160509A, we followed a methodology similar to that used for GRB~160821A, but selecting detectors n0, n3, and b0. Its $T_{90}$ is around 370~s. \cite{2017ApJ...844L...7T} pointed out the observed trend of a soft-to-hard transition in the evolution of the photon index within the LAT band at $T_{0}+40~\rm s$, which we interpreted as indicative of an early afterglow, akin to that seen in GRB~180720B \citep{2020A&A636A55R}. Besides, its the light curve of GBM consisted of a soft precursor peak between $T_{0}+[-5,~5.5]~\rm s$. Hence, we picked up the time interval from $T_{0}+5.5$~s to $T_{0}+40$~s, which we further divided into 3 intervals for detailed analysis and examination. The hightest energy photon is 2.3~GeV during entire interval. The result of $\emph{\text{Fermi}}$ analysis can be obtained in Table~\ref{tab:lat-160509A}.

$\emph{\text{GRB~131231A}}$: GRB~131231A triggered GBM at 04:45:16.08 UT ($T_{0}$) on 2013 December 31. The redshift of this GRB is $\sim$ 0.642 \citep{cucchiara2014grb}. We followed a method similar to that used for GRB~160821A, but selecting detectors n0, n3, and b0. The light curve of the prompt emission exhibits a single large peak profile, with the main flare of $\sim$31~s (starting from $T_{0}+13.3$~s) in the 50-300~keV band \citep{jenke2014grb}. \cite{2014ApJ...787L...6L} plotted the light curve of the high energy range from 0.1 to 100~GeV emission and found the flux decaying following a PL with an index of -1.29, indicating the dominated early afterglow emission. Hence, we picked up the interval from  $T_{0}$~s to $T_{0}+30$~s, which we further divided into 3 intervals. The hightest energy photon is 0.4~GeV during entire interval. The result of $\emph{\text{Fermi}}$ analysis can be obtained in Table~\ref{tab:lat-131231A}.

$\emph{\text{GRB~130427A}}$: GRB~130427A triggered GBM at 47:06.42 UT on 27 April 2013. The redshift of this GRB is $\sim$ 0.34. We followed a method similar to that used for GRB~160821A, but selecting detectors n9, n10, and b1. The overall duration of the prompt emission was $T_{90}$~s (15 to 150 keV) = 276~s, starting from $T_{0}+4.1$~s. \cite{2013ApJ...779L...1K} plotted its light curves and found the early afterglow had dominated the GeV emission from  $\sim T_{0}+18$~s. Many articles held similar views, seeing \cite{2013ApJ...773L..20L} and \cite{2017ApJ...844...92F}. Besides, the data of GBM from $T_{0}+4.9$~s to $T_{0}+11.4$~s was piled-up. Hence, we ignored this interval and picked up the interval from $T_{0}+4.1$~s to $T_{0}+4.9$~s and $T_{0}+11.4$~s to $T_{0}+18$~s. The hightest energy photon is 9.9~GeV during entire interval. The result of $\emph{\text{Fermi}}$ analysis can be obtained in Table~\ref{tab:binsize}.

 $\emph{\text{GRB~090902B}}$: GRB~090902B triggered the GBM instrument at 11:05:08.31 UTC. Detailed analysis of the GBM data for energies 50–300 keV yields a formal $T_{90}$ duration of 21.9~s starting at $T_{90}+2.8$~s \citep{2009ApJ...706L.138A}. \cite{2009ApJ...706L.138A} and \cite{2011ApJ...730..141Z} reported the early afterglow component had appeared at $T_{90}+9$~s. Hence, we selected detectors n0, n1, and b0 for the data processing procedures of GRB~090902B. We picked up the interval from $T_{0}+2.8$~s to $T_{0}+9$~s divided into 3 intervals. The hightest energy photon is 1.3~GeV during entire interval. Different with other selected sources, we found an additional PL component was needed during $T_{0}+[7.0,~9]~\rm s$, then we used SBPL+PL to fit its photon field. The result of $\emph{\text{Fermi}}$ analysis can be obtained in Table~\ref{tab:lat-090902B}.

\section{Calculation of the minimum variability timescale.} \label{sec:variability}
\begin{figure}
\centering
\includegraphics[width=0.45\textwidth]{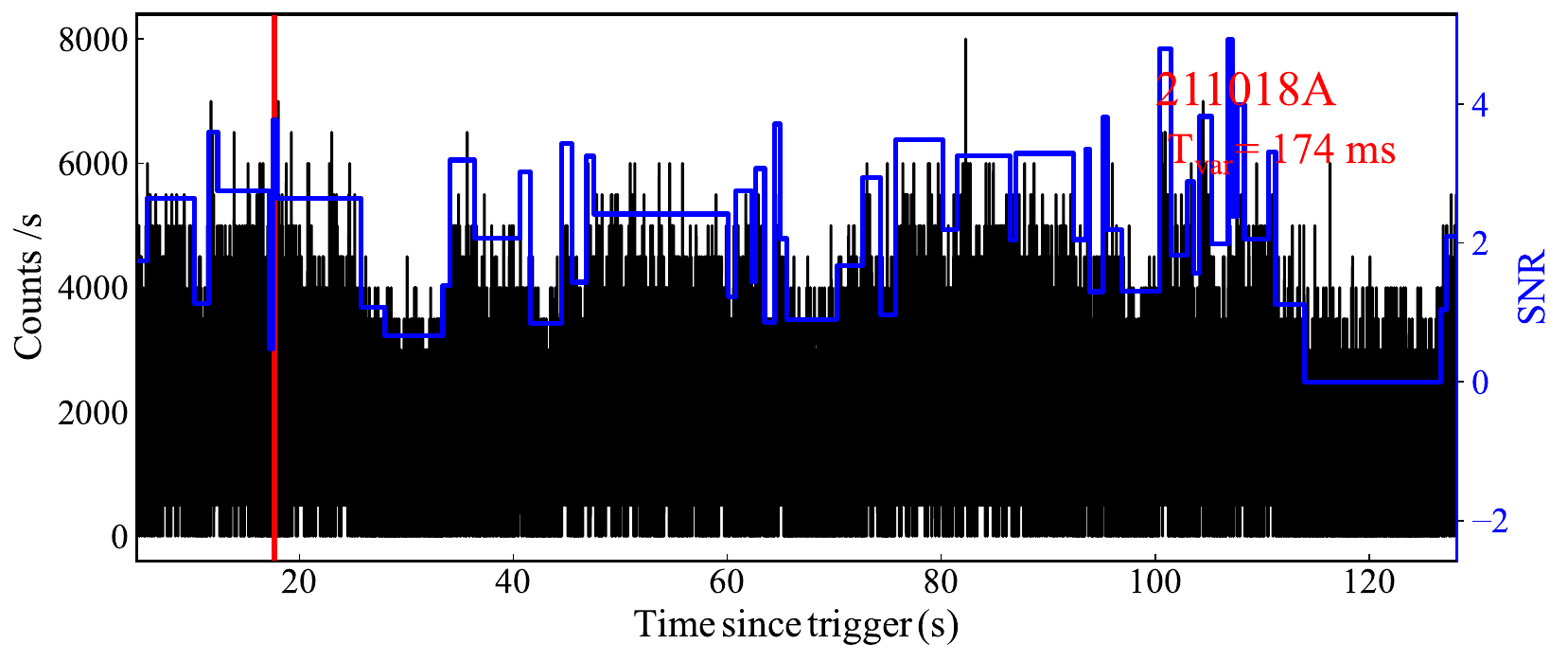}
\includegraphics[width=0.45\textwidth]{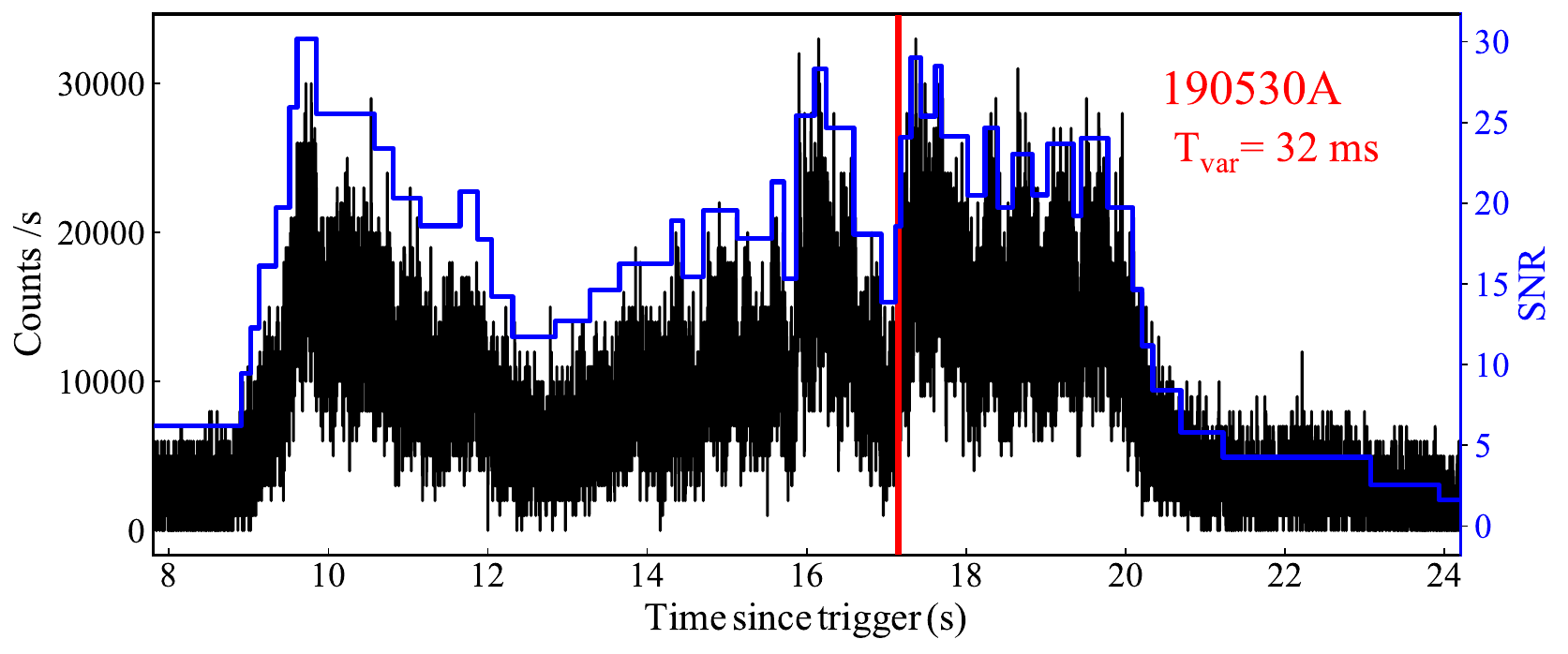}
\includegraphics[width=0.45\textwidth]{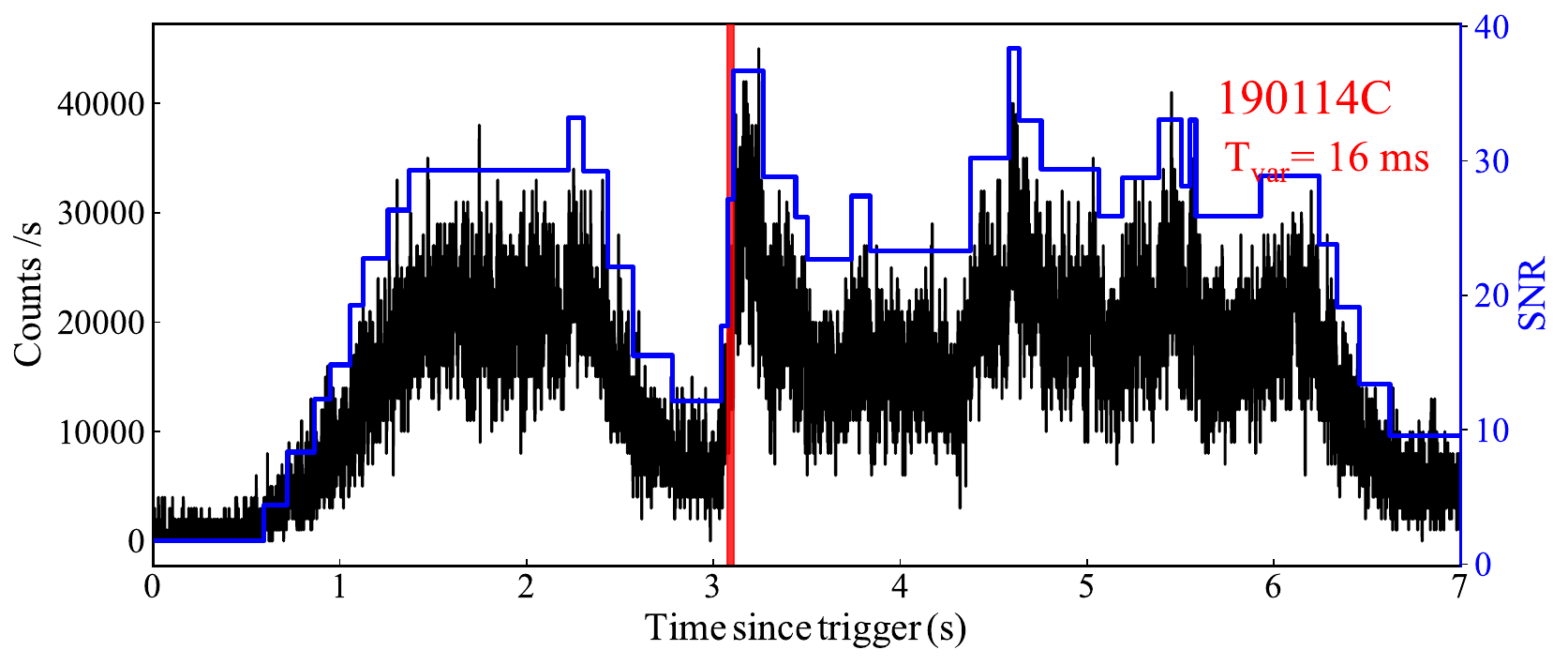}
\includegraphics[width=0.45\textwidth]{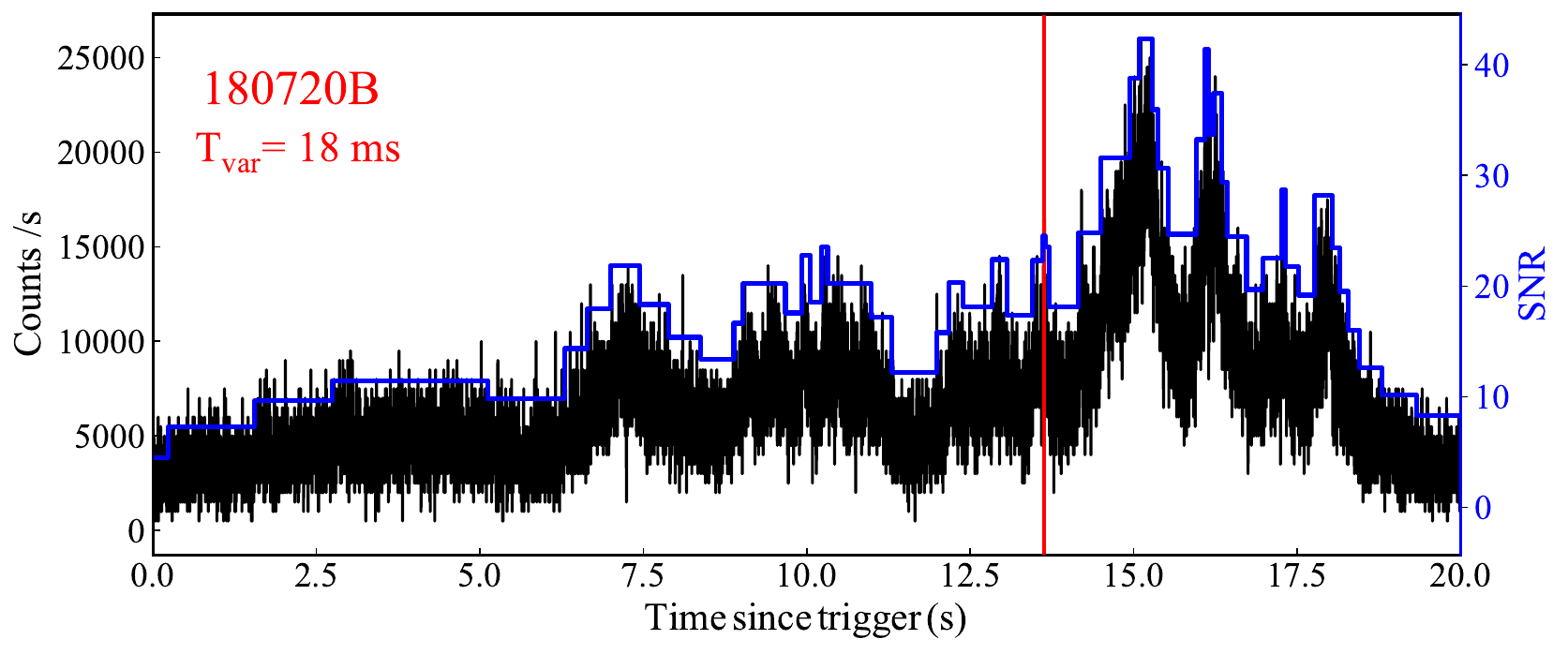}
\includegraphics[width=0.45\textwidth]{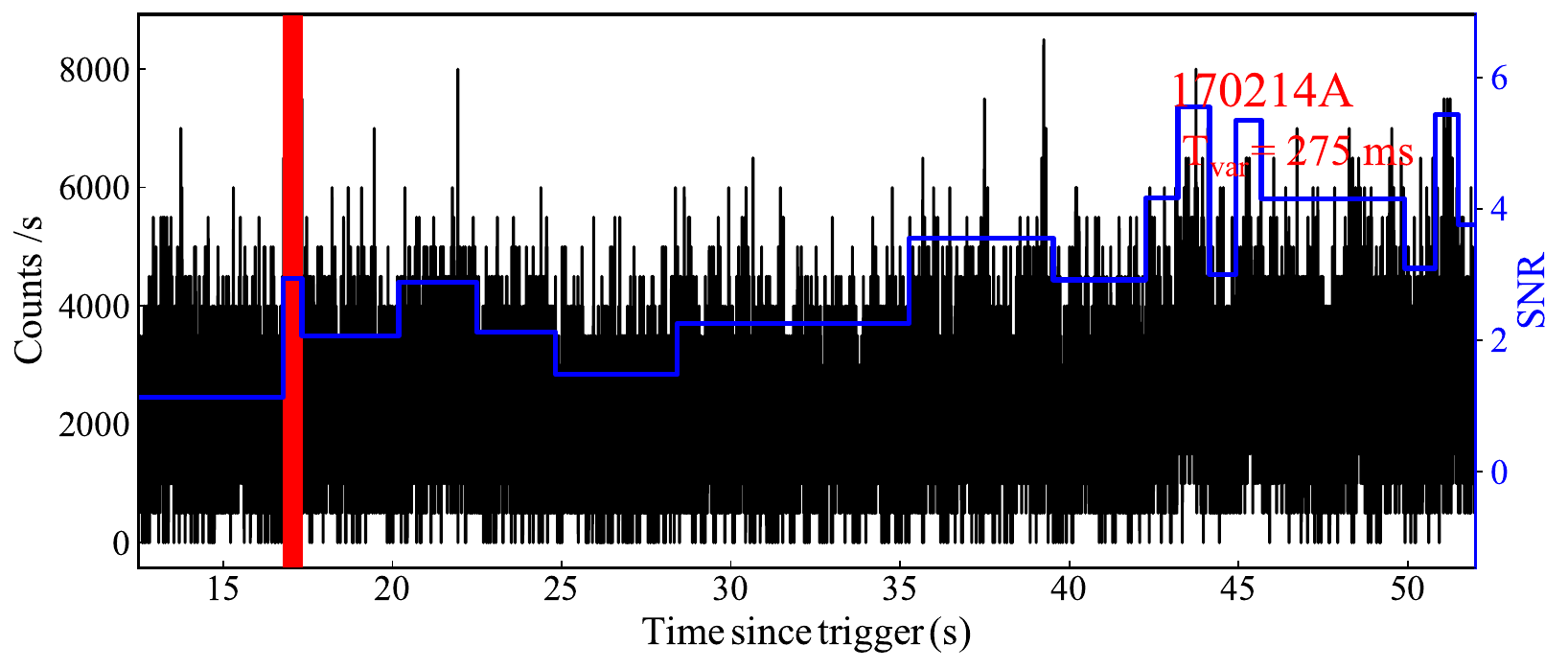}
\includegraphics[width=0.45\textwidth]{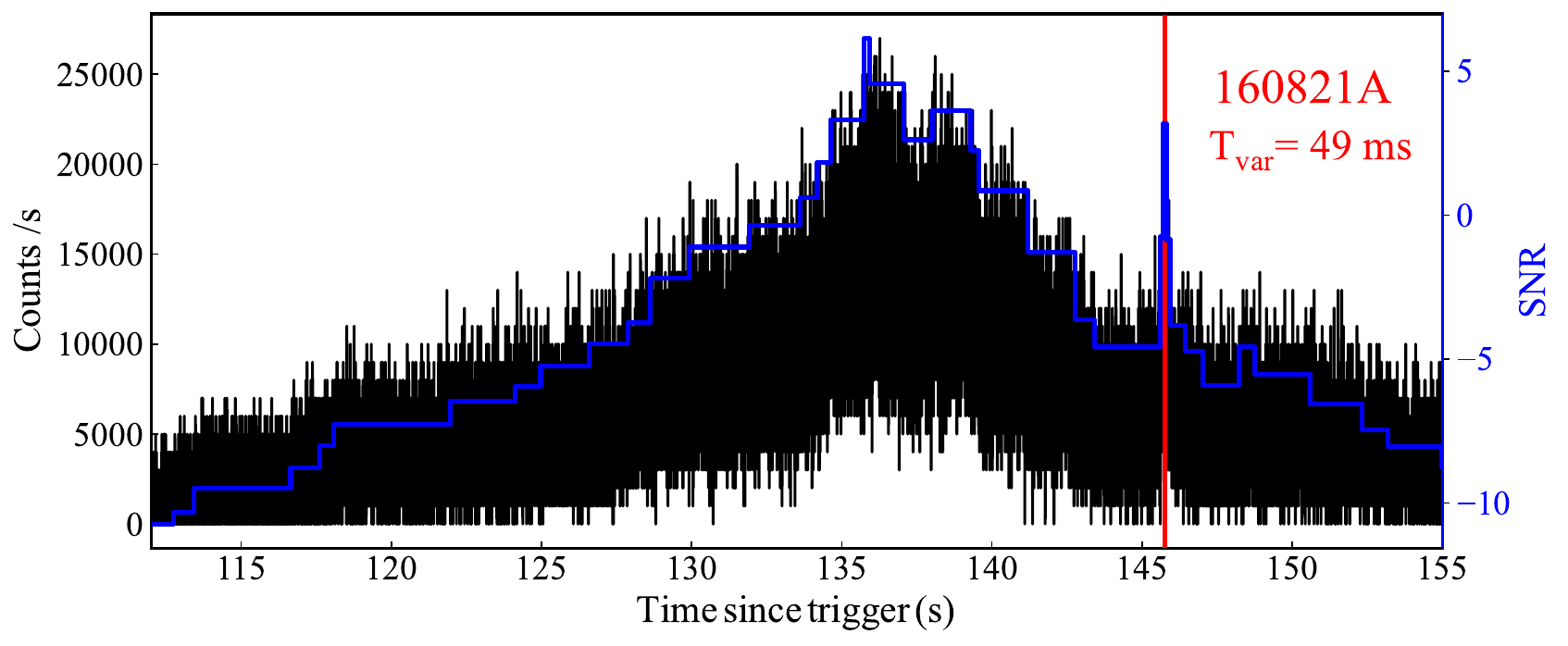}
\includegraphics[width=0.45\textwidth]{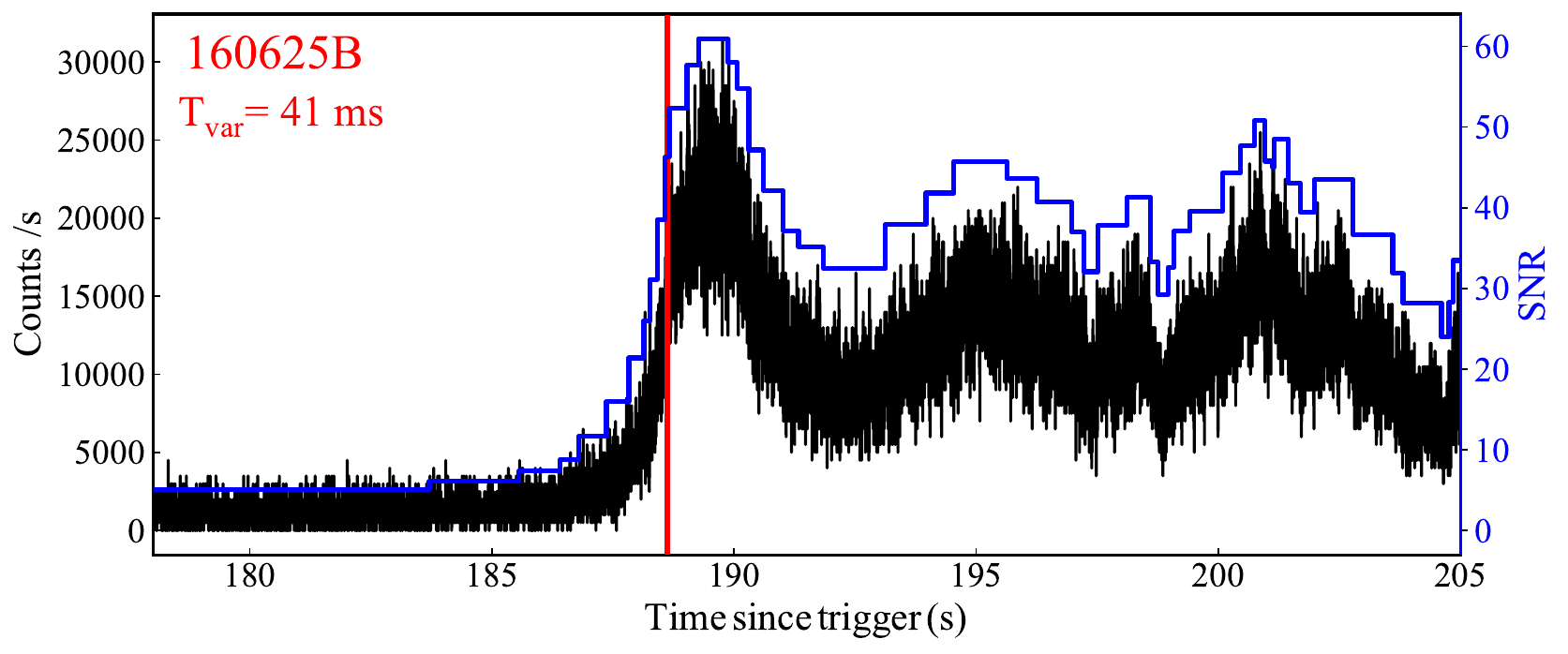}
\includegraphics[width=0.45\textwidth]{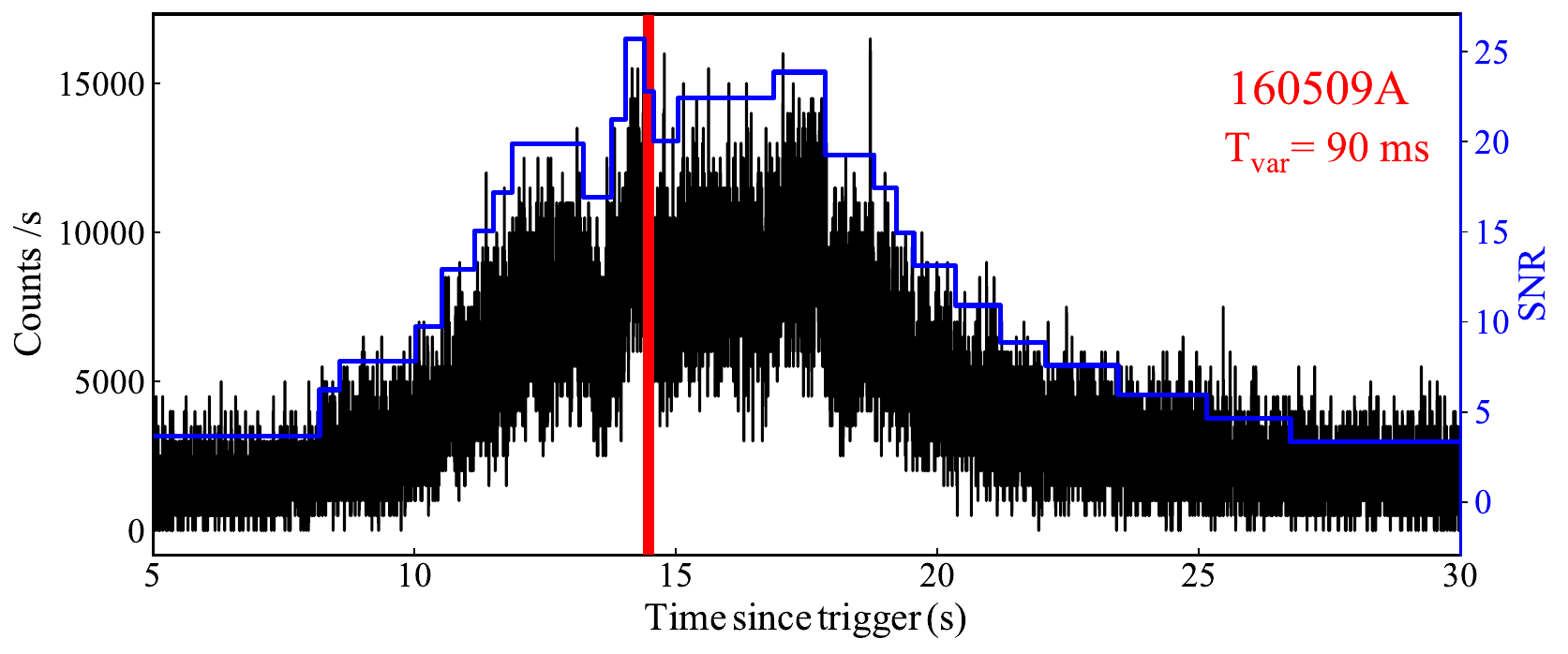}
\includegraphics[width=0.45\textwidth]{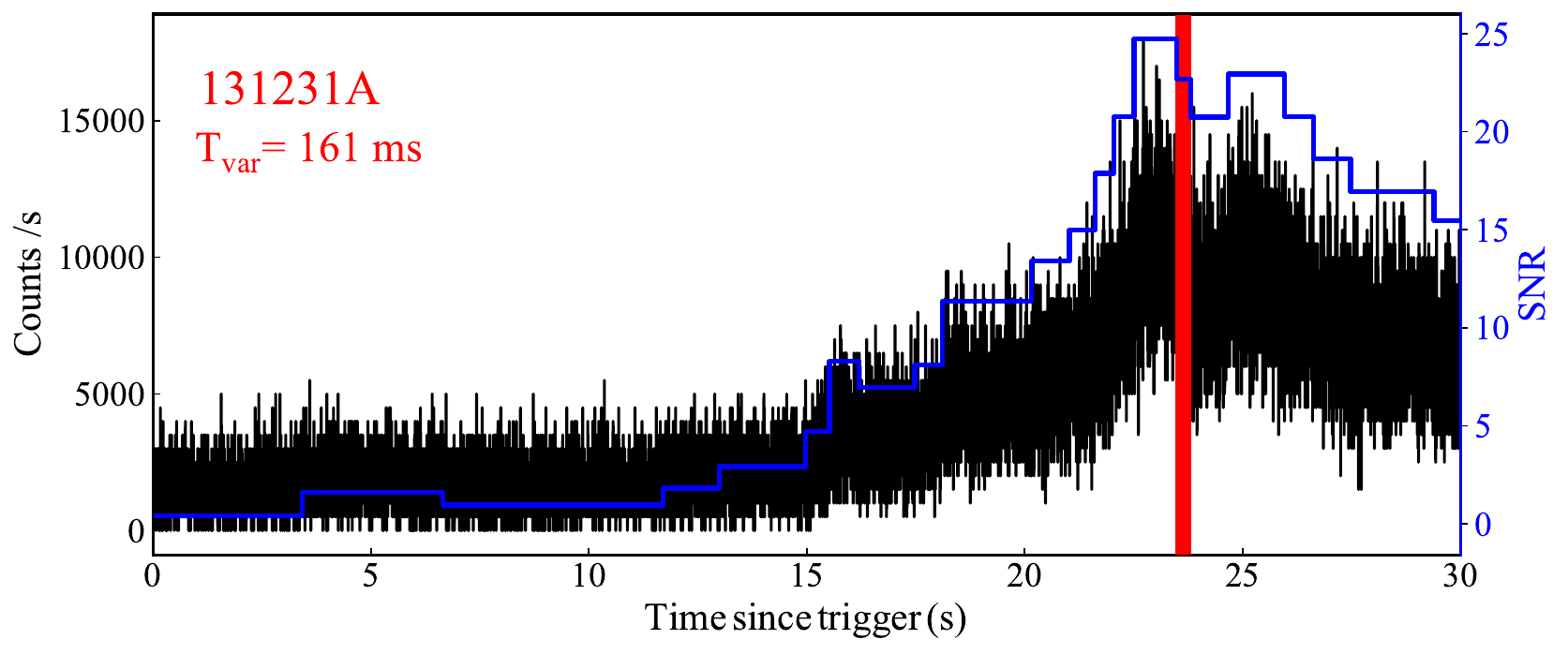}
\includegraphics[width=0.45\textwidth]{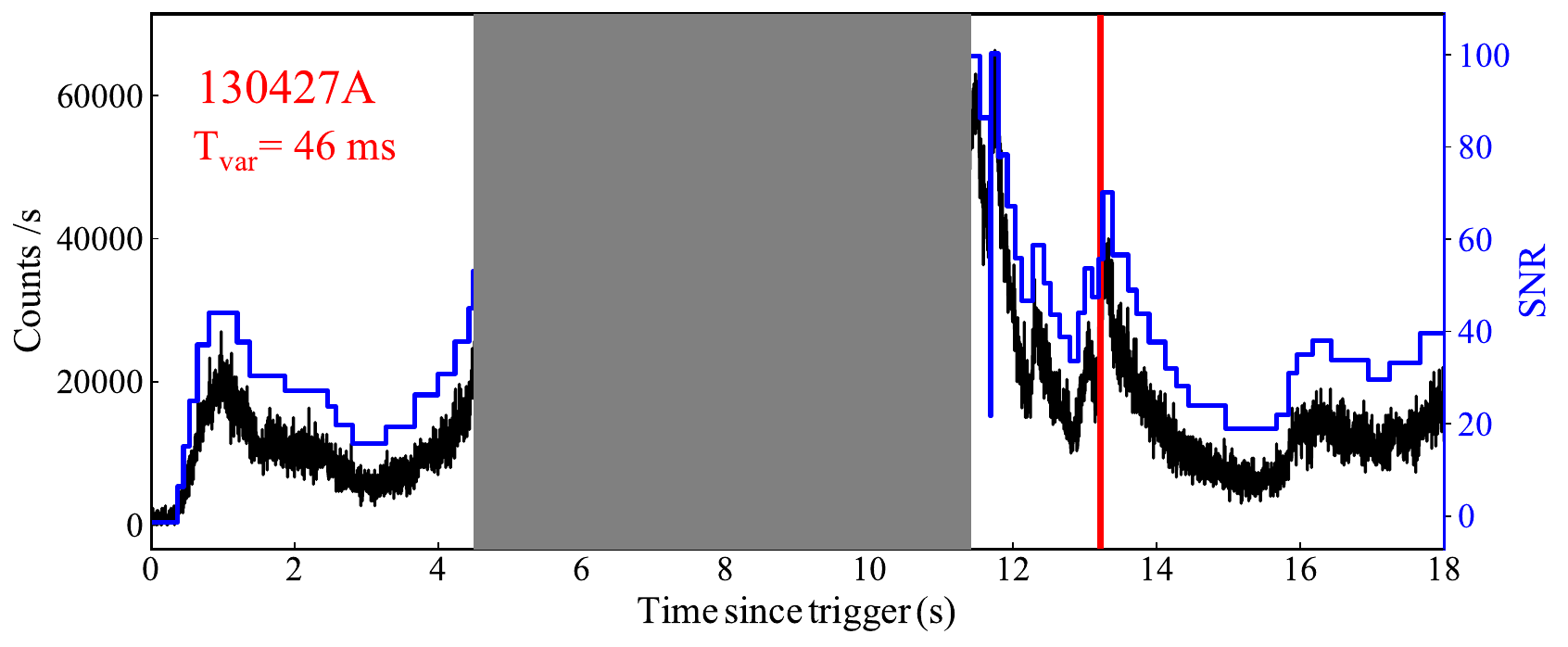}
\includegraphics[width=0.48\textwidth]{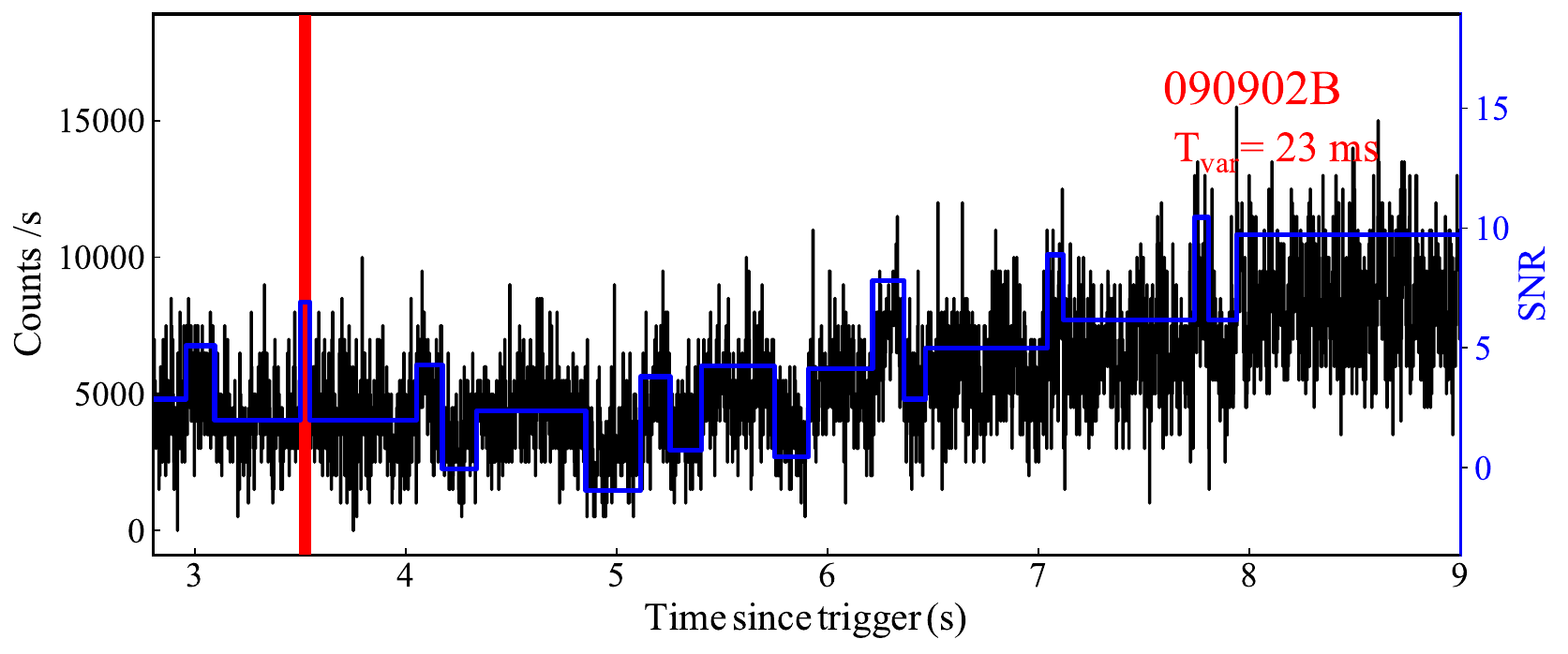}
\caption{The background subtracted light curves in 8.0--900.0 keV extracted from the TTE data of different GRBs. The blue solid lines and red shadows correspond to the Bayesian block light curves and the minimum bin sizes of the obtained blocks, respectively. The grey shadow in GRB~130427A is the bad time interval.}
\label{var-time}
\end{figure}

To determine the timescale of variability in these GRBs, we use the Bayesian block method \citep{scargle2013studies} on TTE data with a time-bin resolution of 1-3~ms for selected GRBs. The results are presented in Figure~\ref{var-time}. The values of $T_{\text{var}}$ correspond to the half of the minimum bin sizes of the obtained blocks ,which are 174~ms, 32~ms, 16~ms, 18~ms, 275~ms, 49~ms, 41~ms, 90~ms, 161~ms, and 23~ms for GRB~211018A, GRB~190530A, GRB~190114C, GRB~180720B, GRB~170214A, GRB~160821A, GRB~160625B, GRB~160509A, GRB~131231A, and GRB~090902B, respectively. It is approximately 82~ms for GRB~221009A \citep{liu2023constraints}. For GRB~130427A, characterized by its remarkable luminosity, we consider half of the average bin size within the selected time interval and obtain $T_{\text{var}}=46$~ms, aligning with previous results, \cite{ackermann2014fermi}.

\bibliography{sample631}{}
\bibliographystyle{aasjournal}

\begin{table*}
\addtolength{\tabcolsep}{0pt}
\caption{Time-resolved spectral fits of LAT and GBM for GRB~221009A. The average baryon factor is 23.2.}  
\label{tab:lat-221009A}
\centering                         
\begin{tabular}{cccccccccc}       
\hline\hline                  
Sr. no. & Intervals & $\Gamma_{\text{LAT}}^{\mathrm{b}}$ & $\rm F_{\text{LAT}}^{c}$& TS$^{\mathrm{a}}$&$\alpha$&$\beta$&$E_{\text{peak}}$&$F_{\text{GBM}}^{d}$&$\eta_{p}^{\mathrm{UL}}$\\
&(s, from $T_{0}$)&&&&&&(keV)&\\
\hline 
1&177.0 - 192.0&$2.0^{*}$&3.27&$\sim0$&-1.12 $\pm$ 0.01&-2.14  $\pm$ 0.02&549  $\pm$ 5&40.4 $\pm$ 1.1&2.8\\
2&192.0 - 205.0&$2.0^{*}$&2.65&0.87&-1.39 $\pm$ 0.01&-3.91  $\pm$ 0.27&354  $\pm$ 4&27.3 $\pm$ 0.5&3.5\\
3&205.0 - 219.0&$2.0^{*}$&81.73&$\sim0$&-1.38 $\pm$ 0.01&-2.03  $\pm$ 0.01&623  $\pm$ 7&73.1 $\pm$ 0.8&42\\
\hline
\end{tabular}
\begin{flushleft}
$^{\mathrm{a}}$ TS value of each interval; the significance of the GRB is approximate to $\sqrt{\mathrm{TS}}~\sigma$ ; the TS value of the interval that is less than 25 will be estimated as a 95\% C.L. UL (labeled with UL)\\
$^{\mathrm{b}}$ The photon index. ULs are calculated with a photon index with 2.0 (labeled with *)\\
$^{\mathrm{c}}$ The unit of $10^{-7}$ erg~cm$^{-1}$~s$^{-1}$. \\
$^{\mathrm{d}}$ The unit of $\times10^{-6}$ erg~cm$^{-1}$~s$^{-1}$. \\
\end{flushleft}
\end{table*}

\begin{table*}
\addtolength{\tabcolsep}{-3pt}
\caption{Similar to Table~\ref{tab:lat-221009A}, but for GRB~211018A. The average baryon factor is 131.}  
\label{tab:lat-211018A}
\centering                         
\begin{tabular}{cccccccccc}       
\hline\hline                  
Sr. no. & Intervals & $\Gamma_{\text{LAT}}^{\mathrm{b}}$ & $\rm F_{\text{LAT}}^{c}$& TS$^{\mathrm{a}}$&$\alpha$&$\beta$&$E_{\text{peak}}$&$F_{\text{GBM}}^{d}$&$\eta_{p}^{\mathrm{UL}}$\\
&(s, from $T_{0}$)&&&&&&(keV)&\\
\hline 
1&4.3 - 45.6&1.71 $\pm$ 0.30&1.85 $\pm$ 0.88&96&-0.54 $\pm$ 0.04&-2.76  $\pm$ 0.45&314  $\pm$ 7&1.4 $\pm$ 0.5&173\\
2&45.6 - 86.9&2.26 $\pm$ 0.49&0.77 $\pm$ 0.41&82&-0.50 $\pm$ 0.04&-3.18  $^{+0.43}_{-6.80}$&469  $\pm$ 9&2.3 $^{+0.42}_{-0.36}$&46.3 \\
3&86.9 - 128.2&1.66 $\pm$ 0.34&2.05 $\pm$ 1.05&66&-0.92 $\pm$ 0.04&-2.74  $^{+0.44}_{-7.26}$&351  $\pm$ 16&1.1 $^{+0.31}_{-0.27}$&253\\
\hline
\end{tabular}
\begin{flushleft}
\end{flushleft}
\end{table*}

\begin{table*}
\addtolength{\tabcolsep}{-3pt}
\caption{Similar to Table~\ref{tab:lat-221009A}, but for GRB~190530A. The average baryon factor is 2.8.}  
\label{tab:lat-190530A}
\centering                         
\begin{tabular}{cccccccccc}       
\hline\hline                  
Sr. no. & Intervals & $\Gamma_{\text{LAT}}^{\mathrm{b}}$ & $\rm F_{\text{LAT}}^{c}$& TS$^{\mathrm{a}}$&$\alpha$&$\beta$&$E_{\text{peak}}$&$F_{\text{GBM}}^{d}$&$\eta_{p}^{\mathrm{UL}}$\\
&(s, from $T_{0}$)&&&&&&(keV)&\\
\hline 
1&7.8 - 11.9&$2.0^{*}$&8.80&18&-1.09 $\pm$ 0.01&-2.85  $\pm$ 0.20&852  $\pm$ 19&29.2 $\pm$ 1.8&5.6\\
2&11.9 - 16.0&-3.91  $\pm$ 0.82&2.5 $\pm$ 0.9&23&-0.90 $\pm$ 0.01&-5.76$\pm$2.02&1010  $\pm$ 14&46.7 $\pm$ 2.3&1.6\\
3&16.0 - 20.2&$2.0^{*}$&5.92&8&-0.74 $\pm$ 0.01&-2.80  $\pm$ 0.10&462  $\pm$ 15&43.2 $\pm$ 0.5&2.3\\
\hline
\end{tabular}
\begin{flushleft}
\end{flushleft}
\end{table*}

\begin{table*}
\addtolength{\tabcolsep}{-3pt}
\caption{Similar to Table~\ref{tab:lat-221009A}, but for GRB~180720B. The average baryon factor is 1.6.}  
\label{tab:lat-180720B}
\centering                         
\begin{tabular}{cccccccccc}       
\hline\hline                  
Sr. no. & Intervals & $\Gamma_{\text{LAT}}^{\mathrm{b}}$ & $\rm F_{\text{LAT}}^{c}$& TS$^{\mathrm{a}}$&$\alpha$&$\beta$&$E_{\text{peak}}$&$F_{\text{GBM}}^{d}$&$\eta_{p}^{\mathrm{UL}}$\\
&(s, from $T_{0}$)&&&&&&(keV)&\\
\hline 
1&4.3 - 14.5&3.70 $\pm$ 1.15&0.19 $\pm$ 0.12&47&-0.96 $\pm$ 0.01&-2.30  $\pm$ 0.06&660  $\pm$ 11&19.3 $\pm$ 1.1&0.2\\
2&14.5 - 24.7&$2.0^{*}$&3.12&24&-1.09 $\pm$ 0.01&-2.32  $\pm$ 0.06&547  $\pm$ 9&17.3 $\pm$ 1.0&3.2\\
3&24.7 - 35.0&3.93 $\pm$ 1.00&0.39 $\pm$ 0.17&111&-1.25 $\pm$ 0.01&$-3.65^{+1.3}_{-6.3}$&402  $\pm$ 15&$4.1^{+1.4}_{-0.6}$&1.7\\
\hline                                   
\end{tabular}
\end{table*}

\begin{table*}
\addtolength{\tabcolsep}{-3pt}
\caption{Similar to Table~\ref{tab:lat-221009A}, but for GRB~170214A. The average baryon factor is 8.4.}  
\label{tab:lat-170214A}
\centering                         
\begin{tabular}{cccccccccc}       
\hline\hline                  
Sr. no. & Intervals & $\Gamma_{\text{LAT}}^{\mathrm{b}}$ & $\rm F_{\text{LAT}}^{c}$& TS$^{\mathrm{a}}$&$\alpha$&$\beta$&$E_{\text{peak}}$&$F_{\text{GBM}}^{d}$&$\eta_{p}^{\mathrm{UL}}$\\
&(s, from $T_{0}$)&&&&&&(keV)&\\
\hline
1&12.5 - 25.6&$2.0^{*}$&0.44&3&-0.94 $\pm$ 0.02&-2.28  $\pm$ 0.18&512  $\pm$ 17&3.4 $\pm$ 0.7&16.0\\
2&25.6 - 38.7&$2.0^{*}$&0.27&$\sim0$&-0.96 $\pm$ 0.02&-2.11  $\pm$ 0.11&425  $\pm$ 14&3.3 $\pm$ 0.6&5.4\\
3&38.7 - 52.0&6.07 $\pm$ 1.00&0.25 $\pm$ 0.09&131&-0.86 $\pm$ 0.02&-2.34  $\pm$ 0.14&597  $\pm$ 15&5.3 $\pm$ 0.8&5.3\\
\hline                                   
\end{tabular}
\end{table*}

\begin{table*}
\addtolength{\tabcolsep}{-3pt}
\caption{Similar to Table~\ref{tab:lat-221009A}, but for GRB~160821A.The average baryon factor is 0.64.}  
\label{tab:lat-160821A}
\centering                         
\begin{tabular}{cccccccccc}       
\hline\hline                  
Sr. no. & Intervals & $\Gamma_{\text{LAT}}^{\mathrm{b}}$ & $\rm F_{\text{LAT}}^{c}$& TS$^{\mathrm{a}}$&$\alpha$&$\beta$&$E_{\text{peak}}$&$F_{\text{GBM}}^{d}$&$\eta_{p}^{\mathrm{UL}}$\\
&(s, from $T_{0}$)&&&&&&(keV)&\\
\hline  
1&118.5 - 132.8&9.44 $\pm$ 1.10&0.05 $\pm$ 0.04&29&-0.91 $\pm$ 0.01&-2.13  $\pm$ 0.03&788  $\pm$ 11&23.2 $\pm$ 1.2&0.12\\
2&132.8 - 146.8&6.05 $\pm$ 0.89&0.99 $\pm$ 0.18&307&-0.95 $\pm$ 0.01&-2.22  $\pm$ 0.03&900  $\pm$ 10&42.7 $\pm$ 1.4&0.70\\
3&146.8 - 161.5&$2.0^{*}$&0.56&7&-1.14 $\pm$ 0.01&-3.17  $\pm$ 0.43&453  $\pm$ 14&4.7 $\pm$ 0.6&2.65\\
\hline                                   
\end{tabular}
\end{table*}

\begin{table*}
\addtolength{\tabcolsep}{-3pt}
\caption{Similar to Table~\ref{tab:lat-221009A}, but for GRB~160625B. The average baryon factor is 12.8.}  
\label{tab:lat-160625B}
\centering                         
\begin{tabular}{cccccccccc}       
\hline\hline                  
Sr. no. & Intervals & $\Gamma_{\text{LAT}}^{\mathrm{b}}$ & $\rm F_{\text{LAT}}^{c}$& TS$^{\mathrm{a}}$&$\alpha$&$\beta$&$E_{\text{peak}}$&$F_{\text{GBM}}^{d}$&$\eta_{p}^{\mathrm{UL}}$\\
&(s, from $T_{0}$)&&&&&&(keV)&\\
\hline 
1&188.4 - 192.3&3.03 $\pm$ 0.24&8.44 $\pm$ 1.32&1373&-0.87 $\pm$ 0.01&-2.26  $\pm$ 0.03&865  $\pm$ 11&71.9 $\pm$ 3.2&16.2\\
2&192.3 - 196.2&2.65 $\pm$ 0.35&3.28 $\pm$ 1.02&142&-0.84 $\pm$ 0.01&-2.68  $\pm$ 0.08&538  $\pm$ 6&39.5 $\pm$ 2.1&14.1\\
3&196.2 - 200.0&3.73 $\pm$ 0.39&1.74 $\pm$ 0.47&130&-0.86 $\pm$ 0.01&-2.63  $\pm$ 0.07&641  $\pm$ 8&45.3 $\pm$ 2.3&6.2\\
\hline  
\end{tabular}
\end{table*}

\begin{table*}
\addtolength{\tabcolsep}{-3pt}
\caption{Similar to Table~\ref{tab:lat-221009A}, but for GRB~160509A. The average baryon factor is 11.2.}  
\label{tab:lat-160509A}
\centering                         
\begin{tabular}{ccccccccccc}       
\hline\hline                  
Sr. no. & Intervals & $\Gamma_{\text{LAT}}^{\mathrm{b}}$ & $\rm F_{\text{LAT}}^{c}$& TS$^{\mathrm{a}}$&$\alpha$&$\beta$&$E_{\text{peak}}$&$F_{\text{GBM}}^{d}$&$\eta_{p}^{\mathrm{UL}}$\\
&(s, from $T_{0}$)&&&&&&(keV)&\\
\hline 
1&8.2 - 18.8&4.12 $\pm$ 0.33&3.50 $\pm$ 0.43&640&-0.90 $\pm$ 0.01&-2.16  $\pm$ 0.04&504  $\pm$ 7&18.2 $\pm$ 1.0&7.6\\
2&18.8 - 29.4&2.70 $\pm$ 0.30&2.03 $\pm$ 0.52&194&-1.00 $\pm$ 0.02&-2.00  $\pm$ 0.00&250  $\pm$ 7&4.7 $\pm$ 0.2&17.8\\
3&29.4 - 40.0&2.54 $\pm$ 0.51&0.62 $\pm$ 0.31&64&-1.11 $\pm$ 0.07&-4.07  $^{+2.05}_{-5.93}$ &181  $\pm$ 25&0.30 $^{+0.50}_{-0.02}$&131\\
\hline                                   
\end{tabular}
\end{table*}

\begin{table*}
\addtolength{\tabcolsep}{-3pt}
\caption{Similar to Table~\ref{tab:lat-221009A}, but for GRB~131231A. The average baryon factor is 2.8.}  
\label{tab:lat-131231A}
\centering                         
\begin{tabular}{cccccccccc}       
\hline\hline                  
Sr. no. & Intervals & $\Gamma_{\text{LAT}}^{\mathrm{b}}$ & $\rm F_{\text{LAT}}^{c}$& TS$^{\mathrm{a}}$&$\alpha$&$\beta$&$E_{\text{peak}}$&$F_{\text{GBM}}^{d}$&$\eta_{p}^{\mathrm{UL}}$\\
&(s, from $T_{0}$)&&&&&&(keV)&\\
\hline
1&13.2 - 18.9&$2.0^{*}$&0.65&$\sim 0$&-0.95 $\pm$ 0.02&-2.03  $\pm$ 0.05&225  $\pm$ 6&4.7 $\pm$ 0.5&3.1\\
2&18.9 - 24.5&$2.0^{*}$&1.43&$\sim 0$&-0.97 $\pm$ 0.01&-3.95  $\pm$ 0.62&328  $\pm$ 4&9.4 $\pm$ 0.6&3.3\\
3&24.5 - 30.0&5.07 $\pm$ 1.87&0.28 $\pm$ 0.16&32&-1.04 $\pm$ 0.01&-3.96  $\pm$ 0.32&165  $\pm$ 2&6.4 $\pm$ 0.2&2.0\\
\hline                                   
\end{tabular}
\end{table*}

\begin{table*}
\addtolength{\tabcolsep}{-3pt}
\caption{Similar to Table~\ref{tab:lat-221009A}, but for GRB~090902B. Note that we used SBPL pulsing PL function to fit the soft photon field for the intervals $T_{0}+[7.0,~9.0]~\rm s$ and the best fitting parameters are $\lambda_{1}=0.14\pm0.02$, $\lambda_{2}=-4.13\pm0.10$, $\Lambda=0.41\pm 0.01$, $E_{b}=911\pm16~\rm keV$ and $\Gamma_{\rm GBM}=1.86\pm 0.02$. The average baryon factor is 59.}  
\label{tab:lat-090902B}
\centering                         
\begin{tabular}{ccccccccccc}       
\hline\hline                  
Sr. no. & Intervals & $\Gamma_{\text{LAT}}^{\mathrm{b}}$ & $\rm F_{\text{LAT}}^{c}$& TS$^{\mathrm{a}}$&$\alpha$&$\beta$&$E_{\text{peak}}$&$F_{\text{GBM}}^{d}$&$\eta_{p}^{\mathrm{UL}}$\\
&(s, from $T_{0}$)&&&&&&(keV)&\\

\hline
1&2.8 - 4.9&1.89$\pm$ 0.26&7.45$\pm$0.41&75&-0.31 $\pm$ 0.04&-5.30$\pm$3.04&640 $\pm$ 14&8.7$\pm$6.9&95.6\\
2&4.9 - 7.0&2.53$\pm$0.40&4.54$\pm$1.74&63&-0.38 $\pm$ 0.02&-3.93$\pm$2.02&953 $\pm$ 20&17.4$\pm$17.0&46.1\\
3&7.0 - 9.0&2.96$\pm$ 0.10&15.0$\pm$2.0&445&-&-&-&33.7$\pm$19.8&56.6\\
\hline                                   
\end{tabular}
\end{table*}

\end{document}